\def\deltabarf{\delta\hspace{-.5mm}\raisebox{4.6mm}{{\special{em:graph
opens.pcx}}\,\,\,}}

\def\deltabarf{{\sla{\hspace{.5pt}\delta}}}
\def\deltabarf{{\bar\delta}}
\def\deltabarf{{\delta^S}}

\def\deltabar{~{\raisebox{.35em}{-}\hspace{-.44em}\delta }}

\def\nablab{{\mbox{\twelvembsy\symbol{'162}}}}
\def\nablab{{\mbox{\boldmath$\nabla$}}}
\def\deltab{{\mbox{\boldmath$\delta$}}}

\def\aut#1{#1}
\def\ins#1{}

\def\rms{}
\def\s{ \sigma}

\def\comment#1{}

\def\cm#1{}

\newcommand{\sfrac}[2]{\raisebox{0.095ex}{\footnotesize${\frac{#1}{#2}}$}}
 \def\lfrac#1#2{{{{#1}/{#2}}}}

\def\>{\rangle}
\def\<{\langle}
\def\comment#1{}
\def\ind#1{#1}

\documentclass[pra,aps]{article}
\usepackage{graphics}

\begin{document}

\title{Nonholonomic Mapping Principle
for Classical and Quantum Mechanics
in Spaces with Curvature and Torsion
\thanks{
kleinert@physik.fu-berlin.de,
{}~ http://www.physik.fu-berlin.de/\~{}kleinert }}
\author{
Hagen Kleinert\\
Institute for Theoretical Physics, FU-Berlin,
Arnimallee 14, D-14195, Berlin, Germany
}
\maketitle
\begin{abstract}
I explain the geometric basis for
the recently-discovered
nonholonomic mapping
principle which permits
deriving laws of nature in spacetimes with curvature and torsion
from those in flat spacetime, thus replacing and
extending
Einstein's equivalence principle.
As an important consequence,
it yields a new
action
principle
for
determining the equation of motion of a free spinless point particle
in such spacetimes. Surprisingly,
this equation
contains a torsion force,
although the
action involves only the metric.
This force
 makes
trajectories
autoparallel rather than geodesic, as a
 manifestation of inertia.
A  generalization
of the mapping principle transforms
path integrals
from flat spacetimes to those
with curvature and torsion,
thus playing the role of
a {\em quantum equivalence principle\/}.
This generalization yields consistent results
only for completely antisymmetric or
for gradient torsion.
\end{abstract}

\section{Introduction}

Present generalizations of Einstein's theory
of
gravity
to spacetimes with torsion proceed by
setting up model actions
in which
gravity is coupled minimally to
matter,
and deriving field equations
from extrema of these actions
\cite{Utiyama,Kibble,Hehl1,hojman,Hehl2,Kleinert1}.
So far, there is no way
of verifying experimentally  the correctness of such theories
due to the smallness of torsion effects
upon  gravitating matter.
The presently popular field equations
are a straightforward extension
of Einstein's
equation in which
the Einstein-Cartan tensor is proprtional to the
energy-momentum tensor of matter.
When forming a spacetime derivative of these
equations,
the purely geometric
Bianchi identity
for the Einstein-Cartan tensor
which expresses the single-valuedness of the connection
is balanced by the conservation law
for the
energy-momentum tensor. For spinless
 point particles, this
law
yields
directly
the trajectories
of such particles, which turn out to be
geodesics \cite{Hehl3}, the shortest paths in spacetime.
The appeal
of the mathematics
and the
success of the original
Einstein equation
left little doubt as to the physical correctness of this
result.

In this paper
I shall try to convince the reader
that the result is nevertheless physically incorrect, and that
spinless
particles move on autoparallels after all, thus calling for a revision of
the field equations.
My conclusions
are
derived from a study of point mechanics in
a given spacetime with curvature and torsion,
leaving the origin of the geometry
open.
The equations of motion
imply
a
simplified covariant conservation law for the energy momentum tensor,
which
is no longer completely analogous to the Bianchi
identity, thus preventing me from writing
down a field equation as usual, a problem which is left to the future.

My conclusions are based on a
careful
reanalysis
of the action
principle
in
spacetimes with torsion.
Due to the fact that
in the presence of torsion,
parallelograms are in general not closed but exhibit a closure
failure proportional to the
torsion, the standard variational procedure
for finding the extrema of the action
must be modified.   Whereas usually, paths
are varied keeping the endpoints fixed, such that
variations form closed paths,
the closure failure
makes the variation at the
final
point nonzero, and this gives rise
to a torsion force.

In quantum mechanics, the nonholonomic mapping principle
was essential for solving  the path integral of the hydrogen atom.
Its time-sliced version has existence problems,
but a nonholonomic coordinate transformation to
a space with torsion makes it harmonic and solvable.
In the absence of truly gravitating
systems with torsion, the hydrogen atom in that description
may serve as a testing ground for theories with torsion.

\section{New Equivalence Principle}

Some time ago it was pointed out
\cite{Kleinert2,Kleinert3,Kleinert4,Kleinert7},
that Einstein's rules
for finding
correct equations of motion
in spacetimes with curvature can be
replaced by
a more efficient {\it nonholonomic mapping principle},
which has additional predictive power by
being applicable also
in the presence of torsion.
This new
principle was originally discovered
for the purpose
of
transforming
 nonrelativistic path integrals correctly from flat spacetime
to spacetimes with torsion \cite{Kleinert4}. In that context it appeared as
a
{\em quantum equivalence
principle\/}.
Evidence for its correctness was derived from
its essential role
in solving the path integral
of the hydrogen atom via a
nonholonomic
Kustaanheimo-Stiefel transformation
\cite{Kleinert4}.   \\

Recall that Einstein
found the laws of nature
in curved space via
the following two steps. First, he went
from
rectilinear
coordinates $x^a \, ( a = 0 , 1 , 2 , 3 )$ to arbitrary curvilinear
ones $q^{\lambda}
\, ( \lambda = 0 , 1 , 2 , 3 )$
by a coordinate transformation
\begin{equation}
\label{GL}
x^a = x^a ( q ).
\end{equation}
This brought
the
flat Minkowski metric
\begin{eqnarray} \label{4.280}
 \eta_{ ab} & = & \left(
\begin{array}{rrrr}
 1 &&&\\
 & -1 &&\\
 && -1 &\\
 &&&-1
\end{array}\right)_{ ab }
\label{Mink}\end{eqnarray}
to the induced metric
\begin{equation}
\label{IN}
g_{\lambda\mu} ( q ) =  e^{\,a}_{\,\,\,\lambda} ( q )
e^{\,b}_{\,\,\,\mu} ( q ) \eta_{ab} \, ,
{}~~~~~e^{\,a}_{\,\,\,\lambda} ( q ) \equiv \partial x^a ( q )
/ \partial q^{\lambda},
\end{equation}
with the same flat geometry as before, only
parametrized in an arbitrary way.
Here Einstein {\em postulated\/}
that when written in such generalized coordinates,
the flat-spacetime laws of nature remain valid
in spacetimes with curvature.

\begin{figure}[tbh]
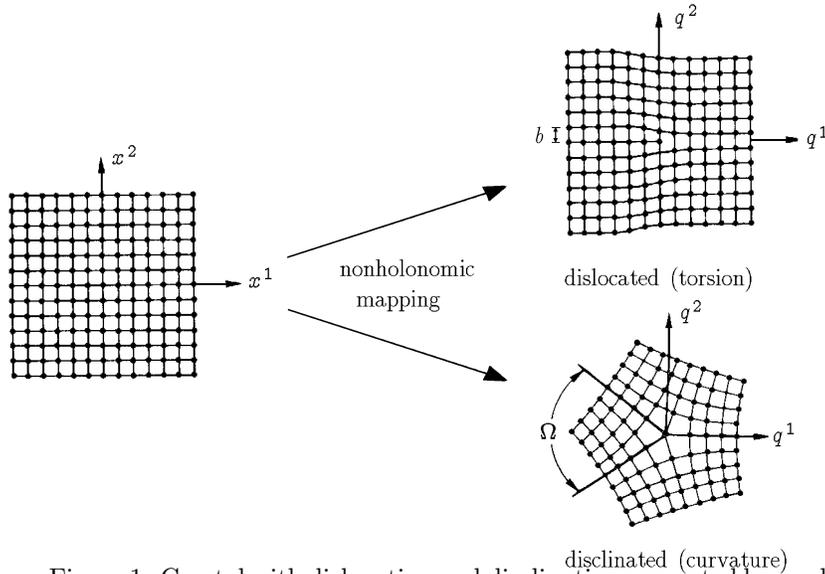

\phantom{xxxxxxxxxxxxxxxxxxxxxx}
\input dislocc2.tdf
\caption[]{Crystal with dislocation and disclination generated by nonholonomic
coordinate transformations from an ideal crystal.
Geometrically, the former transformation introduces torsion and no curvature,
 the latter
curvature and no torsion.}
\label{torcur}\end{figure}
The new formulation and extension of this
procedure \cite{Kleinert4}
was inspired by a standard
technique in describing line-like topological defects in crystals
\cite{Kleinert1,Bilby,Kroener1,Kroener2}.
In that context it was recognized, that
crystalline defects may be generated via a thought experiment,
a so-called {\em Volterra process\/}, in which layers or sections
of matter are cut from
a crystal, with a subsequent smooth rejoining of the cutting surfaces
(see Fig.~\ref{torcur}).

Mathematically, this cutting and joining may be described by
{\em active nonholonomic mappings\/} of the next-neighbor
atomic distance vectors.
Since there are missing or excess atoms in the image space,
the mapping is not integrable
to a global coordinate transformation
(\ref{GL}). Instead, it is described by
a
 local transformation
\begin{equation}
\label{LO}
d x^{a} \, = \, e^{\,a}_{\,\,\,\lambda} ( q ) \, d q^{\lambda} \, ,
\end{equation}
whose
coefficients $e^{\,a}_{\,\,\,\lambda} ( q )$
have a nonvanishing curl
\begin{equation}
\label{SCHWARZ}
\partial_{\mu} \, e^{\,a}_{\,\,\,\lambda} ( q ) \, - \,
\partial_{\lambda}\, e^{\,a}_{\,\,\,\mu} ( q ) \neq  0 \,,
\end{equation}
 implying that
any candidate for a coordinate transformation
$x^a(q)$ corresponding to (\ref{LO})
must
disobey the integrability conditions of Schwarz, i.e.,
its second derivatives do not commute:
\begin{equation}
(\partial _\mu\partial _ \nu-
\partial _\nu\partial _ \mu )
x^{\, a} ( q )\neq0.
\label{nonc}
\end{equation}
The functions $x^a(q)$ must therefore be {\em multivalued\/},
thus being no proper
functions of mathematical textbooks,
which require them to be
single-valued.
We shall see that  such functions
are the ideal tools
for constructing
the
 {\em nonholonomic\/} coordinate transformations
which carry theories
in flat space to spaces with curvature
and torsion.
It is therefore important
to learn how to handle such functions.

As a matter of fact,
the multivaluedness of the coordinate transformations
$x^a(q)$ implied by (\ref{nonc}) is not enough to describe all topological
defects
in a crystal. Also the coefficient functions
$ e^{\,a}_{\,\,\,\lambda} ( q )$ themselves
will    have to
 violate the Schwarz criterion
by having noncommuting derivatives
\cite{Kleinert1}:
\begin{equation}
\label{SCHWARZe}
(\partial_{\mu} \partial _ \nu-
\partial_{\nu} \partial _ \mu )
 e^{\,a}_{\,\,\,\lambda} ( q ) \neq 0.
\end{equation}
They are called {\em multivalued basis tetrads\/} \cite{Kleinert1x}.
The multivaluedness distinguishes them
in an essential way
from the similar-looking objects
 well-known {\em tetrad\/} or  {\em vierbein\/} formalism
used in the standard literature on gravity to be found in all major textbooks
(for instance \cite{Kleinert1x}).
In contrast to our basis tetrads, those are single-valued.
The difference will be explained below and in more detail
in Section \ref{MINF}.


As in the standard tetrad formalism,
the induced metric (\ref{IN})
can be used to introduce
{\em  reciprocal multivalued
tetrads\/}
\begin{equation}
e_  a{}^ \mu(q)\equiv   \eta_{ab}
g^{ \mu\nu}(q)e^ b{}_\nu(q).
\label{recipre}\end{equation}
The satisfy the orthogonality and completeness relations
\begin{equation}
\label{ORTHO}
e_{a}^{\,\,\,\lambda} ( q ) \, e^{\,a}_{\,\,\,\mu} ( q ) \, = \,
\delta^{\,\lambda}_{\,\,\,\mu}\, ,~~~~~~~
e_{a}^{\,\,\,\lambda} ( q ) \, e^{\,b}_{\,\,\,\mu} ( q ) \, = \,
\delta_{\,a}{}^{\,b}\, .
\end{equation}

Parallel
transport of a vector field
is defined by a
vanishing covariant derivative
\begin{equation}
D_\mu v_ \nu(q)=\partial _\mu v_ \nu(q)- \Gamma_{\mu \nu}{}^ \lambda (q)v_
\lambda(q),~~~~~
D_\mu v^  \lambda(q)=\partial _\mu v^ \lambda(q)+\Gamma_{\mu \nu}{}^ \lambda
(q)
v^ \nu(q),~~~~~
\label{covderx}\end{equation}
where $ \Gamma_{\mu \nu}{}^ \lambda(q)$ is the
affine connection
\begin{equation}
\label{connection0}
 \Gamma_{\mu \nu}{}^ \lambda(q)\equiv
 e_{a}^{\,\,\,\lambda} ( q )\partial_{\mu} \, e^{\,a}_{\,\,\, \nu} ( q ) \, =-
  e^{\,a}_{\,\,\, \nu} ( q ) \, \partial_{\mu} \,e_{a}^{\,\,\,\lambda} ( q )  .
\end{equation}
Note that by definition,
the multivalued tetrads themselves
form a parallel field:
\begin{equation}
  D_ \mu e_ a {}^ \lambda(q) = 0,~~~D_ \mu e^ a {}_ \nu(q)
	   = 0,
\label{2.47}\end{equation}
implying that  the induced metric is a parallel tensor field
({\em metricity condition\/}):
\begin{equation}
  D_  \lambda g_{\mu \nu}(q)=0.
\label{2.47g}
\end{equation}

The  antisymmetric  part of the affine connection
$ \Gamma_{\mu \nu}{}^ \lambda(q)$ is defined as the torsion tensor
\begin{equation}
S_{\mu \nu}{}^ \lambda(q) \equiv \frac{1}{2}[
\Gamma_{\mu \nu}{}^ \lambda(q)-
\Gamma_{\nu \mu}{}^ \lambda(q) ].
\label{torsion0@}
\end{equation}
By expressing the right-hand side in terms of the multivalued tetrads
according to (\ref{connection0}),
\begin{equation}
S_{\mu \nu}{}^ \lambda(q)=
\frac{1}{2} e_{a}^{\,\,\,\lambda} ( q ) \left[
\partial_{\mu} \, e^{\,a}_{\,\,\, \nu} ( q ) \, - \,
\partial_{ \nu}\, e^{\,a}_{\,\,\,\mu} ( q ) \right] ,
\label{tors}\end{equation}
we see that it measures directly the violation
of the
integrability condition
as in (\ref{SCHWARZ}),
and thus the noncommutativity (\ref{nonc})
of the derivatives in front of $x^a(q)$.

While torsion measures the degree of violation of the Schwarz
integrability condition of the nonholonomic
coordinate transformations in (\ref{nonc}), the violation
of the condition in (\ref{SCHWARZe})
defines curvature tensor:
\begin{equation}
R_{\mu \nu \lambda }{}^\kappa ( q ) =
e_{\, a}^{\,\,\,  \kappa} ( q )
\left( \partial _\mu\partial _ \nu-
\partial _\nu\partial _ \mu \right)
e^{\, a}_{\,\,\,  \lambda} ( q ) \,.
\label{RC}\end{equation}
Indeed, using
(\ref{connection0}), we find
for
 $R_{\mu \nu \lambda }{}^\kappa ( q )$
the covariant curl of the connection
\begin{equation} \label{10.30}
{R_{\mu\nu\lambda}}^\kappa = \partial_\mu {\Gamma_{\nu\lambda}}^\kappa
- \partial_{\nu}{\Gamma_{\mu\lambda}}^\kappa
-
\Gamma_{\mu  \lambda}{}^  \sigma  \Gamma_{\nu \sigma}{}^\kappa
+\Gamma_{\nu  \lambda}{}^  \sigma  \Gamma_{\mu \sigma}{}^\kappa,
\end{equation}
which is the defining equation for the Riemann-Cartan curvature tensor.
By constructing, the curvature tensor is antisymmetric
in the first index pair.

In spite of the multivaluedness
of the tetrads $e^a{}_\mu(q)$,
the  metric and connection
must be single-valued so that their second derivatives commute:
\begin{eqnarray}
(\partial _\mu\partial _ \nu-
\partial _\nu\partial _ \mu )
 \Gamma_{ \sigma \tau }{} ^ \lambda( q )=0,~~~~~~~~~~
(\partial _\mu\partial _ \nu-
\partial _\nu\partial _ \mu )
g_{ \sigma \tau } ( q )=0.
\label{noncgG}
\label{@}\end{eqnarray}
In fact, these properties are the origin of the
first and second  Bianchi identities
of general relativity, respectively.

{}From the integrability condition for the metric in (\ref{noncgG})
we derive
the antisymmetry
of $R_{\mu \nu  \lambda  \kappa }$  with respect to the second index
pair, namely
\begin{eqnarray}
  R_{\mu \nu \lambda \kappa } = -R_{\mu \nu \kappa \lambda }
\label{@}\end{eqnarray}
where $R_{\mu \nu \lambda \kappa } \equiv  R_{\mu \nu \lambda }{}^\sigma
  g_{\kappa \sigma }$: from
  the definition (\ref{RC}) we calculate directly
\begin{eqnarray}
  R_{\mu \nu \lambda \kappa }+ R_{\mu \nu \kappa \lambda }
     & = & e_{a\kappa } \left( \partial _\mu
     \partial _\nu  - \partial _\nu \partial _\mu \right)
      e^a{}_\lambda  + e_{a\lambda } \left( \partial _\mu
      \partial _\nu  - \partial _\nu \partial _\mu \right)
      e^a{}_\kappa \nonumber \\
      & = & \partial _\mu  \partial _\nu
            \left( e_{a\kappa } e^a{}_\lambda \right) - \partial _\nu
            \partial _\mu \left( e_{a\kappa } e^a{}_\lambda \right)
            \nonumber \\
      & = & \left( \partial _\mu \partial _\nu - \partial _\nu
            \partial _\mu \right) g_{\lambda \kappa } =0 .
\label{2.64}\end{eqnarray}

The second Bianchi identity
follows from
the integrability condition for the affine connection
in (\ref{noncgG}) as follows.
First we simplify the algebra  by using  a vector
notation ${\bf e}_\mu$ for the basis tetrades $e^a{}_\mu$,
and defining a corresponding quantity
\begin{eqnarray}
  {\bf R}_{\sigma \nu \mu } \equiv
     \left( \partial _\sigma \partial _\nu -\partial _\nu
     \partial _\sigma \right) {\bf e}_\mu ,
\label{2.150}\end{eqnarray}
which determines the curvature tensor $R_{\sigma \nu \mu }{}^\lambda $
via the scalar product with ${\bf e}^\lambda $. Applying
 the covariant derivative gives
\begin{eqnarray}
  D_\tau {\bf R}_{\sigma \nu \mu } = \partial _\tau
       {\bf R}_{\sigma \nu \mu } - \Gamma _{\tau \sigma }{}^\kappa
       {\bf R}_{\kappa \nu \mu } -
       \Gamma _{\tau \upsilon }{}^\kappa  {\bf R}_{\sigma \nu \kappa }.
\label{2.151}\end{eqnarray}
Performing cyclic sums over $\tau \sigma \nu $ and
taking advantage of
 the trivial antisymmetry of ${\bf R}_{\sigma \nu \mu }$ in $\sigma \nu $
we find
\begin{eqnarray}
   D_\tau {\bf R}_{\sigma \nu \mu } = \partial _\tau
    {\bf R}_{\sigma \nu \mu } - \Gamma _{\tau \mu }{}^\kappa
    R_{\sigma \nu \kappa } + 2 S_{\tau \sigma }{}^\kappa
    R_{\nu \kappa \mu }.
\label{2.152}\end{eqnarray}
Now we use
\begin{eqnarray}
  \partial _\sigma \partial _\nu {\bf e}_\mu =
   \partial _\sigma  \left( \Gamma _{\nu \mu }{}^\alpha
   {\bf e}_\alpha \right) = \Gamma _{\nu \mu }{}^\kappa
    {\bf e}_\kappa
\label{2.153}\end{eqnarray}
to derive
\begin{eqnarray}
   \partial _\tau \partial _\sigma \partial _\nu e_\mu
   &  = & \partial _\tau \Gamma _{\nu \mu }{}^\kappa
          \partial _\sigma {\bf e}_\kappa + (\tau \leftrightarrow  \sigma ) +
          \partial _\tau \partial _\sigma \Gamma _{\nu \mu }{}^\kappa
          {\bf e}_\alpha + \Gamma _{\nu \mu }{}^\kappa
           \partial _\tau \partial _\sigma {\bf e}_\kappa .
\label{2.154}\end{eqnarray}
Antisymmetrizing this in $\sigma \tau $ gives
\begin{eqnarray}
 \partial _\tau \partial _\sigma \partial _\nu {\bf e}_\mu
    - \partial _\sigma \partial  _\tau \partial _\nu {\bf e}_\mu
     = \Gamma _{\nu \mu }{}^\alpha {\bf R}_{\tau \sigma \alpha }
       + \left[ \left( \partial _\tau \partial _\sigma -
        \partial _\sigma \partial _\tau \right)
        \Gamma _{\nu \mu }{}^\alpha  \right] {\bf e}_\alpha .
\end{eqnarray}\label{2.155}
This is the place where we make use of the integrability
 condition for the connection (\ref{noncgG})
to drop
the last
 term.  Together with (\ref{2.150}), we find
\begin{eqnarray}
  \partial _\tau {\bf R}_{\sigma \nu \mu } - \Gamma_{\nu \mu }{}^\alpha
  {\bf R}_{\tau \sigma \alpha } = 0
\label{2.156}\end{eqnarray}
Inserting this into (\ref{2.152}) and multiplying by ${\bf e}^\kappa$
we obtain an expression involving the covariant
derivative of the curvature tensor
\begin{eqnarray}
  D_\tau R_{\sigma \nu \mu }{}^\kappa
    - 2 S_{\tau \sigma }{}^\lambda R_{\nu \lambda \mu }{}^\kappa = 0.
\label{2.157}\end{eqnarray}
  This is the second {\em Bianchi identity\/},
 guaranteeing the integrability of the connection.

The Riemann connection is given
by the
Christoffel symbol
\begin{eqnarray}
 \bar  \Gamma_{\mu \nu \lambda}\equiv  \left\{ \mu \nu ,\lambda \right\} =
\frac{1}{2}
            \left( \partial _\mu  g_{\nu \lambda }
            + \partial _\nu  g_{\mu \lambda}  - \partial _\lambda
            g_{\mu \nu } \right).
\label{2.9}\end{eqnarray}
It forms part of the affine connection
(\ref{connection0}), as shown
by the decomposition
\begin{eqnarray}
  \Gamma _{\mu \nu \kappa } =
\bar  \Gamma _{\mu \nu \kappa }
            + K_{\mu \nu \kappa },
\label{2.78}\end{eqnarray}
   in which $ K_{\mu \nu \kappa }
$ is the {\em contortion tensor\/}, a combination of
three torsion tensors:
\begin{equation}
      K_{\mu \nu \lambda } =
      S_{\mu \nu \lambda } -
      S_{ \nu \lambda \mu} +
      S_{ \lambda \mu \nu}.
\label{contorttenb}\end{equation}
This decomposition follows directly from the trivially rewritten
expression (\ref{connection0}),
\begin{eqnarray}&&\!\!\!\!\!\!\!\!\!\!\!\!\!\Gamma _{\mu \nu \lambda
}=\frac{1}{2}\left\{
e_{i\lambda }\partial  _\mu e^i{}_\nu
+\partial _\mu e_{i\lambda } e^i{}_\nu
+e_{i\mu      }\partial  _\nu  e^i{}_\lambda
+\partial _\nu e_{i\mu } e^i{}_\lambda
-e_{i\mu     }\partial  _\lambda  e^i{}_\nu
-\partial _\lambda  e_{i\mu  } e^i{}_\nu\right\} \nonumber \\
&&\!\!\!\!\!\!\!\!+\frac{1}{2}\left\{
\left[  e_{i\lambda }\partial  _\mu e^i{}_\nu -e_{i\lambda }\partial  _\nu
e^i{}_\mu\right]
-\left[ e_{i\mu }\partial  _\nu e^i{}_\lambda  -e_{i\mu }\partial  _\lambda
e^i{}_\nu\right]
+\left[ e_{i\nu  }\partial  _\lambda  e^i{}_\mu -e_{i\nu  }\partial  _\mu
e^i{}_\lambda  \right] \right\}
 \label{10.26}
\end{eqnarray}
using
${e^i}_\mu(q) {e^i}_\nu (q)=g_{\mu\nu}(q)$.
The contortion tensor is antisymmetric in the last two indices:
\begin{equation}
K_{\mu \nu \lambda}=-
K_{\mu  \lambda \nu},
\label{antisK@}\end{equation}
this being a direct consequence of the antisymmetry of the torsion tensor
in the first two indices:
\begin{equation}
S_{\mu \nu \lambda}=-
K_{ \lambda\mu  \nu}.
\label{@}\end{equation}

It is useful to state in more detail the
differences between our multivalued
tetrads
$e^{\,a}_{\,\,\,\lambda} ( q )$
and the standard
tetrads or vierbein fields $h^{\, \alpha}_{\,\,\,  \lambda} ( q )$
whose mathematics is described in
\cite{Schouten}. Such tetrads were
introduced
 a long time
ago in gravity
theories of spinning particles
both in purely Riemann \cite{Weinberg2}
as well as in Riemann-Cartan spacetimes
\cite{Utiyama,Kibble,Hehl1,Hehl2,Kleinert1}.
Their purpose was
to define
at every point a
local Lorentz frame
by means of
another set of
coordinate differentials
\begin{equation}
dx^ \alpha=h^{\, \alpha}_{\,\,\,  \lambda} ( q )dq^ \lambda,
\label{4.4}\end{equation}
 which can be
contracted with Dirac matrices $ \gamma^ \alpha$
to form locally Lorentz invariant quantities.
Local Lorentz frames are reached by requiring
the induced metric
in these coordinates
to be Minkowskian:
\begin{equation}
 g_{ \alpha \beta}=h_ \alpha{}^\mu(q) h_ \beta{}^ \nu(q) g_{\mu \nu}(q)=
\eta_{ \alpha \beta}.
\label{ghhx}\end{equation}
Just like $e^a{}_\mu(q)$ in (\ref{recipre}), these vierbeins possess
reciprocals
\begin{equation}
h_  \alpha{}^ \mu(q)\equiv   \eta_{ \alpha \beta}
g^{ \mu  \nu}(q)h^ \beta{}_\nu(q),
\label{recix}\end{equation}
and
satisfy orthonormality and completeness relations as in
(\ref{ORTHO}):
\begin{eqnarray}
   h_\alpha {}^\mu  h^\beta{}_\mu   =  \delta _\alpha {}^\beta,~~~~~~~
   h^\alpha {}_\mu h_\alpha {}^\nu   =  \delta _\mu {}^\nu.
\label{8.7}\end{eqnarray}
They also can be multiplied with each other
as in (\ref{IN}) to yield the metric
\begin{equation}
g_{\mu \nu}(q)=
h^ \alpha{}_\mu(q)
h^  \beta{}_\mu(q) \eta_{ \alpha \beta}.
\label{8.7a@}\end{equation}
Thus they constitute another
``square root" of the metric.
The relation between these square roots
\begin{equation}
e^a{}_\mu(q)=e^a{}_ \alpha(q)h^ \alpha{}_ \mu(q)
\label{@}\end{equation}
is necessarily given by a local Lorentz transformation
\begin{equation}
 \Lambda^a{}_ \alpha(q)=
 e^a{}_ \alpha(q),
\label{loclor@}\end{equation}
since this matrix connects the two Minkowski metrics
(\ref{Mink}) and
(\ref{ghhx}) with each other:
\begin{equation}
 \eta_{ab}   \Lambda^a{}_ \alpha(q)
   \Lambda^b{}_  \beta(q)= \eta_{ \alpha \beta}.
\label{Loren@}\end{equation}
The different
local Lorentz transformations
allow us to choose
different local Lorentz frames which
distinguish
fields with definite spin
by the irreducible representations
of these transformations.
The physical consequences of the
theory must be independent of this local choice, and this is the reason why
the presence of spinning fields
requires the
existence of an additional gauge freedom
under local Lorentz transformations, in addition to
Einstein's invariance under general coordinate transformations.
Since the latter may be viewed as local translations,
the theory with spinning particles are
locally Poincar\'e invariant.

The vierbein fields $h^ \alpha{}_\mu(q)$
have in common with ours
that both violate
the integrability condition
as in (\ref{SCHWARZ}),
thus describing nonholonomic coordinates $dx^ \alpha$
for which there exists only a differential relation  (\ref{4.4}).
However, they differ from ours
by being
single-valued
fields
satisfying the integrability condition
\begin{equation}
( \partial _\mu\partial _ \nu-
\partial _\nu\partial _ \mu )
h^{\, \alpha}_{\,\,\,  \lambda} ( q )=0,
\label{4.7}\end{equation}

in contrast to our multivalued tetrads
 $e^{\, a}_{\,\,\,  \lambda} ( q )$
  in eq.~(\ref{SCHWARZe}).
%
%
%
%

In the local coordinate system $dx^ \alpha$,
curvature
arises from
a violation of the
 integrability condition
of the
local Lorentz transformations
(\ref{loclor@}),
which looks similar to
eq.~(\ref{SCHWARZ}).

Equation (\ref{tors}) for the
torsion tensor in terms of the  multivalued tetrads $e^{\,a}_{\,\,\,\lambda} (
q )$
must be contrasted with
a similar-looking, but geometrically quite different,
quantity formed from the vierbein fields
$h^  \alpha{}_ \lambda(q)$
and their reciprocals,
the
objects of anholonomy \cite{Schouten}:
\begin{equation}
\label{AN}
\Omega_{ \alpha \beta}^{\,\,\,\,\,\, \gamma} ( q ) =
\frac{1}{2}
h_{ \alpha}{}^\mu(q)
h_{ \beta}{}^\nu    (q)
\left[
\partial_{\mu}  h^{ \gamma}_{\,\,\,\nu} ( q ) -
\partial_{\nu}  h^{ \gamma}_{\,\,\,\mu} ( q )
\right].
\end{equation}
A combination of these similar to (\ref{contorttenb}),
\begin{equation}
\label{KO10}
 \mathop K^h{}_{ \alpha \beta}^{\,\,\,\,\,\, \gamma} ( q ) =
 \Omega_{ \alpha \beta}^{\,\,\,\,\,\, \gamma} ( q ) -
 \Omega_{ \beta\,\,\,\,  \alpha}^{\,\,\, \gamma} ( q ) +
 \Omega^{ \gamma}_{\,\,\,\, \alpha \beta} ( q )       ,
\end{equation}
appears in the {\em spin connection\/}
\begin{eqnarray}
   \Gamma _{ \alpha  \beta } {}^ \gamma
=  h^ \gamma {}_ \lambda  h_ \alpha {}^\mu
		 h_ \beta {}^ \nu ( K _{\mu  \nu }{}^ \lambda
		       - \mathop K^{h}{ }_{\mu  \nu }{}^{\!\! \lambda}
			  ),
\label{4.18aa}\end{eqnarray}
 which is
needed
to form a covariant derivative
of local vectors
\begin{equation}
v_ \alpha(q)=v_\mu(q) h_ \alpha{}^\mu(q)
,~~~~~~v^\alpha(q)=v^\mu(q) h^ \alpha{}_\mu(q).
\label{vcpmp}\end{equation}
The spin connection (\ref{4.18aa}) is derived in Section \ref{MINF},
where we shall find that
the covariant derivative of $ v_  \beta(q)$ is
given by
\begin{equation}
D_ \alpha v_  \beta(q)=
 \partial _ \alpha v_ \beta(q)-
 \Gamma
_{ \alpha \beta}^{\,\,\,\,\,\, \gamma}(q)
v_ \gamma(q),~~~~~~~
     D_ \alpha  v ^ \beta  (q)= \partial _\alpha   v ^ \beta (q) +
      \Gamma\rms  _{ \alpha  \gamma }{} ^\beta(q)  v ^ \gamma(q).
\label{covva}\end{equation}
In spite of the similarity between
the defining equations
(\ref{tors})  and (\ref{AN}), the tensor
$ \Omega_{ \alpha \beta}^{\,\,\,\,\,\, \gamma} ( q ) $ bears no relation to
torsion,
and $ \displaystyle\mathop K^h{}_{ \alpha \beta}{}^{\!\!\gamma} ( q )$
is independent of
the contortion $K_{ \alpha \beta}{}^ \gamma$.
In fact, the objects of anholonomy $  \Omega_{ \alpha \beta}^{\,\,\,\,\,\,
\gamma} ( q ) $
are in general nonzero
 in the absence of torsion \cite{moregen},
and may even be nonzero in flat spacetime, where
the matrices $h^ \alpha{}_\mu(q)$ degenerate
to local Lorentz
transformations.
The orientation of the local Lorentz frames
are characterized by
$  \displaystyle\mathop K^h{}_{ \alpha \beta}{}^{\!\!\gamma} ( q )$.

\comment{Seen from the Einstein-Cartan space $q^\mu$,
our coordinates
$dx^a$ are much more nonholonomic then the usual nonholonomic ones $dx^
\alpha$.
The two  are related to each other by a nonholonomic local
Lorentz transformation
\begin{equation}
dx^ \alpha= e^ \alpha{}_ \beta(q)dq^ \beta,
\label{x}\end{equation}
which has noncommuting derivatives:
\begin{equation}
( \partial _\mu\partial _ \nu-
\partial _\nu\partial _ \mu )
  e^ \alpha{}_ \beta(q)   =0.
\end{equation}
}

The nonholonomic coordinates $dx^ \alpha$ transform the metric
to a Minkowskian form at the point $q^\mu$. They
correspond to a small ``falling elevator" of Einstein
in which the gravitational forces vanish
only at the center of mass, the neighborhood still being subject to tidal
forces. In contrast,
the nonholonomic coordinates $dx^ a$ flatten the spacetime
{\em in an entire neighborhood\/}
of the point. This is at the expense of producing defects in spacetime
(like those produced when
flattening an orange peel
by stepping on it),
as will be explained in Section~IV.
The affine connection $ \Gamma_{ab}{}^c(q)$
in the latter coordinates $dx^a$ vanishes identically.

The difference between our
multivalued tetrads and the usual vierbeins is illustrated in
the diagram
of
Fig.~\ref{diagr}.
\begin{figure}[tbhp]
\input boxdia.tdf
\caption[]{The coordinate system $q^\mu$ and the two sets of local nonholonomic
coordinates
$dx^ \alpha$ and $dx^a$. The intermediate coordinates $dx^ \alpha$
have a Minkowski
metric only at the
point $q$, the coordinates $dx^a$ in an entire small neighborhood
(at the cost of a closure failure).
}
\label{diagr}\end{figure}

A long time ago it has been pointed out by Kondo \cite{Kondo}
that a crystal with dislocations and disclinations
may be described geometrically as
a Riemann-Cartan spacetime with curvature and torsion.
Turning the argument around,
active nonholonomic mappings which are used to
produce defects in crystals
may be used to carry us from a flat spacetime
to a Riemann-Cartan spacetime.
This is in contrast to passive nonholonomic coordinate transformation
of Cartesian coordinates, which
are simply an awkward and highly unrecommendable
redescription of flat spacetime.

In the sequel, we shall
use the word ``space" for spaces as well as spacetimes. for brevity.

In order to show that active nonholonomic
transformations
can be well-defined, let us first get some exercise
in using them
by studying some
completely analogous
but much simpler mathematical structures in magnetostatics.

\section[Multivalued Fields in Magnetism]
{Multivalued Fields in Magnetism\label{@MAG}}

To set the stage for the discussion, recall first
the standard treatment of magnetism.
Since there are no
 magnetic monopoles, a magnetic field
$ {\bf B}({\bf x})$ satisfies the
identity $\nablab  \cdot {\bf B}({\bf x})=0$,
implying that only two of the three field components
of $ {\bf B}({\bf x})$ are independent.
To account for this, one usually expresses
a
magnetic field ${\bf B}({\bf x})$
in terms of a vector potential ${\bf A}({\bf x})$, setting
 ${\bf B}({\bf x})=\nablab\times {\bf A}({\bf x})$.
Then Amp\`ere's law, which relates the magnetic field to
the electric current density
${\bf j}({\bf x})$ by $\nablab\times {\bf B}={\bf j}({\bf x})$ (in natural
units with $c=1$),
 becomes a second-order differential  equation for the
vector potential ${\bf A}({\bf x})$
in terms of an electric current
\begin{equation}
\nablab\times [\nablab\times {\bf A}({\bf x})]= {\bf j}({\bf x}).
\label{Amp}\end{equation}
The vector potential ${\bf A}({\bf x})$ is a {\em gauge field\/}. Given ${\bf
A}({\bf x}) $, any
locally gauge-transformed
field
\begin{equation}
{\bf A}({\bf x})\rightarrow {\bf A}'({\bf x})={\bf A}({\bf x})+\nablab
\Lambda({\bf x})
\label{gautr@}\end{equation}
yields the same magnetic field ${\bf B}({\bf x})$.
This reduces the number of physical degrees of freedom
in the gauge field    ${\bf A}({\bf x})$
to two, just as those in
${\bf B}({\bf x})$.
In order for this to hold, the transformation
function must be single-valued, i.e., it must have commuting derivatives
\begin{equation}
(\partial _i\partial _j-
\partial _j\partial _i) \Lambda({\bf x})=0.
\label{@}\end{equation}
The equation for absence of magnetic monopoles
$\nablab  \cdot {\bf B}=0$ is ensured if the vector potential has commuting
derivatives
\begin{equation}
(\partial _i\partial _j-
\partial _j\partial _i) {\bf A}({\bf x})=0.
\label{@}\end{equation}
This integrability property makes
$\nablab  \cdot {\bf B}=0$ the {\em Bianchi identity\/} in this gauge
field representation of the magnetic field.

In order to solve (\ref{Amp}),
we remove the gauge ambiguity by  choosing a particular gauge,
for instance the {\em transverse gauge\/} $\nablab \cdot {\bf A}({\bf x})=0$
in which $\nablab\times [\nablab\times {\bf A}({\bf x})] =
- \nablab^2{\bf A}({\bf x})$, and obtain
\begin{equation}
{\bf A}({\bf x})
=\frac{1}{4\pi}\int d^3x'\frac{{\bf j}({\bf x}')}{|{\bf x}-{\bf x}'|}.
\label{vecpotI}\end{equation}
The associated magnetic field is
\begin{equation}
{\bf B}({\bf x})= \frac{1}{4\pi}
\int d^3x'\,\frac{ {\bf j}( {\bf x}')\times {\bf R}'}{R'{}^3}, ~~~~~{\bf
R}'\equiv {\bf x}'-{\bf x}.
\label{mgf0}\end{equation}

This standard representation of magnetic fields is not the only possible one.
There exists another one in terms of a scalar potential
$ \Lambda({\bf x})$, which must, however, be multivalued to
account for the two physical degrees of freedom
in the magnetic field.

\subsection[Gradient Representation
of Magnetic Field of Current Loop]
{Gradient Representation of Magnetic Field of
Current Loop\label{@GRCL}}

Consider an infinitesimally  thin closed wire carrying an electric current
$I$ along the line $L$.
\begin{figure}[tbhp]
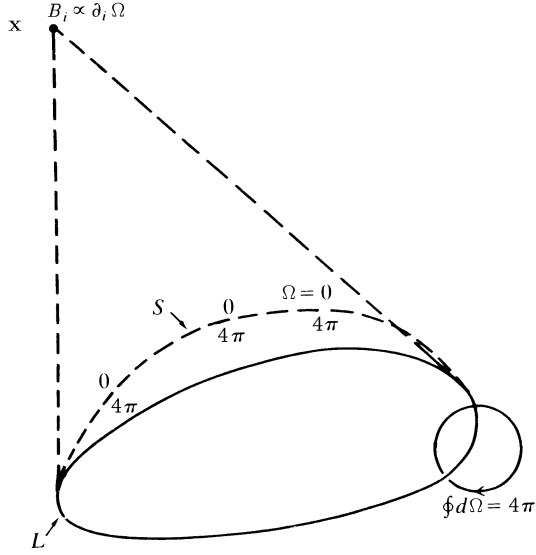

\input loop.tdf
~\\
~\\
\caption[]{Infinitesimally thin   closed current loop $L$. The magnetic field
${\bf B}({\bf x})$ at the point ${\bf x}$ is proportional to the
solid angle $ \Omega({\bf x})$  under
which the loop is seen from ${\bf x}$.
In any single-valued definition of $ \Omega({\bf x})$,
there is some  surface $S$ across which $ \Omega({\bf x})$
jumps by $4\pi$.
In the multivalued definition, this surface is absent.}
\label{loop}\end{figure}
It corresponds to a current density
\begin{equation}
{\bf j}({\bf x})= I\deltab({\bf x};L),
\label{currdensLx}\end{equation}
where $\deltab({\bf x};L) $ is the $ \delta$-function
on the line $L$:
\begin{equation}
\deltab({\bf x};L)=\int_L d {\bf x}'  \delta^{(3)}({\bf x}- {\bf x}').
\label{x}\end{equation}
{}From Eq.~(\ref{vecpotI})
we obtain the associated vector potential
\begin{equation}
{\bf A}({\bf x})
=\frac{I}{4\pi}\int_L d {\bf x}'\frac{1}
{|{\bf x}- {\bf x}'|},
\label{x}\end{equation}
yielding the magnetic field
\begin{equation}
{\bf B}({\bf x})= \frac{I}{4\pi}
\int_L\frac{ d {\bf x}'\times {\bf R}'}{R'{}^3}, ~~~~~{\bf R}'\equiv {\bf
x}'-{\bf x}.
\label{mgf}\end{equation}

Let us now derive the same result
from a multivalued scalar field.
Let $ \Omega({\bf x})$ be the solid
angle under which
the current loop $L$ is seen from the point ${\bf x}$
(see Fig.~\ref{loop}).
If
$S$ denotes an arbitrary smooth
surface enclosed by the loop $L$, and $d{\bf S}'$
a surface element, then $ \Omega({\bf x})$ can be calculated from the
surface integral
\begin{equation}
\Omega({\bf x})=\int_S \frac{d{\bf S}'\cdot {\bf R'}}{R'{}^3}.
\label{OmeS}\end{equation}
We  form
the vector field
\begin{equation}
{\bf b}({\bf x})=\frac{I}{4\pi}\nablab  \Omega({\bf x}).
\label{bomex}\end{equation}
which equal to
\begin{eqnarray}
{\bf b}({\bf x})=
\frac{I}{4\pi}
\int _S
d{ S}'_i\nablab \frac{{ R'}_i}{R'{}^3}.
\label{x}\end{eqnarray}
Using $\partial _k(R'_k/R'{}^3)=- \delta^{(3)}({\bf x}-{\bf x}')$,
it can be rewritten as
\begin{eqnarray}
{b}_i({\bf x})= \frac{I}{4\pi}
\left[
\int _S
\left(d{ S}'_k\, \partial_i\frac{ {R'_k}}{R'{}^3}-
d{ S}_i' \,\partial _k  \frac{{ R'_k}}
{R'{}^3}
\right)
-\int _S d{\bf S}' \delta^{(3)}({\bf x}-{\bf x}')
\right].
\label{x}\end{eqnarray}
With the help of Stokes' theorem
\begin{eqnarray}
\int _S(dS_k\partial _i-dS_i\partial _k)f({\bf x})= \epsilon_{kil}\int
_Ldx_lf({\bf x}),
\label{x}\end{eqnarray}
this becomes
\begin{eqnarray}
{\bf b}({\bf x})= \frac{I}{4\pi}
\left[
\int _L
\frac{d{\bf x}' \times {\bf R'}}{R'{}^3}
-\int _S d{\bf S}' \delta^{(3)}({\bf x}-{\bf x}')
\right].
\label{bomexa}\end{eqnarray}
The first term is recognized
to be precisely the magnetic field   (\ref{mgf})
of
the current $I$.
The second term is the singular magnetic
field of an infinitely thin magnetic dipole layer
lying on the arbitrarily chosen surface $S$
 enclosed by $L$.

This term is a consequence of the fact that the solid angle
$ \Omega({\bf x})$ was defined by
the surface integral (\ref{OmeS}).
If ${\bf x}$ crosses the surface $S$,
the solid angle jumps by $4\pi$.
There exists, however, another possibility of defining
the solid angle $ \Omega({\bf x})$,
namely
by its
analytic continuation from one
side of the surface to the other.
This removes the jump, albeit at the cost of making
$ \Omega({\bf x})$ a {\em multivalued function\/}
defined only modulo $4\pi$.
{}From this multivalued
function, the magnetic field (\ref{mgf})
can be obtained as a gradient:
\begin{equation}
{\bf B}({\bf x})=\frac{I}{4\pi}\nablab  \Omega({\bf x}).
\label{mgF}\end{equation}
Amp\`ere's law
(\ref{Amp}) implies that
the multivalued solid angle
$ \Omega({\bf x})$  satisfies
the equation
\begin{eqnarray}
(\partial _i\partial _j-
\partial _j\partial _i) \Omega({\bf x})=4\pi   \epsilon_{ijk}\delta_k({\bf
x};L).
\label{SCHWO}\end{eqnarray}
Thus, as a consequence of its multivaluedness,
$\Omega({\bf x})$ violates the Schwarz integrability condition
as in (\ref{nonc}).
  This makes it an unusual mathematical object to deal with.
It is, however, perfectly suited
to describe the physics.

In order to see explicitly how
Eq.~(\ref{SCHWO}) is fulfilled by $\Omega({\bf x})$,
let us go
to two dimensions where the loop
corresponds to two points (in which the loop intersects a plane).
For simplicity, we move one of them  to infinity, and place
the other at the coordinate origin.
The role of the solid angle $ \Omega({\bf x})$ is now played by the
azimuthal angle $\phi({\bf x})$ of the point ${\bf x }$:
\begin{equation}
\phi({\bf x})=\arctan \frac{x^2}{x^1}.
\label{10.40}\end{equation}
The function
$\arctan ({x^2}/{x^1})$ is usually made unique
by cutting the ${\bf x}$-plane from the origin along some line $C$ to infinity,
preferably along a straight line
to ${\bf x}=(-\infty,0)$,
and assuming $\phi({\bf x})$ to jump from $\pi$ to $-\pi$
when crossing the cut.
The cut corresponds to the magnetic dipole surface $S$
in the integral (\ref{OmeS}).
In contrast to this, we shall
take $\phi({\bf x})$ to be the {\em multivalued\/}
analytic continuation of this function.
Then
the derivative $\partial _i$ yields
\begin{equation}
 \partial _i \phi({\bf x}) =-\epsilon_{ij}\frac{x _j}{(x^1)^2+(x^2)^2}.
\label{singl}\end{equation}
With the single-valued definition
of $ \partial _i \phi({\bf x})$,
there would have been a $ \delta$-function
$ \epsilon_{ij} \delta_j(C;{\bf x})$ across the cut $C$,
corresponding to the
second term in (\ref{bomexa}).
When integrating the curl of (\ref{singl}) across the surface
$s$
of a   small
 circle $c$ around the origin,
we obtain by Stokes' theorem
\begin{equation}
\int _s d^2x    (\partial _i\partial _j-\partial _j\partial _i)\phi({\bf x})=
\int _c dx_i\partial _i \phi({\bf x}),
\label{SC}\end{equation}
which is equal to $2\pi$
in the  multivalued definition of $\phi({\bf x})$.
This result implies the violation of the integrability condition
as in (\ref{schw}):
\begin{equation}
(\partial _1\partial _2-\partial _2\partial _1) \phi({\bf x})=2\pi  \delta^2({\bf
x}),
\label{schw1}\end{equation}
whose three-dimensional generalization is
(\ref{SCHWO}).
In the single-valued definition with the jump by $2\pi$ across the cut,
the right-hand side of (\ref{SC})  would vanish,
making $\phi ({\bf x})$ satisfy
the integrability condition (\ref{nonc}).

The azimuthal angle
$\phi({\bf x})$ solving the differential equation
 (\ref{schw1}) can be used to construct a Green function for
solving the corresponding differential equation with an
 arbitrary
source, which is a  superposition of infinitesimally
thin line-like currents piercing the two-dimensional space at the points ${\bf
x}_n$:
\begin{equation}
j ({\bf x})=\sum _n I_n  \delta ({\bf x}-{\bf x}_n),
\label{@}\end{equation}
where $I_n$ are currents.
We may then easily solve the differential equation
\begin{equation}
(\partial _1\partial _2-\partial _2\partial _1) f({\bf x})=  j({\bf x}).
\label{schw1f}\end{equation}
with the help of the Green function
\begin{equation}
G({\bf x},{\bf x}')=
\frac{1}{2\pi}\phi({\bf x}-{\bf x}').
\label{schw1fG}\end{equation}
The solution of (\ref{schw1f})
is obviously
\begin{equation}
 f({\bf x})=\int d^2{\bf x}'\,G({\bf x},{\bf x}') j({\bf x}).
\label{superpos1}\end{equation}
The gradient of $ f({\bf x})$ yields the magnetic field of an
arbitrary set of line-like
 currents vertical to the plane under consideration.

It must be pointed out that
the superposition
of line-like currents cannot be smeared out
into a continuous distribution.
The integral
(\ref{superpos1}) yielsd the superposition
of multivalued functions
\begin{equation}
f({\bf x})=\frac{1}{2\pi}\sum _n
I_n\arctan \frac{x^2-x^2_n}{x^1-x^1_n},
\label{@superar}\end{equation}
which is
 properly defined only if one can
clearly continue it analytically into the
all parts of the composite Riemann sheets
defined by the
endpoints of the cut at the rigin.
If we were to replace the sum by an integral,
this possibility would be lost.
Thus it is, strictly speaking,
impossible to represent  arbitrary  continuous
magnetic
fields as gradients of superpositions of scalar potentials $ \Omega ({\bf x})$.
This, however,
is not a severe disadvantage of this representation
since any current
can be approximated
by a superposition
of line-like currents
with any desired accuracy, and the
same will be true for the
associated  magnetic fields.

The arbitrariness of the shape of the jumping surface
is the origin of a further interesting gauge structure
which will be exploited in
Section \ref{GDOFO}.

\subsection[Generating Magnetic Fields
by Multivalued Gauge Transformations]
{Generating  Magnetic Field
by Multivalued Gauge Transformations\label{@GMFMV}}
After this first exercise in multivalued functions,
we now turn to another example in magnetism
which will lead directly to our intended geometric application.
We observed before that
the local gauge transformation (\ref{gautr@})
produces the same magnetic field ${\bf B}({\bf x})=\nablab\times {\bf A}({\bf
x})$
only, as long as
the function $ \Lambda({\bf x})$ satisfies the Schwarz
integrability criterion (\ref{nonc})
\begin{equation}
(\partial _i\partial _j-\partial _j\partial _i) \Lambda({\bf x})=0.
\label{schw}\end{equation}
Any function  $ \Lambda({\bf x})$
violating this condition would
change the
magnetic field by
\begin{equation}
 \Delta{B}_k({\bf x})= \epsilon_{kij} (\partial _i\partial _j-\partial
_j\partial _i) \Lambda({\bf x})
\label{x}\end{equation}
thus being no proper gauge transformation.
The gradient of $ \Lambda({\bf x})$
\begin{equation}
{\bf A}({\bf x})  =\nablab  \Lambda({\bf x}),
\label{nontrivA@}\end{equation}
would be
a {\em nontrivial \/} vector potential.

In analogy with the
multivalued coordinate transformations violating
the integrability conditions  of Schwarz as in
(\ref{nonc}), the function $ \Lambda({\bf x})$
  will be called nonholonomic gauge function.\\

Having just learned how to deal with
multivalued functions we may change our attitude
towards gauge transformations
and decide to generate {\em all\/} magnetic fields
approximately in a
field-free space
by such improper gauge transformations $ \Lambda({\bf x})$.
By choosing for instance
\begin{equation}
 \Lambda({\bf x})=\frac{\Phi}{4\pi} \Omega({\bf x} ),
\label{LamOm}\end{equation}
we see from (\ref{SCHWO}) that this generates a field
\begin{equation}
 {B}_k({\bf x})= \epsilon_{kij} (\partial _i\partial _j-\partial _j\partial _i)
\Lambda({\bf x})
=\Phi \delta_k({\bf x};L).
\label{DelB}\end{equation}
This is a magnetic field of total flux $\Phi$ inside an infinitesimal tube.
By a superposition of such infinitesimally thin flux tubes analogous to
(\ref{superpos1}) we can obviously generate
a dicrete approximation to
any desired magnetic field
in a field-free space.

\subsection{Magnetic Monopoles}
Multivalued fields have also been used
to describe
magnetic monopoles
\cite{Monopol1,Monopol2,Monopol3}. A monopole charge density
$\rho_{\rm m}({\bf x})$
is the source of a magnetic field ${\bf B}({\bf x})$ as
defined by the
equation
\begin{equation}
\nablab \cdot {\bf B}({\bf x})= \rho_{\rm m}({\bf x}).
\label{mgmon@}\end{equation}
If ${\bf B}({\bf x})$ is expressed in terms of a vector potential
${\bf A}({\bf x})$
as ${\bf B}({\bf x})=\nablab\times {\bf A}({\bf x})$, equation
(\ref{mgmon@}) implies the noncommutativity of derivatives
in front of
the vector potential ${\bf A}({\bf x})$:
\begin{equation}
 \frac{1}{2}\epsilon_{ijk}(
\partial _i\partial _j
-\partial _j\partial _i)
A_k({\bf x})= \rho_{\rm m}({\bf x}).
\label{@}\end{equation}
Thus ${\bf A}({\bf x})$ must be multivalued.
Dirac in his famous theory of monopoles \cite{DMON}
made the field single-valued by
attaching to  the
world line
of the particle
a jumping
world surface,
whose intersection with
a  coordinate plane at a fixed time
forms the {\em Dirac string\/},
along which the magnetic field of the monopole
is imported from infinity.
This world surface can be made physically irrelevant
by quantizing it appropriately with respect to the charge.
Its shape in space is just as irrelevant
as that of the jumping surface $S$ in Fig.~\ref{loop}.
The invariance under shape deformations
constitute once more
a second gauge
structure of the type
to be discussed in Section \ref{GDOFO} \cite{Monopol1}.

Once we allow ourselves to work with multivalued fields,
we may easily go one step further
and express also ${\bf A}({\bf x})$ as a gradient of a scalar field
as in  (\ref{nontrivA@}). Then the condition
becomes
\begin{equation}
 \epsilon_{ijk}\partial _i\partial _j \partial _k \Lambda({\bf x})= \rho_{\rm
m}({\bf x}).
\label{@}\end{equation}

There exists by now  a well-developed quantum field theory
for many other systems described by
multivalued fields \cite{Kleinert1,Kleinert1I,Camb}.

\subsection[Minimal Magnetic Coupling of Particles from Multivalued Gauge
Transformations]
{Minimal Magnetic Coupling of Particles from Multivalued Gauge
Transformations\label{Min}}

Multivalued gauge transformations are the ideal
tool to minimally couple electromagnetism
to any type of matter.
Consider for instance a free nonrelativistic point particle
with a Lagrangian
\begin{equation}
 L=\frac{1}{2}\dot {\bf x}^2.
\label{@}\end{equation}
The equations of motion are invariant under a gauge transformation
\begin{equation}
L\rightarrow L'=L+\nablab  \Lambda({\bf x}) \dot {\bf x} ,
\label{63a@}\end{equation}
since this changes the action ${\cal A}= \int_{t_1}^{t_2} dt\, L$ merely by a
surface term:
\begin{equation}
 {\cal A}'\rightarrow{\cal A}={\cal A}+\Lambda({\bf x}_2)-\Lambda({\bf x}_1).
\label{62@}\end{equation}
The invariance is absent if we take
$\Lambda({\bf x}) $ to be a multivalued gauge function.
In this case,
a nontrivial vector potential ${\bf A}
({\bf x})=\nablab \Lambda  ({\bf x})$
 (working in natural units with $e=1$)
is created in the field-free space,
and the
nonholonomically gauge-transformed Lagrangian
corresponding to (\ref{63a@}),
\begin{equation}
 L'=\frac{1}{2}\dot {\bf x}^2+
  {\bf A}({\bf x}) \dot {\bf x},
\label{63@}\end{equation}
describes correctly the dynamics of a free particle
in an external magnetic field.

The coupling derived by multivalued gauge transformations
is automatically invariant under additional ordinary  single-valued
gauge transformations of the vector potential
\begin{equation}
{\bf A}({\bf x})\rightarrow
{\bf A}'({\bf x})=
{\bf A}({\bf x}) +
  \nablab\Lambda({\bf x}) ,
\label{svGT@}\end{equation}
since these
add to the Lagrangian (\ref{63@}) once more the same
pure derivative term which changes the action by an irrelevant surface term
as in (\ref{62@}).

The same procedure leads in quantum mechanics
to the minimal coupling of the Schr\"odinger
field $\psi({\bf x})$.
The Lagrange density is (in natural units with $\hbar=1$)
\begin{equation}
{\cal L}=\psi^*({\bf x})\left(i\partial _t +\frac{1}{2}\nablab^2\right)
 \psi({\bf x})              .
\label{LD@}\end{equation}
The physics described by a Schr\"odinger
wave function $  \psi({\bf x}) $ is
 invariant under arbitrary
U(1) phase changes
\begin{equation}
 \psi({\bf x},t)   \rightarrow
  \psi'({\bf x})= e^{i \Lambda({\bf x})} \psi({\bf x},t)   .
\label{U(1)@}\end{equation}
 This implies that
the Lagrange density (\ref{LD@})
may equally well be replaced by
the gauge-transformed one
\begin{equation}
{\cal L}=\psi^*({\bf x},t)\left(i\partial _t +\frac{1}{2}{\bf D}^2\right)
 \psi({\bf x},t)              ,
\label{LD1@}\end{equation}
where ${\bf D}\equiv \nablab-i\nablab  \Lambda({\bf x})$.
By allowing for nonholonomic gauge functions $ \Lambda({\bf x})$
whose gradient is the vector potential
as in (\ref{nontrivA@}),
the operator
${\bf D}$ turns into
\begin{equation}
{\bf D}=\nablab-i{\bf A}({\bf x}),
\label{@}\end{equation}
which describes correctly the magnetic
coupling in quantum mechanics.

As in the classical case,
the coupling derived by multivalued gauge transformations
is automatically invariant under ordinary  single-valued
gauge transformations under which the vector potential
${\bf A}({\bf x})$ changes as in (\ref{svGT@}),
whereas the Schr\"odinger wave function
undergoes a local U(1)-transformation
  (\ref{U(1)@}).
This invariance is a direct consequence of the simple transformation
behavior
of ${\bf D}\psi({\bf x},t)$
under gauge transformations (\ref{svGT@}) and (\ref{U(1)@})
which is
\begin{equation}
{\bf D} \psi({\bf x},t)   \rightarrow
  {\bf D}\psi'({\bf x},t)= e^{i \Lambda({\bf x})}{\bf D} \psi({\bf x},t) .
\label{@}\end{equation}
Thus
${\bf D} \psi({\bf x},t) $ transforms just like
$\psi({\bf x},t) $ itself, and for this reason, {\bf D} it is called
gauge-covariant derivative.
The generation of magnetic fields by a multivalued
gauge transformation is the simplest example
for the  power of the nonholonomic mapping principle.

We are now prepared to
introduce the same mathematics
into differential geometry, where the role of gauge transformations is played
by
reparametrizations of the space coordinates.
If spins are present, we must
formulate the theory
such as to accommodate also local Lorentz transformations.

\section[Infinitesimal Curvature and Torsion from Active Multivalued Coordinate
Transformations ]
{Infinitesimal Curvature and Torsion from Active Multivalued Coordinate
Transformations\label{seccurtor}}

\comment{In the last two sections we have seen that a Minkowski space
has neither torsion nor curvature. The absence of torsion
 follows from its tensor property, which was a consequence of the
 commutativity of derivatives in front of the infinitesimal
 translation field
\begin{eqnarray}
  \left( \partial _\mu \partial _\nu  - \partial _\nu
    \partial _\mu \right) \xi ^\kappa (x) = 0.
\label{2.71}\end{eqnarray}
The absence of curvature, on the other hand, was a consequence
 of the integrability condition (\ref{2.15}) of the transformation
 matrices
\begin{eqnarray}
  \left( \partial _\mu \partial _\nu - \partial _\nu \partial _\mu
    \right) \alpha ^\kappa{} _\lambda  (x) = 0.
\label{x}\end{eqnarray}
 Infinitesimally, this implies that
\begin{eqnarray}
  \left( \partial _\mu  \partial _\nu -\partial _\nu \partial _\mu
  \right) \partial _\lambda \xi ^\kappa (x) = 0,
\label{2.72}\end{eqnarray}
i.e., that derivatives commute in front of {\em derivatives\/}
 of the infinitesimal translation field. This suggests a simple
 way of constructing general metric affine spaces with torsion or
 curvature or both from a Minkowski space by performing
 {\em singular\/} coordinate transformations which do not
 satisfy (\ref{2.71}), (\ref{2.72}).
}

We are now going to study the properties of a space at which we can arrive
 from a flat space using multivalued tetrad fields $e_a{}^\mu$ and
 $e^a{}_\mu $ which are close to  unit matrices
$ \delta _a{}^\mu$ and
 $ \delta ^a{}_\mu $, respectively.
It is easy to see that these
correspond to a geometric analog ofinfinitesimal
gauge transformations
in magnetostatics
with multivalued gauge functions of the type
(\ref{LamOm}).
\comment{They correspond to
{\em  multivalued transformations\/} of the line elements $dx^a$
to new line elements $dq^\mu$.}
 Because of the nonlinearity of all geometric quantities,
we shall restrict ourselves
to {\em infinitesimal \/} Einstein transformations
\begin{equation}
x^a\mathop{\rightarrow}_E q^{\mu}=x^ {\mu=a}-\xi ^\mu (x),
\label{inftr}\end{equation}
 which play the role of infinitesimal local translations.
According to (\ref{LO}), the associated
multivalued tetrad fields are
\begin{eqnarray}
  e_a{}^\mu  & = & \delta _a{}^\mu  - \partial _a \xi ^\mu \nonumber \\
  e^a{}_\mu & = & \delta ^a{}_\mu  + \partial _\mu  \xi ^a .
\label{2.73}\end{eqnarray}
Thus they are transformed by a gradient
of the  functions $\xi ^\mu (x)$
in complete analogy with
the magnetic vector potential in (\ref{gautr@}). The metric (\ref{IN}) induced
by the infinitesimal local translations
(\ref{inftr})
is
\begin{eqnarray}
  g_{\mu \nu }
       = \eta_{\mu \nu } + \left( \partial _\mu \xi _\nu
          + \partial _\nu \xi _\mu \right) .
\label{2.74}\end{eqnarray}

For small transformation functions $\xi^\mu(x)$,
the affine connection (\ref{connection0}) becomes
\begin{equation}
  \Gamma _{\mu \nu }{}^{\lambda } = \partial _\mu \partial _\nu
         \xi ^\lambda     .
\label{2.77x}\end{equation}
For multivalued transformation functions $\xi _\mu(x)$, the metric
and the affine connection are, in general,
   also multivalued. This could cause difficulties in performing
   consistent length measurements and parallel displacements.
   In order to avoid this, Einstein postulated that the
   metric $g_{\mu \nu }$ and the affine connection $\Gamma _{\mu \nu
}{}^\lambda $
   should be single-valued
and smooth enough to
be differentiated twice.
Because of the single-valuedness, derivatives in front of
  $g_{\mu \nu }$  and $\Gamma _{\mu \nu }{}^\lambda $
should commute with each other [see (\ref{noncgG}),
implying  the infinitesimal
integrability conditions
\begin{eqnarray}
   \left( \partial _\mu \partial _\nu -\partial _\nu \partial _\mu \right)
   \left( \partial _\lambda \xi _\kappa + \partial _\kappa \xi _\lambda
   \right) &  = & 0,
\label{2.75a}\\
   \left( \partial _\mu \partial _\nu  - \partial _\nu \partial _\mu \right)
      \partial _\sigma \partial _\lambda \xi _\kappa & = & 0.
\label{2.75}\end{eqnarray}
Since $\xi ^\mu$ are infinitesimal, we can lower the index
 in both equations (with a mistake which is only of the order
 of $\xi ^2$ and thus negligible) so that
(\ref{torsion0@}) and
(\ref{10.30}) yield
\begin{eqnarray}
     S_{\mu \nu \lambda } & = & \frac{1}{2} \left( \partial _\mu
             \partial _\nu - \partial _\nu \partial _\mu \right)
             \xi _\lambda ,
\label{2.77}                     \\
  R_{\mu \nu \lambda \kappa }& =&
         \left( \partial _\mu \partial _\nu -\partial _\nu
         \partial _\mu \right) \partial _\lambda \xi _\kappa .
\label{2.77b}\end{eqnarray}
Note that the curvature tensor is antisymmetric
 in the last two indices, as an immediate consequence of the
 integrability condition (\ref{2.75}).
This antisymmetry is therefore a Bianchi identity of the gauge field
representation
of the curvature tensor for infinitesimal
deviations from flat space, where it
constitutes
the fundamental or second identity in Schouten's nomenclature \cite{Schouten}.

Let us also calculate the Riemann
part (\ref{2.9}) of the infinitesimal connection (\ref{2.77x}).
Inserting
(\ref{2.74}) into (\ref{2.9}), we find
\begin{equation}
  \bar  \Gamma_{\mu \nu \lambda}=\frac{1}{2}
        \left[ \partial _\mu \left( \partial _\nu \xi _\lambda
        + \partial _\lambda \xi _\nu \right) +\partial _\nu
          \left( \partial _\mu \xi _\lambda +
          \partial _\lambda \xi _\mu \right) - \partial _\lambda
           \left( \partial _\mu \xi _\nu + \partial _\nu
           \xi _\mu \right) \right].
\label{2.79}\end{equation}
The affine connection
(\ref{2.77x}) can then be decomposed
as in (\ref{2.78}),
   with  the contortion tensor
\begin{eqnarray}
      K_{\mu \nu \lambda } & = & \frac{1}{2} \left( \partial _\mu
            \partial _\nu - \partial _\nu \partial _\mu \right)
             \xi _\lambda -\frac{1}{2} \left( \partial _\nu
             \partial _\lambda -\partial _\lambda \partial _\nu \right)
             \xi _\mu  + \frac{1}{2} \left( \partial _\lambda
             \partial _\mu -\partial _\mu \partial _\lambda \right)
             \xi _\nu \nonumber \\
     & = & \frac{1}{2}\left[  \partial _\mu  \left( \partial _\nu
            \xi _\lambda -\partial _\lambda  \xi _\nu \right)
            + \partial _\lambda \left( \partial _\nu  \xi_\mu
            + \partial _\mu \xi _\nu \right) -
             \partial _\nu \left( \partial _\lambda \xi _\mu
             + \partial _\mu \xi _\lambda \right)\right] ,
\label{2.78x}\end{eqnarray}
the first line being  the  combination
(\ref{contorttenb}) of torsion tensors.
By inserting the infinitesimal Riemann connection (\ref{2.79})
into (\ref{10.30}), we find the associated Riemann curvature tensor
\begin{eqnarray}
\bar{R}_{\mu \nu \lambda \kappa }
         &=& \frac{1}{2} \partial _\mu \left[ \partial _\nu
         \left( \partial _\lambda \xi _\kappa +\partial _\kappa
         \xi _\lambda \right) +\partial _\lambda \left( \partial _\nu
         \xi _\kappa +\partial _\kappa \xi _\nu \right)
         -\partial _\kappa \left( \partial _\nu \xi _\lambda +
          \partial _\lambda \xi _\nu \right) \right] \nonumber \\
   & = & -\frac{1}{2}\partial _\nu \left[ \partial _\mu \left( \partial
_\lambda
         \xi _\kappa +\partial _\kappa \xi _\lambda \right) +\partial _\lambda
         \left( \partial _\nu \xi _\kappa +\partial _\kappa
          \xi _\nu \right) - \partial _\kappa \left( \partial _\mu
          \xi_\lambda +\partial _\lambda \xi _\lambda \right) \right] .
\label{x}\end{eqnarray}
Averaging the two
 equal right-hand sides, the integrability condition (\ref{2.75})
for the metric
removes the two first parentheses, and we obtain
\begin{eqnarray}
  \bar{R}_{\mu \nu \lambda \kappa }
   = \frac{1}{2} \left\{ \left[ \partial _\mu \partial _\lambda
         \left( \partial _\nu \xi _\kappa + \partial _\kappa
         \xi _\nu \right) - \left( \mu \leftrightarrow \nu \right) \right]
         -[\lambda\leftrightarrow  \kappa]\right\} .
\label{2.80}\end{eqnarray}

Multivalued coordinate transformations of the type (\ref{inftr})
appear naturally in the
theory of topological defects in three-dimensional crystals.
There one considers
infinitesimal
 displacements of atoms
\begin{eqnarray}
  x_i \rightarrow x_i' = x_i + u_i ({\bf x}) ,~~~~(i=1,~2,~3).
\label{x}\end{eqnarray}
where $x_i'$ are the shifted positions, as seen from an ideal
 reference crystal. If we change the point of view
 to an intrinsic description, i.e., if we measure coordinates
 by counting the number of atomic steps {\em within\/} the distorted
 crystal, then the atoms of the ideal reference crystal are
 displaced by
\begin{eqnarray}
  x_i \rightarrow x_i' = x_i - u_i ({\bf x}).
\label{x}\end{eqnarray}
The displacement field is defined
only modulo
 lattice spacings.
This makes it intrinsically
multivalued, having
noncommuting derivatives which
contain information on the crystalline topological defects.
The physical
 coordinates of material points $x^i$ for $i = 1,2,3$
are identified  with the
 previous
 spatial coordinates\footnote{
  When working with four-vectors, it is conventional to
  consider the upper indices as physical components. In purely
  three dimensional calculations one usually employs the metric
   $\eta_{ab} = \delta _{ab}$ such that $x^{a=i}$ and
   $x_i$ are the same.}   $x^a$ for $a = 1,2,3$
  and $\partial _a = \lfrac{\partial }{\partial x^a}
  (a = i)$ with the previous derivatives $\partial _i$.
  The infinitesimal translation fields in (\ref{inftr})
  are equal to the displacements $u_i ({\bf x})$ such that
  the multivalued tetrads are
\begin{eqnarray}
  e_a^ i = \delta _a^i - \partial _a u_i , ~~~ e^a{}_i = \delta ^a{}_i
      + \partial _i u_a,
\label{2.82}\end{eqnarray}
and all geometric quantities
are defined as  before.

In a crystal, one likes to specify the deformation
by
a
strain tensor
\begin{equation}
u_{kl}=\frac{1}{2}( \partial _k u_l + \partial _l u_k ),
\label{x}\end{equation}
and a local rotation tensor
\begin{equation}
 \omega_{kl}=\frac{1}{2}( \partial _k u_l - \partial _l u_k ).
\label{rottens}\end{equation}
 For these, the
integrability conditions
(\ref{2.75a}), (\ref{2.75})
imply that
\begin{eqnarray}
    \left( \partial _i \partial _j - \partial _j \partial _i \right)
         \left( \partial _k u_l + \partial _l u_k \right)  &=& 0, \\
    \left( \partial _i \partial _j - \partial _j \partial _i \right)
     \partial _n \left( \partial _k u_i + \partial _l u_k\right)
     &=& 0 \\
  \left( \partial _i \partial _j - \partial _j \partial _i
      \right) \partial _k \left( \partial _k u_i -
     \partial _l u_k\right)  &=& 0,
\label{2.75'}\end{eqnarray}
stating
that the strain tensor
\begin{equation}
u_{kl}=\frac{1}{2}( \partial _k u_l + \partial _l u_k ),
\label{x}\end{equation}
its derivative, and the
derivative of the local rotation tensor
\begin{equation}
 \omega_{kl}=\frac{1}{2}( \partial _k u_l - \partial _l u_k ),
\label{rottens}\end{equation}
 are all twice-differentiable
single-valued functions everywhere.
In three dimensions
one often uses the rotation  vector
\begin{equation}
\omega _j = \frac{1}{2} \epsilon _{jmn}  \omega_{mn}=
\frac{1}{2}
    \epsilon _{jmn} \partial _m u_n
\label{@}\end{equation}
instead of the tensor field (\ref{rottens}).

A single-valued distortion field $u_i({\bf x})$ corresponds to an
{\em elastic\/}
deformation,
a multivalued field to a {\em plastic\/} deformation of the crystal.

The
 local vector field   $\omega_j$
has noncommuting derivatives, as
measured by the tensor
\begin{eqnarray}
           G_{ji} =  \epsilon _{ikl} \partial _k \partial _l \omega _j .
\label{2.88}\end{eqnarray}
This is the
 {\em  Einstein
  curvature\/} tensor of the Riemann-Cartan geometry.
 Since
the
derivative of the local rotation tensor
has commuting derivatives,
the Einstein tensor is divergenceless:
\begin{equation}
\partial _iG_{ji}=0.
\label{@}\end{equation}
This corresponds to the famous original Bianchi identity
(the first identity) of Riemann
spaces which has served as a prototype
for all identities expressing
the single-valuedness of physical fields.

Let us prove that $G_{ji}$ coincides
with
the Einstein tensor
in the common definition
as
the combination of Ricci tensor and scalar curvature:
\begin{eqnarray}
      G_{ji} = R_{ji} - \frac{1}{2} g_{ji}
               R_{kk}               .
\label{Gtens}\end{eqnarray}
Returning to the notation
$\xi_i(q)$ for the infinitesimal translations,
and
 taking advantage
of the  integrability condition
(\ref{2.75}), we write the curvature tensor
(\ref{2.78}) as
\begin{eqnarray}
           R_{ijkl} =
\left( \partial _i \partial _j -
      \partial _j \partial _i\right) \frac{1}{2}
      \left( \partial _k \xi_l - \partial _l \xi_k\right) =
\left( \partial _i \partial _j -
      \partial _j \partial _i\right)    \epsilon_{klm} \omega_m(q).
\label{2.86}\end{eqnarray}
 In three dimensions,
 the antisymmetry in $ij$ and $kl$ suggests
the introduction of a  of second-rank tensor
\begin{eqnarray}
           G_{ji} \equiv
          \frac{1}{4}  \epsilon_{ikl}  \epsilon_{jmn} R^{klmn}.
\label{2.87}\end{eqnarray}
 In the full nonlinear Riemann geometry, the $ \epsilon$-tensors are simply
replaced
by their generally covariant versions
where
\begin{eqnarray}
    e_{ijk} = \sqrt{ g}
     \epsilon _{ijk} = g_{ii'} g_{jj'} g_{kk'} e^{i'j'k'}
     = g_{ii'} g_{jj'} g_{kk'} \left( \frac{1}{\sqrt{ g}}
       \epsilon ^{i'j'k' } \right)    .
\label{x}\end{eqnarray}
If we now
 insert into the fully covariant version of (\ref{2.87}) the identity
\begin{eqnarray}
   \!\!\!\!\!\!\!\!\!\!\!e_{ikl} e_{jmn} = g_{ij} g_{km} g_{ln} + g_{im}
              g_{kn} g_{lj} + g_{in} g_{kj} g_{lm}
              -g_{ij} g_{lm} g_{kn} - g_{im} g_{kn}
              g_{kj}  - g_{in} g_{lj} g_{km} ,
\label{x}\end{eqnarray}
 we recover (\ref{Gtens}).

 In
 four dimension, the combination
(\ref{Gtens}) can be rewritten as
\begin{eqnarray}
G^{\nu \mu } =\sfrac{1}{4}  e^{\mu \alpha \beta \gamma }
      e^{\nu}{}_\alpha {}^{ \delta \tau }
 R_{\beta \gamma \delta \tau },
\nonumber
\label{@}\end{eqnarray}
a direct generalization of
(\ref{2.87}).

 Inserting
(\ref{2.86}) into
(\ref{2.87}), we find for small displacements
\begin{eqnarray}
  G_{ij}
= \epsilon _{ikl} \partial _k \partial _l \left( \frac{1}{2}
            \epsilon _{jmn} \partial _m \xi_n\right)
,\label{x}\end{eqnarray}
which coincides with (\ref{2.88}), as we wanted to prove.

Let us also form the Einstein tensor $\bar {G}_{ij}$
associated with the Riemannian curvature tensor $\bar{R}_{
 ijkl}$. Using (\ref{2.80}) we find
\begin{eqnarray}
           \bar{G}_{ji}
            = \epsilon _{ikl} \epsilon _{jmn}
         \partial _k \partial _m \frac{1}{2} \left( \partial _l
         \xi_n + \partial _n \xi_l \right)
            = \epsilon _{ikl} \epsilon _{jmn}
         \partial _k \partial _m \xi_{ln}.
\label{2.89}\end{eqnarray}
 In the theory of crystalline  topological defects one introduces the following
 measures for the noncommutativity of derivatives. The
  dislocation density
\begin{eqnarray}
     \alpha _{ij} = \epsilon _{ikl} \partial _k \partial _l \xi_j   ,
\label{x}\end{eqnarray}
the disclination density
\begin{eqnarray}
   \Theta_{ij} = \epsilon _{ikl} \partial _k \partial _l
    \omega _j        ,
\label{2.90}\end{eqnarray}
and the defect density
\begin{eqnarray}
  \eta_{ij} = \epsilon _{ikl} \epsilon _{jmn}
      \partial _k \partial _m \xi_{lm}.
\label{2.91}\end{eqnarray}
   Comparison with Eq.~(\ref{2.80})
   shows that $\alpha _{ij}$ is directly related to the torsion
   tensor
   $ S_{kl}{}^i = \frac{1}{2} \left( \Gamma _{kl}{}^i - \Gamma _{lk}{}^i
   \right) $:
\begin{eqnarray}
  \alpha _{ij} \equiv \epsilon _{ikl} \Gamma _{klj}
        \equiv \epsilon _{ikl} S_{klj}.
\label{2.92a}\end{eqnarray}
 Hence torsion  is a measure
 of the translational defects
 contained in singular coordinated transformations.
 We can also use the decomposition (\ref{contorttenb}) and
express this in terms of the contortion tensor as
\begin{eqnarray}
  \alpha _{ij} = \epsilon _{ikl} K_{klj}         .
\label{x2.89}\end{eqnarray}
In terms of the strain tensor $ \xi_{kj}=
\frac{1}{2}(
\partial _k\xi_j+
\partial _j\xi_k)
$ and the rotation field $ \omega_l$,
 the contortion tensor becomes
%
\begin{eqnarray}
   K_{ijk} & = & \frac{1}{2} \partial _j \left( \partial _j \xi_k
         - \partial _k \xi_j\right)  -\frac{1}{2}
          \left[ \partial _j \left( \partial _k
          \xi_j + \partial _i \xi_k\right)  - (j\leftrightarrow k)\right]
\nonumber \\
          & = & \partial _i \omega _{jk} - \left[
                 \partial _j \xi_{ki} - (j\leftrightarrow k)\right].
\label{Nyestx}\end{eqnarray}
Since $K_{ijk} $ is antisymmetric in $lj$, it is useful to introduce
 the tensor of second rank,
 called Nye's contortion tensor
\begin{eqnarray}
 K_{ln} = \frac{1}{2} K_{klj} \epsilon _{ljn}.
\label{x}\end{eqnarray}
 Inserting this into (\ref{x2.89}) we see that
\begin{eqnarray}
  \alpha _{ij} = - K_{ji} + \delta _{ij} K_{ll}
\label{x}\end{eqnarray}
For
 Nye's contortion tensor,
the decomposition (\ref{Nyestx})
takes the form
\begin{eqnarray}
  K_{il} = \partial _{i}\omega_l - \epsilon _{lkj} \partial _j
              \xi_{kj}
\label{x}\end{eqnarray}
   Consider now the disclination density $ \Theta_{ij}$.
   Comparing (\ref{2.91}) with (\ref{2.88}) we see that it
   coincides exactly with the Einstein tensor
   $G_{jl}$ formed from the full curvature tensor
\begin{eqnarray}
  \Theta_{ij} \equiv G_{ji}.
\label{2.93}\end{eqnarray}
   The defect density (\ref{2.91}), finally, coincides with the
   Einstein tensor formed from the Riemannian curvature tensor.
\begin{eqnarray}
  \eta_{ij} =  \bar{G}_{ij}
\label{x}\end{eqnarray}

\section[Explicit Multivalued Transformations Producing Curvature
and Torsion]
{Explicit Multivalued Transformations Producing           \\Curvature
and Torsion}

Let us give explicit multivalued functions
$\xi^\mu(q)$ generating infinitesimal
pointlike curvature and torsion in an otherwise flat space.
We may restric ourselves
to two dimensions. The generalization to
$D$ dimensions is straightforward--we may simply deal with
each of the $D(D-1)/2$ coordinate plains
separately, and compose the
results at the end.
In each coordinate plane, we now write down
 transformation functions
which correspond to the fundamental topological defects
pictures in Fig.~\ref{torcur}.

\subsection{Torsion}
Consider first the upper example in Fig.~\ref{torcur},
in where a dislocation is generated
by a Volterra process in which a
layer of atoms is added or removed.
The active nonholonomic
transformation may be described differentially
by
\begin{equation} \label{10.39}
dx^i =
\left\{
\begin{array}{ll}
dq^1 & ~~~\mbox{for $i = 1$,} \\
dq^2  +  \epsilon \partial_\mu \phi(q) dq^\mu & ~~~\mbox{for $i = 2$,}
\end{array}\right.
\end{equation}
where
$ \epsilon$ is a small parameter,
 and
$\phi(q)$ the multivalued function
(\ref{10.40}).
In the two-dimensional
 subspace under consideration,
 the tetrads are dyads with components
\begin{eqnarray} \label{10.41}
{e^1}_\mu(q) & = & {\delta^1}_\mu~~ ,   \nonumber\\
{e^2}_\mu(q) & = & {\delta^2}_\mu + \epsilon \partial_\mu \phi(q)~~,
\end{eqnarray}
yielding for the torsion tensor
the components
\begin{equation} \label{10.42a}
 {S_{\mu\nu}}^1(q)= 0 ,\;\;\;\;\;
 {S_{\mu\nu}}^2(q) = \frac{\epsilon}{4\pi} (\partial_\mu
\partial_\nu - \partial_\nu \partial_\mu) \phi(q),
\end{equation}
Using the noncommutativity (\ref{schw1}), we obtain
a torsion localized at the origin:
\begin{equation} \label{10.44}
 {S_{12}}^2 (q)=
\frac{ \epsilon }{2}\delta^{(2)} (q).
\end{equation}
The mapping introduces no curvature.
When encircling a dislocation
along
a closed path
 $C$,
its counter image
$C'$
in the ideal crystal
does not form a
closed path.
The {closure failure}
is called the {\em Burgers vector}
\begin{equation}
b^ i  \equiv
\oint_{C'} dx^i
=\oint_{C} dq^ \mu e^i{}_ \mu .
\label{10.burg1}\end{equation}
It specifies the direction and thickness
of the layer of additional atoms.
With the help of Stokes' theorem,
it is seen to measure the torsion
contained in any surface  $S$ spanned by $C$:
\begin{eqnarray}
b^i& =&
\oint_Sd^2s^{ \mu   \nu  } \partial  _ \mu  e^i{}_  \nu
=\oint_Sd^2s^{ \mu  \nu }e^i{}_ \lambda S_{ \mu  \nu }{}^  \lambda ,
\label{10.encl1}\end{eqnarray}
where
$
d^2s^{ \mu  \nu }=-
d^2s^{ \nu  \mu }
$
is the projection of an oriented
infinitesimal area element onto the plane $ \mu  \nu $.
The above example has the Burgers vector
\begin{equation}
b^ i =(0, \epsilon ).
\label{x}\end{equation}

A corresponding {closure failure} appears when
 mapping a
closed contour $C$ in the ideal crystal
into a crystal containing
a dislocation. This defines a Burgers vector:
\begin{equation}
b^ \mu  \equiv
\oint_{C'} dq^ \mu
=\oint_{C} dx^ie_i{}^ \mu .
\label{10.burg}\end{equation}
By
Stokes' theorem, this becomes a surface integral
\begin{eqnarray}
b^ \mu& =&
\oint_{S}d^2s^{ij} \partial  _i e_j{}^  \mu
=\oint_Sd^2s^{ij}e_i{}^  \nu
\partial _  \nu   e_j{}^  \mu \nonumber \\
&=&-\oint_Sd^2s^{ij}e_i{}^   \nu   e_j{}^  \lambda   S_{ \nu  \lambda   }{}^
\mu ,
\label{10.encl}\end{eqnarray}
with the
last step following from (\ref{tors}).

Different pointlike torsions
 (\ref{10.44}) can be used
 to generate
a torsion as an arbitrary
superposition
of infinitesimal
point-like torsions
\begin{equation} \label{10.44xy}
 {S_{12}}^2(q) = \frac{\epsilon_n}{2}  \sum _n \delta (q-q_n)
{}.
\end{equation}
We simply have to choose the angular function $\phi(q)$
in (\ref{10.41}) in analogy to (\ref{@superar}) as
\begin{equation}
\phi^f(q)=\sum _n  \epsilon _n
\arctan \frac{q^2-q^2_n}{q^1-q^1_n}.
\label{superpp}\end{equation}
As in the magnetic case,
one is not allowed to replace the sum by an integral
over a continuous distribution of these functions,
since the endpoints of the cuts of the Riemann surfaces must
remain clearly distinguishable
[see the discussion after Eq.~(\ref{@superar})].
In crystal physics,
this means that there is no mathematically
well-defined way of setting up continuous theory of defects.
Fortunately, this
which need not bother us since defects in crystals are discete
objects anyhow.
It is curious to see
how theorists of plastic deformations
have tried to escape this problem verbally.

When applied to spacetime of gravitational physics,
this implies that
it is impossible to generate, even infinitesimally,
a space with a smooth torsion.
We can only generate a space carrying
a superposition of discrete torsion lines (or surfaces in four spacetime
dimensions).
This is similar to the geometry
generated by the Regge calculus  \cite{regge}. For the arguments
to be presented in the sequel,
however, this problem
will be irrelevant. We merely need to
be sure
that
a flat space can be transformed into
spaces with arbitrary
 discrete superpositions of infinitesimal
line- or surface-like curvatures and torsions.
Once we know the transformed laws of nature for such
discrete
superpositions,
we may generalize them
to arbitrary infinitesimal
curvature and torsion. These
 can always be approximated discretely
 to any desired degree of accuracy.

By removing a vertical layer of atoms
in Fig.~\ref{torcur},
we obtain the same result with the superscript $1$ exchanged by $2$.
By going through the same procedure in all coordinate planes,
removing a layer of atoms in each spatial direction,
and forming superpositions, we can generate an arbitrary
superposition of discrete infinitesimal torsions
in the initially flat space.
This procedure can be extended to three and four spacetime
dimensions in an obvious way.

\subsection{Curvature}

The second example is the nonholonomic mapping
in the lower part of Fig.~\ref{torcur}, generating
a disclination
which corresponds to
 an entire section of angle $ \Omega$ missing in an ideal
atomic array.
For an infinitesimal angle $  \Omega$, this
may be described, in two dimensions, by the differential mapping
\begin{equation} \label{10.45}
x^{i } = \delta ^i{}_ \mu \left[q^ \mu -
\frac{ \Omega}{2\pi}
 \epsilon ^{ \mu}{}_  \nu q ^\nu\phi(q)\right],
\end{equation}
with the multivalued function (\ref{10.40}).
The symbol $ \epsilon _{ \mu  \nu } $ denotes the
antisymmetric Levi-Civita
tensor.
The transformed metric
\begin{equation} \label{10.46}
g_{ \mu  \nu }=
 \delta _{ \mu  \nu }+
\frac{ \Omega  }{\pi }
\frac{1  }{q^ \sigma  q_ \sigma  }
 \epsilon {}_{ \mu \lambda }
\epsilon _{ \nu\kappa }
q ^ \lambda q ^ \kappa .
\end{equation}
is single-valued and has commuting derivatives.
The torsion tensor  vanishes since
$(\partial _ 1\partial _ 2 -\partial _ 2 \partial _ 1)x^{1,2}$
is proportional to $q ^{2,1}  \delta ^{(2)}(q)=0$.
The local
rotation
field
$ \omega (q)\equiv \sfrac{1}{2}[ \partial _1x^2(q)-\partial _2x^1(q)]$,
on the other hand,
is equal to the multivalued function
$ \Omega \phi(q)$,
thus having the
 noncommuting derivatives:
\begin{equation}
(\partial _1\partial _2-\partial _2\partial _1)
\omega (q)=  \Omega  \delta ^{(2)}(q).
\label{103}\end{equation}
To lowest order in $ \Omega $, this determines
the curvature tensor,
which in two dimensions possesses
 only one independent component, for instance
$R_{1212}$.
{}From Eq.~\ref{2.86}, (\ref{103}) we see that
\begin{equation}
R_{ 1212 }=(\partial _ 1 \partial _ 2 -\partial _ 2 \partial _ 1 )
 \omega (q) = \Omega  \delta^{(2)}(q).
\label{x}\end{equation}

As in the case of torsion, we may write perform the
active nonholonomic coordinate transformation
with a superposition
of point-like curvatures,
inserting into
(\ref{10.45}) the angular field
\begin{equation}
\phi^f(q)=\sum _n   \Omega  _n
\arctan \frac{q^2-q^2_n}{q^1-q^1_n},
\label{superpp}\end{equation}
and obtain
\begin{equation}
R_{1212}= \sum _n \Omega  _n \delta ^{(2)}(q-q_n).
\label{155x}\end{equation}
 This  forms an approximation to an
arbitrary infinitesimal continuous
curvature in the 12-plane.
Again, we cannot take the continuum limit, but
for the derivation
of structure of the physical laws,
the restricted point-like distributions
of curvature and torsion are perfectly sufficient.

By cutting a sector of atoms from
all possible coordinate planes and choosing different directions of the
sector we can generate a four-dimensional spacetime
 with an arbitrary superposition of discrete
infinitesimal
curvatures from an initially flat space.

We conclude:
A space with infintesimally small
 torsion and curvature can be generated from a flat
space via multivalued
coordinate transformations, and
is completely equivalent to a crystal which has undergone
plastic deformation and is filled with dislocations
and disclinations. The nonholomonic mapping principle
has produced a Riemann-Cartan space with
 infintesimal line-like curvature and torsion from
a flat space.
We must emphasize the infinitesimal nature of the line like
torsion and curvature. It is mathematically inconsistent
to generate the structure to a full geometry of defects
as proposed in Refs.~\cite{Bilby}-\cite{Kroener2}.
The reason is that this would prouce higher
powers of $ \delta $-functions (\ref{10.44}), (\ref{155x}),
which are mathematically undefined.

In a Minkowski space,  trajectories of free point
particles are straight lines. A space with curvature and torsion
may be viewed as a "world crystal"
with topological defects. In it, the preferred paths
are no longer
straight since
 defects may lie in their way.
Translating this into Einstein's theory,
 mass points in a gravitational field will run along
the geometrically preferred
path in the
space with defects.
The defects
in the ``world crystal''
 explain all
gravitational effects.

In Subsection~\ref{ssRiemCart} we shall demonstrate that
the nonholonomic mapping principle will turn
straight lines in flat space into
the correct particle trajectories.
There are
autoparallel,  forming the
straightest possible paths
in the metric-affine space.

The natural length scale of gravity is the Planck length
\begin{eqnarray}
   l_p = \left( \frac{c^3}{8\pi k \hbar}\right) ^{-1/2}
\label{x}\end{eqnarray}
 where $c$ is the light velocity ($\approx 3 \times 10^{10}$ cm/s),
 $\hbar$ is Planck's constant ($\approx 1.05459\times 10^{-27}$
 erg/s) and $k$ is Newton's gravity constant
 ($\approx6.673\times 10^{-8}$ cm$^3$/ (g $\cdot$ s$^2$)).
 The Planck length is an extremely small quantity
 ($\approx8.09 \times 10^{-33}$ cm) which at present is
 beyond any experimental resolution. This may be imagined as
 the lattice constant of the world crystal with defects.
 %


\section[The New Action Principle in the Presence of Torsion]
{The New Action Principle in the Presence of Torsion}

In 1993, Fiziev and I \cite{Kleinert5}
applied the  nonholonomic mapping principle
to the variational
derivation of equations of motion
from the extremum of an action. We observed
that variations of paths
in spaces with torsion
should reflect the closure failure of parallelograms
and can therefore not be
performed
with both ends of the paths simultaneously held fixed.
This has the important and surprising consequence that
an action
involving only the metric of
space can produce equations of motion
containing a torsion force.

The new variational
procedure
was simplified by Pelster and myself
\cite{Kleinert6}
by
introducing modified
variations
which
do not
commute with the proper-time derivative of the trajectory.
The simplified procedure
has the
advantage of being applicable to a larger variety of actions, in particular
to particles in external fields.

\subsection{Minkowski Spacetime}

Starting point is
the standard action principle for the free motion of a
spinless point particle of mass $M$ in a flat
space
with  Minkowski metric $ \eta_{ab}$.
Introducing some parameter $\tau $ to describe the path
$x^a ( \tau  )$ of the point particle, the infinitesimal
proper distance $ ds$ is given by
\begin{equation}
\label{PT1}
d s ( \tau  ) =  \sqrt{dx^2} \equiv
\sqrt{ \eta_{ab} d x^a ( \tau  ) d x^b ( \tau  )} \, .
\end{equation}
The associated time $d \s=ds/c$ is the proper time.
The action of the point particle
\begin{equation}
\label{AC1}
{\cal A} [ x^a ( \tau  ) ] = \int\limits_{\tau_1}^{\tau_2} d \tau
L ( \dot{x}^a ( \tau  ) )
\end{equation}
is proportional to the proper time spent by the particle
moving from
$\tau_1$ to $\tau_2$, i.e., the Lagrangian reads \cite{Weinberg2}
\begin{equation}
\label{AC2}
L ( \dot{x}^a ) =- M  \,\sqrt{\dot{x}^2} \, .
\end{equation}
By construction,
the action (\ref{AC1})
is invariant
with respect to arbitrary reparametrizations
$\tau \rightarrow \tau  ' = \tau  ' ( \tau  )$.\\

The Hamiltonian
action principle states that the physically realized trajectory is found
by extremization, requiring the vanishing of the variation
\begin{equation}
\label{VAR1}
\delta {\cal A} [ x^a ( \tau  ) ] = 0
\end{equation}
with respect to all variations $\delta x^a ( \tau  )$ which vanish at the
end points $\tau_1$ and $\tau_2$:
\begin{equation}
\label{VAR2}
\delta x^a ( \tau_1 ) = \delta x^a ( \tau_2 ) = 0 \, .
\end{equation}
The geometric meaning of a variation
implies that they are
independent of changes in the $\tau $-parameter,
i.e.,
that they satisfy
the following commutation relation with the derivative
$d_{\tau }\equiv d / d \tau $:
\begin{equation}
\label{VAR3}
\delta d_{\tau } x^a ( \tau  ) - d_{\tau } \delta x^a ( \tau  ) = 0 \, .
\end{equation}
Under such variations, the extremization of the action (\ref{AC1}) leads
immediately to the Euler-Lagrange equation
\begin{equation}
\label{EL1}
 \frac{d}{d \tau } \frac{\partial L}{\partial \dot{x}^a ( \tau  )} = 0\, .
\end{equation}
Inserting the Lagrangian (\ref{AC2}), and remembering
the proper distance
$ds$ in Eq.~(\ref{PT1})], we
end up with
the equation of motion
\begin{equation}
\label{M1}
\ddot{x}^a ( \tau  ) = f(\tau )
\, \dot{x}^a ( \tau  ) \, ,
\end{equation}
where $f ( \tau  )$ is determined by a
relation between
the proper distance $s$ and
the trajectory parameter $\tau $:
\begin{equation}
\label{F1}
f ( \tau  ) = \ddot s ( \tau  )/\dot s ( \tau  ).
\end{equation}
Just as the action (\ref{AC1}),the equation of motion
(\ref{M1}) is invariant
with respect to arbitrary reparametrizations
$\tau \rightarrow \tau  ' = \tau  ' ( \tau  )$. Under these
\begin{equation}
\label{F2}
f(\tau )\rightarrow
f ' ( \tau  ' ) = \ddot{s} ( \tau ' ) / \dot{s} ( \tau ' ) \, .
\end{equation}
The particular reparametrization
\begin{equation}
\label{PT}
\tau  \, ' ( \tau  ) = \int\limits^{\tau } d u \exp \left[
\int\limits^u d v  f ( v ) \right]
\end{equation}
leads to a vanishing of $f\,'(\tau \, ')$,
implying that $\tau \, '$ coincides with the proper time $ \s=s/c$.
Then the equation of motion (\ref{M1}) simply reduces to
\begin{equation}
\ddot x^a ( \s )=0                    .
\end{equation}

It is useful to realize that the above
relativistic treatment can be reduced to a nonrelativistically
looking procedure by not using (\ref{AC2})
as a Lagrangian but, instead, the completely equivalent one
\begin{eqnarray} \label{19.245}
L(\dot x,)=  -
     \frac{M}{2\rho (\tau  )}  \dot x ^2 (\tau  )
    - \frac{M }{2}\rho (\tau   ) .
\end{eqnarray}
This
contains  the particle orbit
 quadratically, looking like a free nonrelativistic Lagrangian
action, but at the expense of an extra dimensionless
variable $ \rho (\tau )$.
At the extremum, the new
 action coincides
with the
initial
one (\ref{AC1}). Indeed, extremizing $ {{\cal A}}$  in $ \rho (\tau )$  gives
the relation
\begin{equation} \label{19.246}
\rho (\tau ) =  \sqrt{\dot x{}^2 (\tau )} /c.
\end{equation}
Inserting this back into ${\cal A}$ renders the classical action
\begin{equation} \label{19.247}
{\cal A}=- M \int_{\tau_1}^{\tau_2}
 d\tau  \sqrt{  \dot x{} ^2 (\tau )},
\end{equation}
which is the same as (\ref{AC1}).

The new action
shares with the
 old action (\ref{AC1})
the
\ind{reparametrization invariance}
$\tau \rightarrow \tau  ' = \tau  ' ( \tau  )$.
We only have to assign an appropriate transformation
behavior to the extra variable $\rho (\tau )$.
If $ \tau $
is replaced
by a new parameter $\bar\tau =f(\tau )$,
the action remains invariant,
if $\rho (\tau )$ is simultaneously changed
as follows:
\begin{equation} \label{19.258}
\rho  \rightarrow  \rho / f'.
\end{equation}
For the proper time
\begin{equation}
\tau = \s,
\label{propt@}\end{equation}
the extremal variable $ \rho(s)$ is
identically equal to unity.
Thus we can use the Lagrangian (\ref{19.245})
for $ \rho\equiv 1$ to find the correct relativistic particle trajectories
parametrized with the proper time $ \s$. Moreover,
as long as we do not need the numerical value of the action
but only its functional dependence on the paths $x(s)$,
we may drop
the trivial constant last term
$- M/2$ in (\ref{19.245}),
and the action looks exactly
 like a nonrelativistic one, except for the
overall sign:
\begin{equation} \label{nonrelac}
{\cal A}=- \int_{\s_1}^{\s_2}
 d\s  \frac{M}{2}{ \dot x ^2 (\s)}.
\end{equation}
The negative sign ensures that the spatial part of
$\dot x^2 ( \sigma )$ appears with the usual positive
sign.

\subsection[Riemann-Cartan Spacetime]
{Riemann-Cartan Spacetime\label{ssRiemCart}}

In Subsection~\ref{Min} we have learned
how to find the action of a point particle in the presence
of  a magnetic field by simply applying a nonholonomic gauge transformation
to the field-free action.
In the presence of curvature and torsion,
the nonholonomic mapping principle instructs
us to transform the action
(\ref{19.247}), or equivalently,
the actions
(\ref{19.247}) and (\ref{nonrelac})
via the infinitesimal coordinate transformations
(\ref{inftr})  to curvilinear coordinates.
After this we assume the transformation functions $\xi^ \lambda(q)$
to be multivalued.
For finite transformations we use the
mapping
 (\ref{LO}) to transform the action to
an arbitrary metric affine space.
For the paths of the particles, this
implies the mapping
\begin{equation}
\dot q ^\mu=e_i{}^ \mu(q) \dot x^i     ,
\label{10.diffbez}\end{equation}
by which the
 Lorentz-invariant proper time increment
(\ref{PT1}) is mapped into
\begin{equation}
ds=\sqrt{g_{\mu \nu}(q(\tau ))\dot q^\mu(\tau )\dot q^\nu(\tau )} .
\label{prtinc@}\end{equation}
The
action
(\ref{AC1}) and (\ref{AC2})
becomes therefore
\begin{equation}
{\cal A}=-M\int_{\tau _1}^{\tau _2} d\tau  \sqrt{g_{\mu \nu}(q(\tau ))\dot
q^\mu(\tau )\dot q^\nu(\tau )} .
\label{relact@}\end{equation}
whereas the nonrelativistic-looking form
 (\ref{nonrelac})
goes over into
\begin{equation} \label{nonrelac0}
{\cal A}=- \int_{\s  _a}^{\s  _b}
 d\s   \frac{M}{2}{ g_{\mu \nu}(q(\s)) \dot q^\mu (\s  )\dot q^\nu (\s  )}.
\end{equation}

Before proceeding with our main argument
we first observe a general feature of all
actions generated from
flat-space actions
by means of nonholonomic transformations: They are
trivially invariant under ordinary holonomic coordinate transformations.
In the context of multivalued gauge transformations in magnetostatics,
this was
seen before in Subsection~\ref{Min}, where gauge
invariance was automatic.
For the actions (\ref{relact@})
and  (\ref{nonrelac0}) the coordinate
invariance is obvious:
Under a coordinate transformation $q^\mu\rightarrow q'{}^\mu$,
the coordinate differentials
transform like
\begin{eqnarray}
&&dq^\mu\rightarrow dq'{}^\mu=\alpha^\mu{}_ \nu{} dq^\nu  ;~~~~~
\alpha^\mu{}_ \nu{}\equiv \frac{\partial q'{}^\mu}{\partial q{}^\nu},
\label{coortrf@a}\\
&&dq_\mu\rightarrow dq'_\mu= \alpha_\mu{}^ \nu{} dq_\nu;~~~~~
\alpha_\mu{}^ \nu{}\equiv \frac{\partial q'{}_\mu}{\partial q_\nu},
\label{coortrf@}\end{eqnarray}
where
\begin{eqnarray}
  \alpha ^\nu {}_\lambda  \alpha _\nu {}^\mu  =
 \delta _\lambda {}^\mu ,
{}~~~~~~~  \alpha _\nu {}^\mu  \alpha ^\lambda {}_\mu
 = \delta_\nu{}^\lambda .
\label{2.27}\end{eqnarray}
The new coordinate differentials
are related to  flat ones
by a relation like
(\ref{LO}):
\begin{equation}
\label{LOP'}
d x^{a} \, = \, e'{}^{\,a}_{\,\,\,\lambda} ( q' ) \, d q'{}^{\lambda} \, .
\end{equation}
Inserting  (\ref{coortrf@a}),
we obtain
the transformation law for the multivalued tetrads:
\begin{eqnarray}
  e_a{}^\mu  (q) =
        \frac{\partial q{}^\mu }{\partial x^a}\rightarrow  e'{}_a{}^\mu  (q')
\equiv
        \frac{\partial q'{}^\mu }{\partial x^a} & = &
         \frac{\partial q'{}^\mu }{\partial q^\nu }
         \frac{\partial q^\nu }{\partial x^a}
 \,= \alpha ^\mu {}_\nu (q) e_a{}^\nu (q)          ,
\label{2.25a}\\
  e^a{}_\mu (q)
= \frac{\partial x^a}
         {\partial q{}^\mu }\rightarrow e'{}^a{}_\mu (q') \equiv
\frac{\partial x^a}
         {\partial q'{}^\mu } & = & \frac{\partial q^\nu }
           {\partial q'{}^\mu} \frac{\partial x^a}{\partial q{}^\nu }
  = \alpha _\mu {}^\nu (q) e^a{}_\nu  (q).\nonumber
\label{2.25}\end{eqnarray}
Inserting this into (\ref{IN}),
we find
the corresponding transformation law for the metric tensor
\begin{eqnarray}
  g_{\mu \nu}(q) \rightarrow g'_{\mu' \nu'} (q')
=
 \alpha _{\mu'}{}^\mu(q)
 \alpha _{ \nu'}{}^ \nu(q)
 g_{\mu \nu}(q) .
\label{2.25}\end{eqnarray}
Using this and (\ref{coortrf@a}), (\ref{coortrf@}), we readily prove
the invariance
of the proper time increment (\ref{prtinc@}),
and thus of the actions
(\ref{relact@}) and (\ref{nonrelac0})
under arbitrary coordinate transformations.

An arbitrary vector field $v_\mu(q)$ transforms
like
\begin{eqnarray}
v_{\mu}(q) \rightarrow v'_{\mu'} (q')
=
 \alpha _{\mu'}{}^\mu(q)
v_{\mu }(q),~~~~~~
v^{\mu}(q) \rightarrow v'{}^{\mu'} (q')
=
 \alpha ^{\mu'}{}_\mu(q)
v^{\mu }(q),~~~~~~
\label{2.25}\end{eqnarray}
as follows directly from a comparison of the two local representations
for a vector field in flat space
$v_a(q)=
e_a{}^\mu (q)v_\mu(q)=
e'{}_a{}^\mu (q')v_\mu(q')$ and
$v^a(q)=
e^a{}_\mu (q)v^\mu(q)=
e'{}^a{}_\mu (q')v^\mu(q')$.

As announced before,
the closure failure
of parallelograms in a space with torsion
forces us
to
reexamine the variational procedure
in the
action principle
for spinless point particles.
To be consistent, the same nonholonomic mapping which generates
the Riemann-Cartan space
 requires that
the variations in the transformed $q^\mu$-coordinates
are performed as gauge images of the variations in the
euclidean
\begin{figure}[tbh]
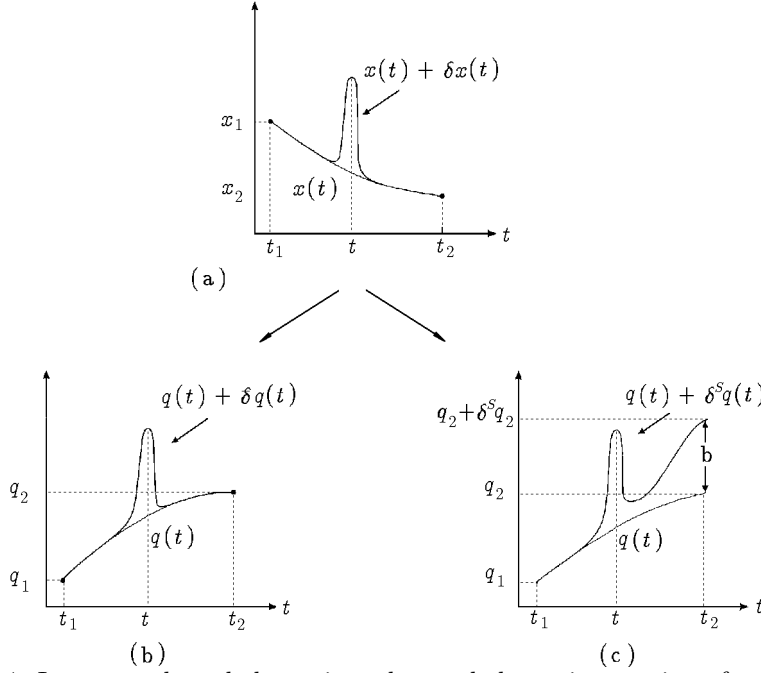

{}\input nonholm3.tdf
\caption[]{Images
under a holonomic and a nonholonomic mapping
of a fundamental path variation.
In the holonomic case,
the paths $x(\tau  )$ and $x(\tau  )+\delta x(\tau  )$  in (a)
turn into the
paths $q(\tau )$ and $q(\tau ) + \deltabar q(\tau )$
in (b). In the
nonholonomic case with $S_ {\mu \nu} {}^{\lambda  } \neq 0$,
they go over into
$q(\tau  )$ and $q(\tau  )+\deltabarf q(\tau  )$
shown in (c) with a {closure failure} $ \deltabarf q_2=b^ \mu $ at
$\tau _2$  analogous
to the Burgers vector $b^ \mu$ in a solid with dislocations.}
\label{act}\end{figure}
$x^i$-space, to be found via
 (\ref{10.diffbez}).
It is easy to see that
the images of variations $ \delta x^i(\tau )$
 are quite different
from ordinary variations as illustrated in Fig.~\ref{act}(a).
The variations  of the Cartesian coordinates $\delta x^i(\tau )$
are performed
at fixed end points  of
the paths. Thus
they form {\em closed paths} in  the $x^i$-space.
Their images, however,
 lie in a space with defects
and thus  possess a \ind{closure failure}
indicating the amount of torsion
introduced by the mapping.
This property will be emphasized by writing the images
 $\deltabarf q^\mu (\tau )$ and calling them {{\em nonholonomic variations}}.

Let  us calculate them explicitly. The paths in the two spaces
   are related by the integral equation
\begin{equation}
    q^\mu (\tau ) = q^\mu (\tau_1) + \int^{\tau }_{\tau_1}
     d\tau '  \,e_i{}^\mu (q(\tau ')) \dot{x}^i(\tau ') .
\label{10.pr2}\end{equation}
%
Note that the left-hand side is well defined even though $ e_i{}^\mu (q(\tau
'))
$ is a multivalued function. When performing the integral
along a specific path $q^\mu(s)$, we may continue
$e_i{}^\mu (q(\tau '))$ analytically through any jumping surface
of the type sketched in Fig.~\ref{loop}.

If a path $x^i( \tau)$-space
is varied by
$ \delta x^i(\tau )$,
equation (\ref{10.pr2}) determines the
associated change in the image path
$q^\mu (\tau )$ by
\begin{eqnarray}
 \deltabarf    q^\mu (\tau )
&=&  \int^{\tau }_{\tau_1}   d\tau '
      \deltabarf [  e_i{}^\mu (q(\tau ')) \dot{x}^i(\tau ')]
\nonumber \\&=&  \int^{\tau }_{\tau_1}   d\tau '
     \{ [\deltabarf   e_i{}^\mu (q(\tau ')) ]\dot{x}^i(\tau ')+
         e_i{}^\mu (q(\tau ')) \delta \dot{x}^i(\tau ')\},
\label{10.pr2p}\end{eqnarray}
which will be referred to as
a {\em nonholonomic variation\/} of the image path $q^\mu(\tau)$.
The superscript $S$ indicates that the properties of
this change
depend crucially
on the torsion in $q$-space.
A comparison with (\ref{10.diffbez}) shows that
the variations $ \deltabarf q^ \mu  $ and the $ \tau$-derivative
of $q^ \mu$
are independent of each other
\begin{equation}
 \deltabarf \dot{q}^\mu (\tau ) = \frac{d}{d\tau } \deltabarf q^\mu (\tau ),
\label{10.pdelta}\end{equation}
just as for ordinary variations $ \delta x^i$ [recall (\ref{VAR3})].

It will be useful to introduce
in addition a
further quantity to be called {\em auxiliary nonholonomic variation}
in
$q^\mu$-space by the relation:
\begin{equation}
  \deltabar q^\mu ( \tau) \equiv e_i{}^\mu (q( \tau)) \delta x^i( \tau).
\label{10.delq}\end{equation}
In contrast to $\deltabarf q^\mu (\tau )$,
these vanish at the endpoints:
\begin{equation}
\deltabar q(\tau_1)=
 \deltabar q(\tau_2)=0,
\label{10.versch}\end{equation}
i.e., they form closed paths in $q^\mu$-space.

With the help of (\ref{10.delq})
we derive from (\ref{10.pr2p}) the relation
\begin{eqnarray}
 \frac{d}{d\tau }\deltabarf    q^\mu (\tau ) &=&
      \deltabarf  [e_i{}^\mu (q(\tau ))] \dot{x}^i(\tau )
      +  e_i{}^\mu (q(\tau ))\delta \dot x^i(\tau )\nonumber \\
     &=& \deltabarf [ e_i{}^\mu (q(\tau )) ]\dot{x}^i(\tau )
      +  e_i{}^\mu (q(\tau )) \frac{d}{d\tau }[e^i{}_ \nu (q(\tau) ) \deltabar
q^ \nu (\tau )].
\label{x}\end{eqnarray}
After inserting
\begin{equation}
\deltabarf  e_i{}^\mu= -\Gamma_{ \lambda  \nu }{}^ \mu \deltabarf q^ \lambda
e_i{}^ \nu  ,
{}~~~~~
\frac{d}{d\tau  } e^i{}_\nu = \Gamma_{ \lambda   \nu  }{}^ \mu  \dot q^ \lambda
e^i{}_\mu,
\label{x}\end{equation}
this becomes
\begin{equation}
\!\!\!\!\!\!\!\!\frac{d}{d\tau } \deltabarf    q^\mu =  -\Gamma_{ \lambda  \nu
}{}^ \mu \deltabarf q^ \lambda \dot q^ \nu
      +
\Gamma_{ \lambda    \nu  }{}^ \mu  \dot q^ \lambda  \deltabar q^ \nu
+ \frac{d}{d\tau } \deltabar  q ^\mu           .
\label{10.prneu}\end{equation}
It is useful to introduce the difference
between the nonholonomic
variation
$ \deltabarf q^\mu $
and the auxiliary nonholonomic variation $ \deltabar q^ \mu $:
\begin{equation}
   \deltabarf b ^\mu\equiv \deltabarf q^\mu -\deltabar q^\mu.
\label{10.deldel}\end{equation}
Then we can rewrite
(\ref{10.prneu})
as a first-order
differential equation for $\deltabarf b^ \mu $:
\begin{equation}
   \frac{d}{d\tau } \deltabarf b^ \mu  = -
            \Gamma _{\lambda \nu }{}^\mu  \deltabarf b ^\lambda
\dot{q}^\nu
             + 2S _{ \lambda \nu  }{}^\mu
           \dot{q}^ \lambda      \deltabar q^ \nu  .
\label{10.p83}\end{equation}

Under an arbitrary nonholonomic
variation   $\deltabarf q^\mu = \deltabar q  ^ \mu+
\deltabarf b ^ \mu  $, the action (\ref{nonrelac0})
changes by
\begin{eqnarray}
  \deltabarf {\cal A}
= -M\int^{\s_2}_{\s_1} d\s  \left( g_{\mu \nu }
               \dot{q}^\nu \deltabarf \dot{q}^\mu + \frac{1}{2}
              \partial _\mu g_{\lambda \kappa }
             \deltabarf q^\mu   \,  \dot{q}^\lambda \dot{q}^\kappa \right),
\label{10.pvar0}\end{eqnarray}
where $\s$ is now the proper time.
Using
(\ref{10.pdelta}) and (\ref{10.versch}) we
partially integrate of the $ \delta \dot q$-term, and apply
the identity
$
 \partial _\mu g_{ \nu  \lambda  }
\equiv \Gamma _{\mu \nu  \lambda  }+
\Gamma _{\mu  \lambda  \nu  }$,
which follows from the definitions
$g_{ \mu  \nu  }\equiv e^i{}_ \mu  e^i{}_ \nu  $ and
 $\Gamma _{\mu \nu }{}^\lambda  \equiv e_i{}^\lambda
 \partial _\mu e^i{}_\nu $, to obtain
\begin{equation}
 \deltabarf {\cal A}
=- M\!\!\int^{\s_2}_{\s_1} d\s  \bigg[ \!-\!g_{\mu \nu } \left(
               \ddot{q}^\nu  + \bar \Gamma _{ \lambda  \kappa }{} ^\nu
                \dot{q}^\lambda \dot{q}^\kappa \right)\deltabar q^ \mu
+\left(g_{ \mu  \nu }
\dot q^ \nu \frac{d}{d\s } \deltabarf b ^ \mu +
\Gamma _{ \mu  \lambda   \kappa  }
\deltabarf b^ \mu \dot{q}^ \lambda \dot{q} ^ \kappa
\right)\bigg].
\label{10.pvar}\end{equation}

To derive the equation of motion we first
vary the action
in a space without torsion.
Then we have
$\deltabarf b^\mu (\s ) \equiv 0$, and we obtain
\begin{equation}
   \deltabarf{\cal A}= \delta {\cal A} =
M\int _{\s_1}^{\s_2}d\s
g_{\mu \nu }(\ddot{q}^\nu + \bar{\Gamma }_{\lambda \kappa }{}^\nu
        \dot{q}^\lambda \dot{q}^\kappa ) \deltabar q^ \nu .
\label{10.delact}\end{equation}
Thus, the action principle $\deltabarf{\cal A}=0$ produces
the equation for the
 geodesics
\begin{equation}
 \ddot{q}^\nu  +
          \bar\Gamma _{\lambda \kappa }{}^\nu \dot{q}^\lambda
           \dot{q}^\kappa =0.
\label{geodes}\end{equation}
This describes
 the correct particle
 trajectories in the absence of torsion.

In the presence of torsion where
$\deltabarf b^ \mu \neq 0$, the equation of motion
receives a contribution
from the second parentheses in
(\ref{10.pvar}).
After inserting
(\ref{10.p83}), the
terms proportional to
$\deltabarf b^ \mu $ cancel
and
the total nonholonomic variation of the action becomes
\begin{eqnarray}
  \deltabarf{\cal A} & = & M \int _{\s_1}^{\s_2}d\s  g_{\mu \nu }\left[
\ddot{q}^\nu +
     \left( \bar{\Gamma }_{\lambda \kappa }{}^\nu +2S^ \nu {} _{\lambda \kappa
}
 \right) \dot{q}^\lambda \dot{q}^\kappa \right] \deltabar q^ \mu  \nonumber \\
   & = & M \int _{\s_1}^{\s_2}d\s  g_{\mu  \nu }\left( \ddot{q}^\nu  +
          \Gamma _{\lambda \kappa }{}^\nu \dot{q}^\lambda
           \dot{q}^\kappa \right) \deltabar q^ \mu  .
\label{10.delat}\end{eqnarray}
The second line follows from the first after using the identity
${\Gamma }_{\lambda \kappa }{}^\nu  = \bar\Gamma _{\{\lambda \kappa \}}{}^\nu
 + 2 S ^\nu {}_{\{\lambda \kappa \}}$.
The curly brackets indicate the symmetrization
of the enclosed indices.
Setting $\deltabarf{\cal A}=0$ and using
(\ref{10.versch}) gives the autoparallel
equation
of motion
\begin{equation}
 \ddot{q}^\nu  +
          \Gamma _{\lambda \kappa }{}^\nu \dot{q}^\lambda
           \dot{q}^\kappa =0.
\label{autop}\end{equation}
Physically, autoparallel trajectories are a manifestation of inertia,
which makes particles run along the straightest lines
rather than the shortest ones.
In the absence of torsion, the
two types of curves happen to coincide.
In the presence
of torsion the autoparallel trajectory is more natural
than the geodesic. It
 is hard to conceive, how a particle should know where to go
to make the trajectory
the shortest curve
to a distant point.
This seems to contradict our concepts of locality.


%
%
In order appreciate the
geometric significance of the
differential equation (\ref{10.p83}), we
introduce the matrices
\begin{eqnarray}
 \label{10.p15}
G {} ^\mu{}_ \lambda  (\tau )  \equiv   \Gamma _{ \lambda  \nu }{}^\mu
               (q(\tau ))\dot{q}^\nu (\tau )
\end{eqnarray}
and
\begin{eqnarray}
    \Sigma  ^\mu{}_\nu (\tau ) \equiv  2   S_{\lambda \nu } {}^\mu
                 (q(\tau )) \dot{q}^ \lambda  (\tau )
                ,
\label{10.p2equ}\end{eqnarray}
and rewrite Eq.~(\ref{10.p83}) as a
differential
equation for a vector
\begin{equation}
   \frac{d}{d\tau } \deltabarf b  = - G\deltabarf b
            +  \Sigma (\tau )\,\deltabar q^ \nu  (\tau ).
\label{10.p83p}\end{equation}
The solution is
\begin{eqnarray}
   \deltabarf b (\tau ) = \int^{\tau }_{\tau_1} d\tau '
          U({\tau ,\tau '})~ \Sigma  (\tau ')~\deltabar q  (\tau '),
\label{10.pdel}\end{eqnarray}
with the matrix
\begin{equation}
  U({\tau ,\tau '}) = T \exp \left[ -\int^{\tau }_{\tau '} d\tau '' G(\tau
'')\right].
\label{10.pum}\end{equation}
In the absence
of torsion,
 $\Sigma (\tau )
 $ vanishes identically and $\deltabarf b (\tau ) \equiv 0$,
and the
variations
 $\deltabarf q^\mu (\tau )$
coincide with the
holonomic
 $\deltabar  q^\mu (\tau )$ [see Fig.~\ref{act}(b)].
In a space with torsion,
the
variations
 $\deltabarf q^\mu (\tau )$ and $\deltabar  q^\mu (\tau )$
are different
from each other
[see
 Fig.~\ref{act}(c)].

The above variational treatment of the action
is somewhat complicated and
calls for
a simpler
procedure \cite{Kleinert6}.
The extra term arising from the second parenthesis
in the variation (\ref{10.pvar})
can traced to a simple property
of the
auxiliary nonholonomic variations
(\ref{10.delq}).
To find this we
form the $ \tau$-derivative
$d_ \tau\equiv d/d \tau$
of the defining equation
(\ref{10.delq})  and find
\begin{equation}
d_ \tau \deltabar q^\mu( \tau)= \partial _ \nu e_a{}^\mu (q(\tau))\,\dot q^
\nu(\tau) \delta x^a(\tau)
+e_a{}^\mu(q(\tau))     d_ \tau  \delta x^a(\tau).
\label{@}\end{equation}
Let us now perform variation $ \deltabar $ and $ \tau$-derivative
in the opposite order
and
calculate
$d_ \tau \deltabar q^\mu( \tau)$. From (\ref{LO}) we have
the relation
\begin{equation}
\label{LO1}
d_ \tau q^{\lambda}(\tau) \,
 = e_{\,a}^{\,\,\,\lambda} ( q (\tau)) \,d_ \tau x^{a}(\tau) \, .
\end{equation}
Varying this gives
\begin{equation}
\deltabar d_ \tau  q^\mu( \tau)=
\partial _ \nu e_a{}^\mu (q(\tau))\, \deltabar
 q^ \nu d_  \tau  x^a(\tau)
+e_a{}^\mu(q(\tau))  \deltabar    d_ \tau x^a.
\label{@}\end{equation}
Since the variation
 in $x^a$-space commute with the $ \tau$-derivatives
[recall (\ref{VAR3})],
we obtain
\begin{equation}
 \deltabar d_ \tau q^\mu( \tau)-d_ \tau \deltabar q^\mu( \tau)
=\partial _ \nu e_a{}^\mu (q(\tau))\, \deltabar
 q^ \nu d_  \tau  x^a(\tau)- \partial _ \nu e_a{}^\mu (q(\tau))\,
\dot q^ \nu( \tau) \delta x^a(\tau)
{}.
\label{@}\end{equation}
After reexpressing
$\delta x^a(\tau) $  and $d_  \tau  x^a(\tau)$   back in terms of
$\deltabar q^\mu( \tau)
$  and
$d_  \tau  q^\mu(\tau)=\dot q^\mu(\tau)$,
this becomes {using (\ref{connection0})
\begin{equation}
  \deltabar d_ \tau q^\mu( \tau)-d_ \tau \deltabar q^\mu( \tau)
= 2S_{ \nu \lambda}{}^\mu
\dot q^ \nu( \tau) \deltabar q^ \lambda(\tau).
\label{COMMUTE}\end{equation}
Thus, due to the closure failure
in spaces with torsion,
the operations  $d_ \tau$ and $ \deltabar$
do not commute in front of the path $q^\mu(\tau)$, implying that
  in contrast to variations $ \delta$, the auxiliary
nonholonomic variations $ \deltabar$ of
velocities $\dot q^\mu(\tau )$
no longer coincide with
the velocities of variations.

This property
 is responsible for
shifting the trajectory from
geodesics to
autoparallels.
Indeed, let us vary an action
\begin{eqnarray}
\label{ACTION}
{\cal A}
= \int\limits_{\tau_1}^{\tau_2} d\tau
L
\left( q^{\mu} (\tau ), \dot{q}^{\mu} ( \tau ) \right)
\end{eqnarray}
directly by $ \deltabar q^ \mu(\tau )$ and impose
(\ref{COMMUTE}), we find
\begin{eqnarray}
\deltabar {\cal A} =  \int\limits_{\tau_1}^{\tau_2}d\tau
 \left\{ \frac{\partial L}{\partial q^{\mu}}
 \deltabar  q^{\mu}
+ \frac{\partial L}{\partial \dot{q}^{\mu}}  \frac{d}{d \tau}  \deltabar
q^{\mu} \right.
 \left. +  2 \,S^\mu{}_{ \nu \lambda}
 \frac{\partial L}{\partial \dot{q}^{ \mu } } \,
\dot{q}^{ \nu}  \deltabar q^{ \lambda} \right\}  .
\label{VARIATION}
\end{eqnarray}
After a partial integration of the second term
using the vanishing $ \deltabar q^ \mu(\tau )$
at the endpoints,
we obtain
the
Euler-Lagrange equation
\begin{eqnarray}
&&
\frac{\partial L}{\partial q^{\,\mu} } -
\frac{d}{d \tau} \frac{\partial L}{\partial
\dot{q}^{\mu} }
=- 2  S_{\mu \nu}{}^ \lambda
\dot{q}^{\nu} \frac{\partial
L}{\partial \dot{q}^{ \lambda} }
.\label{EL}
\end{eqnarray}
This differs from the standard Euler-Lagrange equation
by
an additional contribution due to the torsion tensor.
For the action (\ref{nonrelac0}) with the proper time $\s$
as a path parameter,
we thus obtain the equation of motion
\begin{equation}
M \, \Big[\ddot q^ \mu(\s)+ g^{\mu \kappa}  \Big( \partial_{\nu}
g_{\lambda\kappa}  - \frac{1}{2} \, \partial_{\kappa}
g_{\nu\lambda}  \Big)
-  2 S^ \mu{}_{\nu\lambda}
\Big]\dot{q}^{\,\nu} (\s) \dot{q}^{\lambda}(\s) =
0 ,
\label{autopx}
\end{equation}
which is once more Eq.~(\ref{autop}) for autoparallels.

\section{Compatibility with Conservation Law of Energy Momentum Tensor}
An important consistency check for
the correct equations of motion is based on their rederivation
from the covariant conservation law
for the energy momentum tensor which, in turn,
is a general property of any theory which is invariant
under arbitrary (single-valued) coordinate transformations
(\ref{coortrf@a}),
(\ref{coortrf@}).

To derive this law, we express the  reparametrization  invariance once
more in another way
by studying the behavior of the
relativistic action (\ref{relact@}) under
infinitesimal versions of the coordinate transformation
(\ref{coortrf@}), which we shall write as
local translations
\begin{equation}
q^\mu\rightarrow q'{}^\mu(q)=q^\mu-\xi^\mu(q).
\label{abprel}
\end{equation}
This looks like the previous infinitesimal transformations
(\ref{inftr}), but
now we deal with ordinary coordinate transformation,
where the transformation functions
$-\xi^\mu(q)$ are single-valued and possess commuting derivatives.
As a further difference, the initial space
possesses curvature and torsion.

Inserting (\ref{abprel}) into
(\ref{coortrf@a}) and (\ref{coortrf@}), we have
\begin{eqnarray}
  \alpha ^\lambda{}_\nu & \approx & \delta^\lambda{} _\nu
             - \partial _\nu  \xi ^\lambda  (q),\nonumber \\
   \alpha _\mu {}^\nu  & \approx & \delta _\mu {}^\nu  + \partial _\mu
          \xi ^\nu  (q)                         ,
\label{2.30}\end{eqnarray}
and find from (\ref{2.25a}) and (\ref{2.25})
the infinitesimal transformations
of the multivalued tetrads
$e_a^\mu (q)$:
\begin{eqnarray}
   &&e_a{}^\mu  (q)\rightarrow
   e'{}_a{}^\mu  (q)
+
 \xi ^\lambda \partial _\lambda e_a{}^\mu (q)
                 - \partial _\lambda  \xi ^\mu e_a{}^\mu
                 (q),\\
 &&e^a{}_\mu (q)\rightarrow
 e'{}^a{}_\mu (q)+
  \xi ^\lambda \partial _\lambda
               e^a{}_\mu (q) + \partial _\mu  \xi ^\lambda
               e^a{}_\lambda (q).
\label{2.31}\end{eqnarray}
To save parentheses, differential operators are supposed to act only on the
expression after it.
Inserting (\ref{2.31}) into (\ref{IN}),
we obtain the corresponding transformation law for the metric tensor
\begin{equation}
g_{\mu \nu}(q)\rightarrow
g'_{\mu \nu}(q)
+\xi^ \lambda\partial _ \lambda g_{\mu \nu}(q)
+\partial _\mu\xi^ \lambda  g_{ \lambda \nu}(q)
+\partial _\nu\xi^ \lambda  g_{ \mu\lambda }(q)  .
\label{mettr@}\end{equation}
For an arbitrary vector field $v_\mu(q)$, the transformation laws
(\ref{2.25}) become
\begin{eqnarray}
&&v_{\mu}(q)\rightarrow
v'{}_{\mu }(q)
+\xi^ \lambda\partial _ \lambda v_{\mu }(q)
+\partial _ \mu\xi^  \lambda v_{ \lambda }(q),~~~~  \nonumber \\
&&v^{\mu}(q)\rightarrow
v'{}^{\mu }(q)
+\xi^ \lambda\partial _ \lambda v^{\mu }(q)
-\partial _ \lambda\xi^ \mu  v^{ \lambda }(q).
\label{@}\end{eqnarray}

Recalling (\ref{2.9}), the change of the metric
can be rewritten as
\begin{equation}
 \delta_E
g_{\mu \nu}(q)
=\bar D_\mu\xi_ \nu(q)
+\bar D_\nu\xi_ \mu(q),
\label{deltaeg@}\end{equation}
where  $\bar D_\mu$ are covariant derivatives
defined as in (\ref{covderx}), but with the
Riemann connection  (\ref{2.9}) instead of the affine connections:
\begin{equation}
\bar D_\mu v_ \nu=\partial _\mu v_ \nu- \bar\Gamma_{\mu \nu}{}^ \lambda v_
\lambda,~~~~~
\bar D_\mu v^  \lambda=\partial _\mu v^ \lambda+\bar \Gamma_{\mu \nu}{}^
\lambda
v^ \nu.~~~~~
\label{covderxb}\end{equation}
The subscript of $\delta_E$ indicates that these are the general coordinate
transformations
introduced by Einstein.
With this notation, the change of a vector field is
\begin{equation}
 \delta_E v_{\mu}(q)
=\xi^ \lambda\partial _ \lambda v_{\mu }(q)
+\partial _ \mu\xi^  \lambda v_{ \lambda }(q),~~~~
 \delta_E
v^{\mu}(q)  =
\xi^ \lambda\partial _ \lambda v^{\mu }(q)
-\partial _ \lambda\xi^ \mu  v^{ \lambda }(q).
\label{chv@}\end{equation}
Inserting for $v^\mu(q)$ the coordinate $q^\mu$ themselves, we
see that
\begin{equation}
\delta_E q^\mu=-\xi^\mu(q),
\label{deltaeq@}\end{equation}
which is the initial transformation (\ref{abprel}) in this notation.

We now calculate the change of the action (\ref{relact@})
under infinitesimal Einstein transformations:
\begin{equation}
 \delta_E
{\cal A}=  \int d^4q
\frac{ \delta{\cal A}}{ \delta g_{\mu \nu}(q)} \delta_E    g_{\mu \nu}(q)
+\int d\s
\frac{ \delta{\cal A}}{ \delta q^{\mu}(\s )} \delta_E    q^{\mu }( \s ).
\label{deltaeA@}\end{equation}
The functional derivative
 $ \delta{\cal A}/ \delta g_{\mu \nu}(q)
$ is the general definition of the energy momentum tensor
of a system:
\begin{equation}
\frac{ \delta{\cal A}}{ \delta g_{\mu \nu}(q)} \equiv - \frac{1}{2}\sqrt{-g}\,
T^{\mu \nu}(q),
\label{@}\end{equation}
where $g$ is the determinant of $g_{\mu \nu}$.
 For the spinless particle at hand, the energy momentum tensor
becomes
\begin{equation}
T^{\mu \nu}(q)=\frac{1}{ \sqrt{-g} } M \int d\s \,\dot
q^\mu (\s )
q^\nu (\s ) \,\delta^{(4)}(q-q(\s )),
\label{explenm@}\end{equation}
where $\s$ is the proper time.
This and the explicit variations
(\ref{deltaeg@}) and (\ref{deltaeq@}),
bring
(\ref{deltaeA@})
to the form
\begin{equation}
 \delta_E
{\cal A}= -\frac{1}{2} \int d^4q    \sqrt{-g} T^{\mu \nu}(q)
[\bar D_\mu\xi_ \nu(q)
+\bar D_\nu\xi_ \mu(q)]-\int d\tau
\frac{ \delta{\cal A}}{ \delta q^{\mu}(\tau )} \xi^{\mu }( q(\tau) ).
\label{deltaeA3@}\end{equation}
A partial integration of the derivatives yields (neglecting boundary terms at
infinity)
\begin{eqnarray}
 &&\delta_E
{\cal A}=  \int d^4q  \,\left\{
 \partial _ \nu
[  \sqrt{-g} T^{\mu \nu}(q)]
\!+ \!\sqrt{-g}
\bar  \Gamma_{ \nu \lambda}{}^\mu(q)  T^{ \lambda \nu}(q)
\right\}
\xi_ \mu(q)
\!-\!\int d\tau
\frac{ \delta{\cal A}}
{ \delta q^{\mu}(\tau )} \xi^{\mu }( \tau).
\label{deltaeA4@}
\end{eqnarray}
Because of the  manifest invariance of the action under
general coordinate transformations, the
left-hand side has to vanish
for
arbitrary infinitesimal functions
$\xi^\mu(\tau )$.
We therefore obtain
\begin{eqnarray}
 &&\left\{ \partial _ \nu
[ \sqrt{-g} T^{\mu \nu}(q)]
+ \sqrt{-g} \bar \Gamma_{ \nu \lambda}{}^\mu  T^{ \lambda \nu}(q)
\right\}
\xi_\mu(q)
\!-\!\!\int d\tau
\frac{ \delta{\cal A}}
{ \delta q^{\mu}(\tau )}
 \delta^{(4)}(q\!-\!q(\tau )) \xi^{\mu }( \tau )=0.
\label{194@}\end{eqnarray}
To find
the physical content of this equation
we consider first a space
without torsion.
On a particle trajectory, the action is extremal,
so that
the second term vanishes, and
 we obtain the
 covariant conservation law:
\begin{equation}
 \partial _ \nu
[ \sqrt{-g} T^{\mu \nu}(q)]
+ \sqrt{-g}\bar \Gamma_{ \nu \lambda}{}^\mu (q) T^{ \lambda \nu}(q)
=0.
\label{consLAW@}\end{equation}
Inserting
(\ref{explenm@}), this becomes
\begin{eqnarray}
&& M \int d\s \,[\dot
q^\mu (\s)
\dot q^\nu (\s)\partial _\nu \delta^{(4)}(q-q(\s))
+
 \bar\Gamma_{ \nu \lambda}{}^\mu (q)\dot q^\nu (\s)\dot q^ \lambda(\s)
 \,\delta^{(4)}(q-q(\s))]=0
{}.
\label{explenm1}\end{eqnarray}
A partial integration turns this into
\begin{eqnarray}
&& M \int d\s \,[-
\ddot q^\mu (\s)
+
\bar \Gamma_{ \nu \lambda}{}^\mu(q) \dot q^\nu (\s)\dot q^ \lambda(\s)
]\,\delta^{(4)}(q-q(\s))=0
{}.
\label{explenm2}\end{eqnarray}
Integrating this over a small volume around any
 trajectory point
$q^\mu(s)$, we obtain the equation (\ref{geodes}) for the geodesic trajectory.

This technique
was used by Hehl in his derivation of
 particle trajectories
in the presence of torsion.
Since torsion does not appear in the
action, he found that the trajectories to be geodesic.

The conservation  law (\ref{consLAW@})
can be written more covariantly as
\begin{equation}
 \sqrt{-g} \bar D_ \nu T ^{\mu \nu}(q)=0.
\label{covLAW1@}\end{equation}
This follow directly from the identity
\begin{equation}
\frac{1}{ \sqrt{-g} }\partial _ \nu \sqrt{-g}   =\frac{1}{2}g^{ \lambda \kappa
}\partial _ \nu g_{ \lambda \kappa}
=\bar  \Gamma_{ \nu \lambda}{}^ \kappa,
\label{@}\end{equation}
and is a consequence of the
rule of partial integration applied to
(\ref{deltaeA3@}), according to which
a
covariant derivative  can be treated in a volume integral
$\int d^4 \sqrt{-g}f(q)\bar D g(q)$, just like an ordinary derivative in
an
euclidean integral $\int d^4 x f(x) \partial _a g(x)$
[see Appendix A]. After a partial integration, neglecting surface terms,
Eq.~(\ref{deltaeA3@})
goes over into
\begin{equation}
 \delta_E
{\cal A}=\frac{1}{2}  \int d^4q    \sqrt{-g}\left[
 \bar D_ \nu T^{\mu \nu}(q) \xi_ \nu(q)+(\mu\leftrightarrow   \nu)\right]
-\int d\tau
\frac{ \delta{\cal A}}{ \delta q^{\mu}(\tau )} \xi^{\mu }( q(\tau) ).
\label{deltaeA3@p}\end{equation}
whose vanishing for all $\xi^\mu(q)$
yields
directly
(\ref{covLAW1@}).

Our theory does not lead to this
conservation law.
In the presence of torsion,
the particle trajectory does not satisfy
${ \delta{\cal A}}/
{ \delta q^{\mu}(\tau  )} =0$, but according to
(\ref{EL}):
\begin{equation}
\frac{ \delta{\cal A}}
{ \delta q^{\mu}(\tau )}
=
\frac{\partial L}{\partial q^{\,\mu} } -
\frac{d}{d \tau } \frac{\partial L}{\partial
\dot{q}^{\mu} }
= 2  S_{\mu \nu}^{\,\,\,\,\,\, \lambda}
\dot{q}^{\nu} \frac{\partial
L}{\partial \dot{q}^{ \lambda} }.
\label{198@}\end{equation}
For the Lagrangian in the action (\ref{relact@}),
parametrized with the the proper time
$\s$, the right-hand side becomes
\begin{equation}
 2  S_{\mu \nu}^{\,\,\,\,\,\, \lambda}
\dot{q}^{\nu} \frac{\partial
L}{\partial \dot{q}^{ \lambda} }=-M \, 2  S_{\mu \nu \lambda}
\dot q^\nu(\s)
\dot q^ \lambda(\s)       .
\label{198@}\end{equation}
Inserting this into
(\ref{deltaeA3@p}),
equation
(\ref{explenm2}) receives an extra term and becomes
\begin{eqnarray}
&& M \int d\s \,\left\{ -
\ddot q^\mu (\s)
+
[\bar\Gamma_{ \nu \lambda}{}^\mu (q)
+2S^\mu{}_{ \nu \lambda} (q)
]\dot q^\nu (\s)\dot q^ \lambda(\s)
\right\} \,\delta^{(4)}(q-q(\s))       =0
{}.
\label{explenm2n}\end{eqnarray}
yielding
the correct autoparallel trajectories
(\ref{autop})
for spinless point particles.

Observe that the extra term (\ref{198@})
can be expressed in terms of the energy momentum tensor
(\ref{explenm@})  as
\begin{equation}
 \sqrt{-g} 2S^\mu{}_{ \nu \lambda}  T^{ \lambda \nu}(q)
\xi_\mu(q)                        .
\label{@}\end{equation}
We may therefore rewrite the
change of the action (\ref{deltaeA3@})
as
\begin{equation}
 \delta_E
{\cal A}= -\frac{1}{2} \int d^4q    \sqrt{-g} T^{\mu \nu}(q)
[\bar D_\mu\xi_ \nu(q)
+\bar D_\nu\xi_ \mu(q)-4S^ \lambda{}_{\mu \nu}\xi_ \lambda(q)].
\label{deltaeA3p@}\end{equation}
The quantity in brackets will be denoted by
$\deltabar_E g_{\mu \nu}$, and is equal to
\begin{equation}
\deltabar_E g_{\mu \nu} = D_\mu\xi_ \nu(q)
+ D_\nu\xi_ \mu(q)                      ,
\label{@}\end{equation}
where $D_\mu$ is the
covariant derivative
(\ref{covderx}) involving the full affine connection.
Thus we have
 \begin{equation}
 \delta_E
{\cal A}= - \int d^4q    \sqrt{-g}\, T^{\mu \nu}(q)
 D_\nu\xi_ \mu(q).
\label{deltaeA3pp@}\end{equation}
Integrals over invariant
expressions containing the
covariant derivative $D_\mu$ can be
integrated by  parts
according to a rule very similar to that
for the Riemann covariant derivative
$\bar D_\mu$, which is derived in Appendix A. After neglecting
surface terms we find
 \begin{equation}
 \delta_E
{\cal A}=  \int d^4q    \sqrt{-g}\, D_\nu^*T^{\mu \nu}(q)
 \xi_ \mu(q),
\label{deltaeA3pp@}\end{equation}
where
$D_\nu^*=D_\nu+2S_{\nu  \lambda}{}^ \lambda$.
Thus, due to the closure failure in spaces with torsion,
the energy-momentum tensor
of a free spinless point particles
satisfies the conservation law
\begin{equation}
D_\nu^*T^{\mu \nu}(q)=0.
\label{covLAWx}\end{equation}
This is to be contrasted
with the conservation law
(\ref{covLAW1@}). The difference between the two laws
can best be seen by rewriting
(\ref{covLAW1@})
as
\begin{equation}
D_\nu^*T^{\mu \nu}(q)+2S_{ \kappa }{}^ {\mu }{}_{\lambda}(q) T^{ \kappa
\lambda}(q)=0.
\label{covLAW2@}\end{equation}
This is the form in which the conservation law
has usually been stated
in the literature \cite{Utiyama,Kibble,Hehl1,Hehl2,Kleinert1}.
When written in the form
(\ref{covLAW1@}) it is obvious that
(\ref{covLAW2@}) is satisfied only by geodesic trajectories,
in contrast to (\ref{covLAWx}) which is satisfied by autoparallels.

The variation $\deltabar _Eg_{\mu \nu}(q)$ plays a
similar role
in deriving the new conservation law (\ref{covLAW2@})
as the nonholonomic variation
$\deltabar q(s)$ of Eq.~(\ref{10.pr2p})
does in deriving equations of motion
for point particles.
Indeed, we
may rewrite
the transformation
(\ref{deltaeA@}) formally as
\begin{equation}
 \deltabar_E
{\cal A}=  \int d^4q
\frac{ \delta{\cal A}}{ \delta g_{\mu \nu}(q)} \deltabar_E    g_{\mu \nu}(q)
+\int d\tau
\frac{ \delta{\cal A}}{ \delta q^{\mu}(\tau )} \deltabar_E    q^{\mu }( \tau ).
\label{deltaeAx@}\end{equation}
Now the last term vanishes
according to the new action principle $\deltabar {\cal A}=0$
from which we derived the
autoparallel trajectory (\ref{autopx})
by setting (\ref{VARIATION}) equal to zero.

\comment{
This is easy to understand.
We simply observe that a theory which is invariant under general
coordinate transformations
is also invariant under
the nonholonomic image
of general coordinate transformations $dx^a\rightarrow x'{}^a=dx^a-d\xi^a(x)$.
Under these transformations,
in the flat space.  The In fact, performing the variations}


The question arises whether the new conservation law
(\ref{covLAWx}) allows for the construction of an extension of
Einstein's field equation
\begin{equation}
\bar G^{\mu \nu}= \kappa T^{\mu \nu}
\label{EEQ@}\end{equation}
to spaces with torsion,
where $ \bar G^{\mu \nu}$
 is the Einstein tensor formed from the
Ricci tensor $\bar R_{\mu \nu}\equiv \bar R_{ \lambda\mu \nu}{}^ \lambda$
in Riemannian spacetime [$ \bar R_{ \mu \nu \lambda}{}^  \kappa$
being the same covariant curl of $\bar  \Gamma_{\mu \nu}{}^ \lambda$
as $R_{ \mu \nu \lambda}{}^  \kappa$
is of
$  \Gamma_{\mu \nu}{}^ \lambda$ in Eq.~(\ref{RC})].
The standard extension of (\ref{EEQ@}) to spacetimes with torsion
replaces the left-hand side
by the  Einstein-Cartan tensor
 $  G^{\mu \nu}\equiv  R_{\mu \nu}-\frac{1}{2}g_{\mu \nu} R_ \s{}^ \s$
and becomes
\begin{equation}
 G^{\mu \nu}=\kappa T^{\mu \nu}
\label{EEQC}\end{equation}
The Einstein-Cartan tensor $G^{\mu \nu}$
satisfies a Bianchi identity
\begin{equation}
D_\nu^*G_{ \mu}{}^\nu +2S_{ \lambda\mu}{}^ \kappa G_ \kappa{}^ \lambda
-\frac{1}{2}S^{ \lambda}{}_{ \kappa}{}^{; \nu} R_{\mu  \nu \lambda}{}^
\kappa=0,
\label{Bianchid1@}\end{equation}
where $S^{ \lambda}{}_{ \kappa}{}^{; \nu} $
is the Palatini tensor defined by
\begin{equation}
S_{ \lambda  \kappa}{}^{;  \nu}\equiv 2(S_{ \lambda \kappa}{}^  \nu
+ \delta_{ \lambda}{}^ \nu S_{ \kappa  \s}{}^ \s
- \delta_{  \kappa}{}^  \nu S_{ \lambda  \s}{}^ \s).
\label{palat@}\end{equation}
It is then
concluded
that the energy momentum tensor
satisfies
the conservation law
\begin{equation}
D_\nu^*T_{ \mu}{}^\nu +2S_{ \lambda\mu}{}^ \kappa T_ \kappa{}^ \lambda
-\frac{1}{2 \kappa}S^{ \lambda}{}_{ \kappa}{}^{; \nu} R_{\mu  \nu \lambda}{}^
\kappa =0.
\label{cons2l@}\end{equation}
For standard field theories of matter, this is indeed true
if the Palatini tensor satisfies the second Einstein-Cartan field
equation
\begin{equation}
S^{ \lambda \kappa;\nu} = \kappa \Sigma ^{ \lambda \kappa;\nu} ,
\label{@}\end{equation}
where
$ \Sigma^{ \lambda \kappa;\nu} $ is the canonical spin density of the matter
fields.
A spinless point particle contributes
only to the first two terms in (\ref{cons2l@}),
in accordance with
(\ref{covLAW2@}).

What tensor will stand on the left-hand side of the field equation
(\ref{EEQC}) if the energy momentum tensor satisfies the conservation law
(\ref{covLAWx}) instead of (\ref{covLAW2@})?
At present, we can give an answer
  \cite{Peln}
 only for the case of a pure
gradient torsion  which has the general form \cite{hojman}
\begin{equation}
S_{\mu \nu}{}^ \lambda                      =
\frac{1}{2}[
\delta_{\mu}{}^ \lambda\partial _ \nu   \s
-\delta_{\nu}{}^ \lambda\partial _ \mu   \s].
\label{gradtor@}\end{equation}
Then
we
may simply replace (\ref{EEQC}) by
\begin{equation}
e^{\s }G^{\mu \nu}= \kappa T^{\mu \nu}.
\label{EEQCn}\end{equation}
 Note that for gradient torsion, $G^{\mu \nu}$ is symmetric
as can be deduced from the fundamental identity (which expresses merely the
fact that
the Einstein-Cartan tensor $R_{\mu \nu \lambda}{}^ \kappa$ is the covariant
curl
of the affine connection)
\begin{equation}
{D^*}_ \lambda S_{\mu \nu}{}^{; \lambda}=G_{\mu \nu}-G_{ \nu\mu}.
\label{@}\end{equation}
Indeed, inserting
 (\ref{gradtor@})
into
(\ref{palat@}),
we find the
Palatini tensor
\begin{equation}
S_{ \lambda  \mu}{}^{;\kappa}\equiv -2[ \delta_ \lambda{}^ \kappa
\partial _\mu  \s -( \lambda\leftrightarrow  \mu)].
\label{palatgr@}\end{equation}
This has a vanishing covariant derivative
\begin{equation}
{D^*}_ \lambda S_{\mu \nu}{}^{; \lambda}=
-2[{D^*}_ \mu \partial _ \nu   \s- {D^*}_  \nu \partial _ \mu   \s]
=2[
S_{\mu \nu}{}^ \lambda \partial _  \lambda   \s
-2S_{\mu \lambda}{}^  \lambda  \partial _  \nu   \s
+2S_{\nu \lambda}{}^  \lambda  \partial _  \mu   \s]
, \label{@}\end{equation}
since  the terms on the right-hand
side cancel
after using
(\ref{gradtor@}) and
 $S_{\mu   \lambda}{}^  \lambda\equiv S_\mu
=-\frac{3}{2}\partial _\mu \s$.
Now
we insert
(\ref{gradtor@}) into the Bianchi identity
(\ref{Bianchid1@}), with the result
\begin{equation}
\bar D_\nu^*G_{  \lambda}{}^\nu
+\partial _ \lambda \s G_ \kappa{}^  \kappa
-\partial _  \nu \s G_  \lambda{}^  \nu
+2\partial _\nu \s R_{ \lambda}{}^ \nu=
0.
\label{Bianchidgr11@}\end{equation}
Inserting
here $R_{ \lambda \kappa}=G_{ \lambda \kappa}-\frac{1}{2}g_{ \lambda \kappa}G_
\nu{}^ \nu$,
this becomes
\begin{equation}
D_\nu^*G_{ \lambda}{}^\nu
+\partial _  \nu \s G_  \lambda{}^  \nu
=
0.
\label{Bianchidgr12@}\end{equation}
Thus we find the Bianchi identity
\begin{equation}
D^*_ \nu (e^{ \s}G_{ \lambda}{}^ \nu)=0.
\label{@}\end{equation}
This makes the
 left-hand side
of the new field equation (\ref{EEQCn})
compatible
with the
covariant new conservation law (\ref{covLAWx}), just as in Einstein's theory.

The field equation for the $ \s$-field is still unknown.

\section[Gauge Field Representation of Particle Orbits]
{Gauge Field Representation of Particle Orbits\label{GDOFO}}

In Section~\ref{@MAG} we have given two examples
for the use of multivalued fields in describing magnetic phenomena.
Up to now, we have only transferred the second example
in Subse.~\ref{@GMFMV} to geometry
by generating nontrivial gauge fields from
multivalued
gauge transformations.

The exists an equally important geometric version also
for the mathematical structure in the
first example in \ref{@GRCL}, the gradient representation
of the magnetic field, as we shall elaborate in this section.

\subsection{Current Loop with Magnetic Forces}
To prepare the grounds for this
we pose ourselves the problem of calculating
the magnetic energy of current loop
from the gradient representation of the magnetic field.
Since this will provide us with an example for the construction of
field actions, we shall consider the
 energy as a euclidean action and denote it by ${\cal A}$.
In this sense, the magnetic ``action" reads
\begin{equation}
{\cal A}=\frac{1}{2}\int d^3 x\, {\bf B}^2({\bf x}).
\label{@}\end{equation}
Remembering the gradient representation
(\ref{mgF}) of the magnetic field, this becomes
\begin{equation}
{\cal A}=\frac{I^2}{2(4\pi)^2}\int d^3 x \,[ \nablab  \Omega({\bf x})]^2.
\label{dipolelx@}\end{equation}
This holds for the multivalued
solid angle $ \Omega({\bf x})$.
In order to perform field theoretic calculations, we go over to the
single-valued representation
used in Eqs.~(\ref{OmeS}) and (\ref{bomex}). Recalling (\ref{bomexa}), the
action becomes
\begin{equation}
{\cal A}=\frac{I^2}{2(4\pi)^2}\int d^3 x \,[ \nablab  \Omega({\bf x})-4\pi
\deltab({\bf x};S)
]^2,
\label{dipolel@}\end{equation}
where we have expressed the integral over the
magnetic dipole surface in (\ref{bomexa})
with the help of the $ \delta$-function on the surface $S$:
\begin{equation}
\deltab({\bf x};S)\equiv \int _S d{\bf S}' \delta^{(3)}({\bf x}-{\bf x}').
\label{@}\end{equation}
The $ \delta$-function is essential in removing the unphysical
field energy on the artificial magnetic dipole layer
on $S$ which is only serves to make the solid angle single-valued.
Its unphysical nature can be exhibited in the action
(\ref{dipolel@}) as follows:
Suppose we move the surface
$S$ to a new location $S'$, while keeping its boundary
anchored on the current loop $L$.
Under this move, the $ \delta$-function on the surface changes as follows (see
\cite{Kleinert1I,Camb}):
\begin{equation}
    \deltab ({\bf x};S) \rightarrow  \deltab({\bf x};S')
 =  \deltab({\bf x};S) + \nablab \delta ({\bf x};V).
\label{linegau@}\end{equation}
Here
\begin{equation}
 \delta ({\bf x};V) = \int d^{3}  x'  \delta ^{(3)} ({\bf x} -
 {\bf x}')
\label{splusV@}\end{equation}
is the $ \delta$-function on the volume $V$ over which $S$ has swept
in moving to $S'$.
Thus the $ \delta$-function on the surface $S$
 is a  {\em gauge field
of the current loop\/}, and (\ref{linegau@})
is a gauge transformation which leaves the boundary of $L$ unchanged.
The action
(\ref{dipolel@}) is also invariant, since
the gradient of the $ \delta$-function in (\ref{linegau@})
can be absorbed into $ \Omega({\bf x})$:
\begin{equation}
 \Omega({\bf x})\rightarrow
 \Omega'({\bf x})=
 \Omega({\bf x})+4\pi  \delta ({\bf x};V) .
\label{@}\end{equation}
The gauge invariance makes the field energy independent
of the
position of the artificial magnetic dipole layer for a
current flowing
along the fixed loop $L$.
This gauge invariance has its root in
the fact that
$\Omega$ is defined only up to integer multiples of
 $4\pi$ --- it is a {\em cyclic\/} field.

 We are now ready to calculate the magnetic field energy of
the current loop.
For this we rewrite the action (\ref{dipolel@})
in terms of an {\em auxiliary\/} vector field
$ {\bf B}({\bf x})$ as
\begin{equation}
{\cal A}=\int d^3 x   \left\{ -\frac{1}{2}{\bf B}^2({\bf x})
+ {\bf B}({\bf x})\cdot[ \nablab  \Omega({\bf x})/4\pi -I \deltab({\bf
x};S)]\right\} ,
\label{dipoleln@}\end{equation}
A partial integration brings the middle term to
\begin{eqnarray}
-\int d^3 x
 [ \nablab \cdot {\bf B}({\bf x})] \Omega({\bf x})/4\pi.\nonumber
\end{eqnarray}
Extremizing this in $\Omega({\bf x})$
yields the equation
\begin{equation}
 \nablab\cdot  {\bf B}({\bf x})=0,
\label{@}\end{equation}
implying that the field lines of ${\bf B}({\bf x})$ form closed loops.
This equation may be enforced identically (as a Bianchi identity)
by expressing ${\bf B}({\bf x})$
as a curl of an
auxiliary vector potential ${\bf A}({\bf x})$,  setting
\begin{equation}
{\bf B}({\bf x})\equiv \nablab\times {\bf A}({\bf x}).
\label{@}\end{equation}
With this ansatz, the
equation which brings the action (\ref{dipoleln@})
to the form
\begin{equation}
{\cal A}=\int d^3 x   \left\{ -\frac{1}{2}[\nablab\times {\bf A}({\bf x})]^2
- [\nablab\times {\bf A}({\bf x})]\cdot I \deltab({\bf x};S)\right\} .
\label{dipolel2@}\end{equation}
A further
  partial integration leads to
\begin{equation}
{\cal A}=\int d^3 x   \left\{ -\frac{1}{2}[\nablab\times {\bf A}({\bf x})]^2
-{\bf A}({\bf x})\cdot I[ \nablab\times
 \deltab({\bf x};S)\right\} ,
\label{dipolel2'@}\end{equation}
and we identify in the linear term in ${\bf A}({\bf x})$
the {\em auxiliary current\/}
\begin{equation}
 {\bf j}({\bf x})\equiv I\,\nablab\times \deltab({\bf x};S) .
 \label{auxc@}
\end{equation}
This current is conserved
for closed loops $L$.
This follows from the property
of the $ \delta$-function on an arbitrary line $L$ connecting the points
${\bf x}_1$ and ${\bf x}_2$:
\begin{equation}
\nablab\cdot  \deltab({\bf x};L)=
 \delta({\bf x}_2)
- \delta({\bf x}_1)
\label{currc@}\end{equation}
For closed loops, the right-hand side vanishes.

We now observe that Stokes' theorem
can be rewritten as
an identity for $ \delta$-functions
\begin{equation}
 \nablab\times   \deltab ({\bf x};S) =
  \deltab ({\bf x};L).
\label{stokesr}\end{equation}
 This shows that the auxiliary current
(\ref{auxc@}) is equal to
(\ref{currdensLx}).
The field equation following from the action
(\ref{dipolel2@})  is Amp\`ere's law
(\ref{Amp}).
Thus the auxiliary quantities
${\bf B}({\bf x})$
${\bf A}({\bf x})$, and
${\bf j}({\bf x})$ coincide with the usual magnetic quantities
with the same name.

By inserting the explicit solution  (\ref{vecpotI}) of Amp\`ere's
law into the energy, we obtain the {\em Biot-Savart\/} enery for an arbitrary
current distribution
\begin{equation}
{\cal A}=\frac{1}{4\pi}\int d^3 x
d^3x'\,
{\bf j}({\bf x})\frac{1}{|{\bf x}-{\bf x}'|}
{\bf j}({\bf x}')   .
\label{dipolel2BS@}\end{equation}

The relations
 (\ref{auxc@})
implies that the
 $ \delta$-function on the surface $S$
is a gauge  field whose curl produces a unit current loop.
Thus
the action
(\ref{dipolel2@}) is invariant under two mutually dual gauge
transformations, the usual magnetic one
\ref{gautr@})
by which the vector potential receives a gradient of an arbitrary scalar field,
and the transformation  gauge transformation
(\ref{splusV@}), by which the irrelevant surface $S$ is moved
to another
configuration $S'$.

Thus we have proved the complete equivalence
of the gradient representation of the magnetic field to
the usual gauge field representation.
In the gradient representation, there exists a new type of gauge invariance
which
expresses the physical irrelevance of  the  jumping surface
appearing when using single-valued solid angles.

The action (\ref{dipolel2'@}) describes magnetism
in terms of a {\em double gauge theory\/}, in which the gauge of ${\bf A}({\bf
x})$
and the shape of $S$ can be changes arbitrarily.

\subsection{Particle World Lines with Gravitational Forces}

It is possible to transfer the
entire
double-gauge structure to geometry. In this way we
can derive
a theory in which not only the
gravitational forces are represented
by a  metric affine geometry, but also the
particle orbits.
The latter can be
reexpressed in terms of particle world lines, more specifically,
the
Einstein tensor of the second gauge structure
becomes the energy momentum tensor of the
particle world line.
It is the analog of the auxiliary current
(\ref{auxc@}). The conservation law
(\ref{currc@}) which is satisfied automatically by the Einstein tensor
turns into the conservation law of the energy-momentum
tensor for the world lines.

We shall present such a construction
only for a system without torsion.
For simplicity, we assume the world as a crystal in four
Riemannian spacetime dimensions.
If the crystal is distorted by a displacement field
\begin{equation}
 q^\mu \rightarrow q'{}^\mu
  = q^\mu +u^\mu(q),
\label{@}\end{equation}
it has a strain energy
\begin{equation}
{\cal A}=\frac{M}{4} \int d^4 q\,
 \sqrt{-g}  (\bar D_\mu u_ \nu
      +\bar  D_ \nu u_\mu )^2 ,
\label{4}\end{equation}
where $M$ is some elastic modulus.
If part of the distortions are of the plastic type,
the world crystal contains defects defined by
 Volterra surfaces, where crystalline layers or sections have been cut out.
The displacement field is multivalued,
and the action (\ref{4}) is the analog of the magnetic action
(\ref{dipolelx@}) in the presence of a current loop.
In order to do field theory with this action, we have to make
the displacement field single-valued
with the help of $ \delta$-functions describing the
jumps across the Volterra surfaces,
in complete analogy with
the magnetic energy (\ref{dipolel@}):
\begin{equation}
{\cal A}=
M\int  d^4x \sqrt{-g}  ( u_{\mu \nu}
      - u_{\mu \nu}^P)^2,
\label{4pr}\end{equation}
 where
$ u_{\mu \nu} =   (\bar D_\mu u_ \nu +  \bar D_ \nu u_\mu)/2$ is
the elastic strain tensor and
$ u_{\mu \nu}^P$ the gauge field of plastic deformations
describing the Volterra surfaces via
 $ \delta$-functions on these surfaces \cite{Kleinert1}.
  The energy density is invariant under the single-valued
{\em defect gauge transformations\/} [the analogs of (\ref{linegau@})]
\begin{equation}
 u_{\mu \nu} {}^P \rightarrow u_{\mu \nu}{}^P + (\bar D_\mu \lambda_ \nu
+ \bar D_ \nu \lambda_\mu)/2 ,~~~~~~~~~u_\mu\rightarrow u_\mu+ \lambda_\mu.
\label{5}\end{equation}
  Physically, they express the fact that defects are
not affected by elastic
distortions of the crystal. Only  multivalued
gauge functions $ \lambda_\mu$ would change
the defect content in $u_{\mu\nu}^P$.

We now introduce an auxiliary
symmetric tensor field
$G_{\mu\nu}$ and rewrite the action (\ref{4pr})
in a first-order form [the analog of (\ref{dipoleln@})] as
\begin{equation}
{\cal A}=
\int  d^3 q \sqrt{g}  \left[\frac{1}{4\mu}
G_{\mu\nu}
G^{\mu\nu}
+iG ^{\mu\nu}(
u_{\mu \nu}
 - u_{\mu\nu}^P)\right].
\label{4pr}
\end{equation}
After a partial integration and extremization in $u_\mu$,
the middle terms yield the
equation
\begin{equation}
\bar D _\nu G^{\mu\nu}=0.
\label{@}\end{equation}
This may be guaranteed identically, as a Bianchi identity,
by an ansatz
\begin{equation}
           G^{\nu\mu} =
e^{\nu  \kappa  \lambda  \s}
e^{\mu   \kappa \lambda'   \s'}
\bar D _ \lambda
\bar D _{  \lambda'}
 \chi_{ \s \s'}.
\label{2.87b}
\label{@}\end{equation}
The field $\chi_{ \sigma  \sigma '}$ plays the role
of an elastic gauge field.
Inserting this into (\ref{4pr})
we obtain the analog of
(\ref{dipolel2@}):
\begin{equation}
{\cal A}=
 \int  d^4 q \sqrt{-g}
 \left\{\frac{1}{4M}
\left[ e^{\nu  \kappa  \lambda  \s}
e^{\mu   \kappa \lambda'   \s'}
\bar D _ \lambda
\bar D _{  \lambda'}
 \chi_{ \s \s'}
\right]^2 +{i}
 e^{\nu  \kappa  \lambda  \s}
e^{\mu   \kappa \lambda'   \s'}
\bar D _ \lambda
\bar D _{  \lambda'}
 \chi_{ \s \s'}
 u_{\mu\nu}^P
\right\}.
\label{4pr2}\end{equation}
A further partial integration brings this to the form
\begin{equation}
{\cal A}=
\int  d^4 q \sqrt{-g}  \left\{\frac{1}{4M}G_{\mu \nu}G^{\mu \nu}
 +{i}
 \chi_{\mu\nu}
  T^{\mu\nu}\right\},
\label{4pr3}\end{equation}
where $ T  _{\mu\nu}$ is the
defect density defined in analogy to $ \eta_{ij}$ of Eq.~(\ref{2.91}):
\begin{equation}
     T^{\mu\nu} =
e^{\nu  \kappa  \lambda  \s}
e^{\mu   \kappa \lambda'   \s'}
\bar D _ \lambda
\bar D _{  \lambda'}
 u^P_{ \s \s'}.
\label{6}\end{equation}
It is
  invariant under defect gauge transformations (\ref{5}), and satisfies the
conservation law
\begin{equation}
\bar D_\nu  T^{\mu\nu}  =0.
\label{Tconsl@}\end{equation}

Although we have written (\ref{6}) and
(\ref{Tconsl@})  covariantly,
they are only applicable in their
linearized approximations to infinitesimal defects,
as emphasized in the discussion after
eq.~(\ref{x}).
By identifying  $\chi_{\mu\nu}$ with
half an elastic metric field $g_{\mu\nu} $ [generalizing the linearized
expression in terms of the strain field in Eq.~(\ref{2.89}),
where the metric
is
$g_{\mu\nu}= \delta_{\mu\nu}+2\xi_{\mu\nu}$],
the tensor
$G_{\mu\nu}$
is recognized as the
Einstein tensor associated with
the metric tensor $g_{\mu \nu}$.
The defect density $ T  _{\mu\nu}$
is formed in the same way from  the plastic strain $u^P_{\mu \nu}$.

For small deviations
 $\chi'_{\mu \nu}$
of $\chi_{\mu \nu}$ from flat space limit $
 \eta_{\mu \nu}/2$,
we can
 linearize $G  _{\mu\nu}$
in $\chi'_{\mu \nu}$
and find
\begin{eqnarray}
 \bar G_{\mu \nu}~&\approx&~
 \epsilon^{\nu  \kappa  \lambda  \s}
 \epsilon^{\mu   \kappa \lambda'   \s'}
\partial  _ \lambda
\partial  _{  \lambda'}
  \,\chi'_{ \s \s'}.
 \nonumber \\
&=&-(\partial ^2  \chi'_{\mu \nu}+\partial _\mu\partial _ \nu  \chi'_{
\lambda}{}^ \lambda
-\partial _\mu\partial _ \lambda  \chi'_\mu{}^{ \lambda}
-\partial _\nu\partial _ \lambda  \chi'_\mu{}^{ \lambda})
 + \eta_{\mu \nu}(\partial ^2 \chi'_{ \lambda}{}^ \lambda-\partial _{
\lambda}\partial _ \kappa \chi'{}^{ \lambda \kappa}).
\label{@}\end{eqnarray}
Introducing the
field $\phi_{\mu \nu}\equiv \chi'_{\mu \nu}-\frac{1}{2}
\eta_{\mu \nu}\chi'_ \lambda{}^ \lambda$,
and going to the
Hilbert gauge $\partial _\mu\phi^{\mu\nu}=0$,
the Einstein tensor reduces to
\begin{eqnarray}
 \bar G_{\mu \nu}= -\partial ^2\phi_{\mu \nu},
\label{@}\end{eqnarray}
and the
interaction energy of an arbitrary distribution of defects
[the analog of (\ref{dipolel2BS@})]
\begin{equation}
{\cal A}\approx{M}\int d^4q d^4q' \,
T  _{\mu\nu}(q)\,
 \Delta({q}-{q}')\,
 T _{\mu\nu}(q').
\label{elastM@}\end{equation}
where
\begin{equation}
\Delta(q)=\int\frac{ d^3p}{(2\pi)^4}\,\frac{e^{ipq}}{({ p}^2)^2}
\label{@}\end{equation}
  is the Green function of the differential operator
$(\partial ^2)^2$.

The interaction (\ref{elastM@}) gives the elastic energy
of matter in the world crystal.
The
defect density
$ T _{\mu\nu}(q)$
 plays a similar role as the energy-momentum
tensor $\displaystyle\mathop T ^m{} _{\mu\nu}(q)$ of matter
in gravity.
Indeed, it satisfies the same conservation law (\ref{@}).
The interaction does not, however, coincide with the
gravitational energy
for which the Green function should be
that of the Laplacian
$ \partial ^2$ rather than $(\partial ^2)^2$
to yield Newton's gravitational potential $\propto r^{-1}$ [as
in the magnetic Biot-Savart energy (\ref{dipolel2BS@})].

There is no problem in modifying our world crystal
to achieve this.
We merely have to replace the action (\ref{4pr3})
by
\begin{equation}
{\cal A}=      \int  d^4q \sqrt{-g}  \left[- \frac{1}{2 \kappa}  \bar R
-\frac{1}{2}
 g_{\mu\nu}
  T ^{\mu\nu}\right],
\label{4pr4}\end{equation}
where $ \kappa$ is the gravitational constant.
Indeed, the Einstein action in the first term has the
 linear approximation
\begin{eqnarray}
\frac{1}{4 \kappa}\int d^4q \,g_{\mu \nu}G^{\mu \nu}\approx
\frac{1}{2 \kappa}\int d^4q\, \phi^{\mu \nu}(-\partial ^2)\phi_{\mu \nu}
\label{@}\end{eqnarray}
which leads to the
 field equation
\begin{equation}
-\partial ^2\phi^{\mu \nu}= \kappa  T^{\mu\nu},
\label{@}\end{equation}
and thus
to the correct gravitational interaction energy.

It is easy to verify that the energy
(\ref{4pr4}) is invariant
under defect gauge transformations (\ref{5}), just as
the elastic action (\ref{4pr2}).

A similar construction exists for a full
nonlinear Einstein-Cartan theory of gravity \cite{KlEIC}.
\comment{ There the first-order
elastic action (\ref{4pr}) reads
\begin{eqnarray}
{\cal A}&=&\int d^4 q \sqrt{-g} \left\{
M G_{\mu}{}^ \alpha G^ \nu{}_ \alpha
+M' S_{\mu \alpha;}{}^ \beta S^{\mu \alpha;}{}_ \beta\right.\nonumber \\
&&~~~~~~~~~~~~~+ \left.
G^\mu{}_\alpha\left[
 D_\mu u^ \alpha - \omega_\mu{}^ \alpha
-(A_{ \beta\mu}{}^ \alpha-2S_{ \beta\mu}{}^ \alpha)
u^ \beta-h^P{}^\alpha{}_\mu)
\right]
\right.\nonumber \\&&
{}~~~~~~~~~~~~~+\left.
S^{\mu \alpha;}{}_ \beta\left[
D_\mu  \omega_{ \alpha}{}^ \beta -A^P_{\mu \alpha}{}^ \beta+D_\mu(u^ \gamma A_{
\gamma \alpha}{}^ \beta
)-u^ \gamma   F_{\mu  \gamma \alpha}{}^ \beta
 \right] \right\}
\label{@}\end{eqnarray}
where
the plastic gauge fields
are $ h^P{}^\alpha{}_\mu
$ and $ A^P_{\mu \alpha}{}^ \beta$
are the gauge fields representing the defects.
Their density is characterized by the
field strengths
\begin{equation}
  \displaystyle \mathop F^P{}_{\mu  \nu \alpha}{}^ \beta=
D_\mu  A^P_{\nu   \alpha}{}^ \beta
-D_\nu  A^P_{\nu   \alpha}{}^ \beta-
A^P_{\mu \alpha}{}^  \gamma
A^P_{\nu \gamma}{}^ \beta
-
A^P_{\nu \alpha}{}^  \gamma
A^P_{\mu \gamma}{}^ \beta
\label{@}\end{equation}
for the disclinations and
\begin{eqnarray}
S^P_{ \alpha \beta;}{}^\mu=
2(S^P_{ \alpha \beta}{}^\mu
+h_ \alpha{}^\mu S^P_ \beta-
h_  \beta{}^\mu S^P_  \alpha)
\label{@}\end{eqnarray}
for the dislocations,
where $\tilde D_ \alpha$ is the covariant derivative
formed with the hlep of the gauge field $A_{ \alpha \beta}{}^ \gamma$
and $S^P_{ \alpha \beta}{}^\mu$ is the torsion tensor
obtained from the connection
$ \Gamma^P_{ \alpha \beta}{}^\mu=
h^P_ \gamma{}^{\mu}
h^P_ \beta{}^{ \lambda}
\tilde D_ \alpha h^P{}^{ \gamma }{}_{ \lambda}$
of local Lorentz
transformations the dislocations
and $\displaystyle h^m{}^\alpha{}_\mu
$ is used to
relate covariant vector indices to local Lorentz indices.
The field $ \omega_\mu{}^ \alpha$ is the
rotational part of the distortion tensor $D_\mu u^ \alpha$.
With the help of auxiliary fields
$ G_\mu{}^ \alpha=h^ \alpha_ \nu  \s_\mu{}^ \nu$
and $S _{\mu \alpha;}{}^ \alpha$, the action can be brought to the first-order
form
By extremizing this in $u^ \alpha $ and $ \omega_ \alpha{}^ \beta$, we obtain
the
covariant
conservation laws
\begin{eqnarray}
\label{@}\end{eqnarray}
}
\subsection{Field Representation for Ensembles of Particle World Lines}
To end this section let us mention that
a grand-canonical ensemble of world lines can be transformed into
a quantum field theory \cite{Kleinert1I}.
In this way, we convert the double gauge theory into a field theory
with a single gauge field.
This construction may eventually be helpful in finding the correct theory of
gravitation
with torsion.

\section{Embedding}

The readers who feel uneasy in dealing
with the unfamiliar multivalued tetrads $e^a{}_\mu(q)$ in (\ref{LO})
may be convinced that autoparallels are the
correct particle trajectories of spinless point particles
in another way:
by the special geometric role
of autoparallels in a Riemann-Cartan space
generated
by embedding.
It is well known, that
a $D$-dimensional space with curvature can be produced
by embedding it into a flat space
of a sufficiently large dimension $\bar D>D$
via some functions $x^A(q)$ ($A=1,\dots,\bar D$).
The metric  $\eta_{AB}$
 in this flat space
is pseudo-Minkowskian, containing
only
 diagonal elements $\pm 1$.
The mapping
$x^A(q )$ is smooth, but cannot be inverted to $q^\mu
 (x)$. Let
${\bf E}_A$ be the $\bar D$ fixed
basis vectors in the
 embedding space, then the functions $x^A(q)$
define
$D$ local tangent vectors to the submanifold:
\begin{eqnarray}
 {\bf E}_\lambda (q ) = {\bf E}_A E^A{}_\lambda
        (q );~~~~~~~ { E}^A{}_ \lambda(q)\equiv     \frac{\partial x^A(q)}
{\partial q^\lambda}
\label{2.166}\end{eqnarray}
  They induce a metric
\begin{eqnarray}
  g_{\lambda  \kappa } (q ) = { E}^A{}_\lambda
          (q) { E}^B{}_\kappa  (q) \eta_{AB},
\label{2.167}\end{eqnarray}
which can be used to define the reciprocals
\begin{eqnarray}
 E^{A\lambda } \left( q \right) = g^{\lambda \kappa}
     (q ) E^A {}_ \kappa( q) .
\label{2.168}\end{eqnarray}
Note that in contrast to our multivalued tetrads in (\ref{ORTHO}),
the tangent vectors
satisfy only the orthogonality relation
\begin{eqnarray}
 E^{A\mu}(q) E_{A \nu }(q)= \delta^\mu{}_ \nu,
\label{2.169o}\end{eqnarray}
but not the completeness relation
\begin{eqnarray}
 E^{A\lambda }(q) E_{B\lambda }(q) \neq \delta ^A{}_B,
\label{2.169}\end{eqnarray}
the latter being obvious since the sum over $ \lambda=1,\dots,D<\bar D$
is too
small to span a $\bar D$-dimensional space.
The embedding induces an affine connection in $q$-space
\begin{equation}
\label{connection0E}
 \Gamma_{\mu \nu}{}^ \lambda(q)\equiv
E_{A}{}^{\lambda} ( q )\partial_{\mu} \, E^{\,A}{}_{\nu} ( q ) \, =-
  E^{\,A}{}_{\nu} ( q ) \, \partial_{\mu} \,E_{A}{}^{\lambda} ( q )  .
\end{equation}

Since ${ E}^A{}_ \lambda(q)\equiv {\partial x^A(q)}/
{\partial q^\lambda}  $ are derivatives
of  single-valued embedding functions $ x^A(q)$,
they satisfy a Schwarz integrability condition
[in contrast to (\ref{SCHWARZ})]:
\begin{equation}
\label{SCHWARZE}
\partial_{\mu} \, E^{\,A}_{\,\,\,\lambda} ( q ) \, - \,
\partial_{\lambda}\, E^{\,A}_{\,\,\,\mu} ( q ) =  0 \,.
\end{equation}
The torsion
(\ref{torsion0@}) is therefore
necessarily zero.

Because of their single-valuedness, derivatices commute in front of
the tangent vectors $E^{A}{}_{\lambda} ( q )$,
so there exists no
formula of the type
(\ref{RC})
to calculate the
curvature:
\begin{equation}
R_{\mu \nu \lambda }{}^\kappa ( q ) \neq
E_{A}{}^{ \kappa} ( q )
\left( \partial _\mu\partial _ \nu-
\partial _\nu\partial _ \mu \right)
E^{A}{}_{\lambda} ( q ) =0.
\label{RCE}\end{equation}
In order to derive the curvature tensor
(\ref{10.30}) from (\ref{RC}), we needed the property
\begin{eqnarray}
  \partial _\mu e^a{}_\nu = \Gamma _{\mu \nu }{}^\lambda
     e^a{}_\lambda ,
\label{2.181}\end{eqnarray}
which was deduced from (\ref{connection0})
 using the completeness relation
$e^a{}_\mu(q)e_b{}^\mu(q)= \delta^a{}_b$.
Since such a relation does not exist now [see (\ref{2.169})],
we have
\begin{eqnarray}
  \partial _\mu E^A{}_\nu(q) \neq \Gamma _{\mu \nu }{}^\lambda (q)
     E^A{}_\lambda (q),
\label{2.181E}\end{eqnarray}
and a formula of the type (\ref{RCE})
cannot be used to find $R_{\mu \nu \lambda}{}^ \kappa(q)$.

It is possible to introduce torsion
in the embedded $q$-space \cite{kshab}
 by allowing the tangent vectors to disobey
the Schwarz  integrability  condition (\ref{SCHWARZE}).
In contrast to the multivalued tetrads $e^a{}_\mu(q)$, however,
the functions  $E^A{}_\mu(q)$ possess commuting derivatives.
This brings them in spirit close to the ordinary tetrads or vierbein fields
$h^ \alpha{}_\mu(q)$, except that there are more of them.
For nonintegrable functions $E^A{}_\mu(q)$, the embedding is not defined
pointwise
but only differentially:
\begin{equation}
dx^A=E^A{}_\mu(q)dq^\mu.
\label{mappt@}\end{equation}
For any curve $x^A(\tau)$, we can find a curve in $q$-space which is defined up
to a free choice of the initial point:
\begin{equation}
\dot q^\mu(\tau)=
\dot q^\mu(\tau_1)+\int _{\tau_1}^\tau d\tau'\,E_A{}^\mu(q(\tau '))dx^A(\tau').
\label{@}\end{equation}
In contrast to
(\ref{10.pr2}), the integrand does not require an analytic continuation through
cuts.

A straight line in the embedding $x$-space
has a constant velocity $v^A(s)=\dot x^A(s)$.
Its image in the embedded space
via the mapping (\ref{mappt@}) satisfies
\begin{equation}
E^A{}_\nu(q)\ddot q^\nu(s) + \dot E^A{}_\nu(q(s))\dot q^\nu(s)
     =0.
\label{@}\end{equation}
Multiplying this equation by
$E_A{}^ \mu(q)$ and using
the orthogonality relation
(\ref{2.169o}) as well as
the defining equation (\ref{connection0E}),
we find the Eq.~(\ref{autop}),
so that the straight line
goes over into
an autoparallel trajectories.
Geodesic trajectories, on the other hand,
correspond to complicated curves in $x^A$-space
under this mapping, a fact
which makes them once more
unappealing candidates for physical trajectories
of spinless point particles, apart from the
inertia argument
given after Eq.~(\ref{autop}).

\section[Coulomb System as an Oscillator in a Space with Torsion]
{Coulomb System as an Oscillator in a Space with Torsion}
\label{@Coul}As an application of the new action principle
with the ensuing autoparallel trajectories,
consider
the famous
Kustaanheimo-Stiefel transformation in celestial mechanics
\cite{KUST,Kleinert4}.
For a spinless point particle
orbiting
around a central mass in a three-dimensional space,
the Lagrangian reads:
\begin{equation}
L (x, \dot x) = \frac{M}{2} \vert \dot {\bf x} \vert^2 +
  \frac{ \alpha }{r} ,~~~~~~~~r=|{\bf x}|,
\label{@}\end{equation}
where ($ \alpha  =$ const), yielding upon variation
the
 equation of motion
\begin{equation}
M \ddot{\bf x} +  \alpha  {\bf x}
 /r^3 = 0.
 \label{@}\end{equation}
Let us perform the Kustaanheimo-Stiefel
 transformation in two steps: First we map
the ${\bf x}$-space into a four-dimensional ${\vec u}$-space by setting
\begin{equation}
\begin{array}{ll}
   x^1 = 2 \left(u^1 u^3 + u^2 u^4\right),
 &
\\
   x^2 = 2 \left(u^1 u^4 - u^2 u^3\right),&\\
   x^3 = \left(u^1\right)^2 + \left(u^2 \right)^2 -
	 \left(u^3\right)^2  - \left(u^4\right)^2,&
\end{array}
   ~~~\mbox{for}~~~\vec{ u} = \left(
\begin{array}{l}
   u^1\\
    u^2\\
     u^3\\
      u^4
\end{array}  \right)
{}.
\label{@}\end{equation}
 Every point ${\bf x}$ has infinitely many image points ${\vec u}$.
We restrict this freedom by
extending
 ${\bf x}$-space
to four dimensions with the help of
an artificial forth coordinate $x^4$,  which we map nonholonomically
into ${\vec u}$-space by an equation
\begin{equation}
dx^4=
   u^2du^1 -u^1du^2+ u^4 du^3-u^3du^4.
\label{@}\end{equation}
The combined anholonomic coordinate transformation
reads
 $d x^a = e^i{}_ \mu  (\vec u) du^\mu
   ,~~a,\mu = 1,2,3,4,\,$ with the matrix:
\begin{equation}
   e^a{}_ \mu (\vec u)  =   \left(
\begin{array}{rrrr}
   u ^3 & u^4  & u^1 & u^2 \\
   u ^4 &  -u^3 & -u^2 & u^1 \\
   u^1  & u^2 & -u^3  & -u^4 \\
   u^2 & -u^1  & u^4  & -u^3
\end{array}            \right)  .
\label{@}\end{equation}
The metric induced in $\vec u$ space is
\begin{equation}
g_{\mu \nu}(\vec u)= {\vec u }^2 \delta_{\mu \nu}.
\label{@}\end{equation}
It is easy to check that
derivatives in front of $x^4(\vec u)$
do not commute:
\begin{equation}
(\partial _{u_1}\partial _{u_2}-\partial _{u_2}\partial _{u_1})x^4=2,~~~~~~~
(\partial _{u_3}\partial _{u_4}-\partial _{u_4}\partial _{u_3})x^4=2,
\label{@}\end{equation}
implying the multivaluedness
of $x^4(\vec u)$
and the presence
of torsion,
whose nonzero components are
\begin{equation}
S_{1 \mu 2} =
    S_{ 3 \mu 4} = 4 ( u^2, -u^1, u^4, -u^3).
\label{@tor}\end{equation}
The fourth, anholonomic coordinate $x^ 4$
is assumed to have a trivial dynamics. By
adding only a kinetic term
to  the original Lagrangian,
the new one is defined by
\begin{equation}
  L' (\vec x, \dot {\vec x}) =
 \frac{M}{2}  \vert \dot{\vec x}\vert ^2+\frac{ \alpha}{r},
,~~~~~~~~r=|{\bf x}|
\label{extendL@}\end{equation}
By extremizing this
we obtain the correct three-dimensional orbits
by imposing the constraint
$x^4 =$ const.
 This system is now mapped to $\vec u$-space using
$d x^a = e^i{}_ \mu  (\vec u) du^\mu$,
and we obtain the transformed
Lagrangian
\begin{equation}
 L(\vec u, \dot {\vec u}) = 2 M\vec {u }^2       \dot{\vec u}^2 +
     \frac{ \alpha }{\vec u^2}
{}.
 \label{@}\end{equation}
An arbitrary orbit in he four-dimensional $x^4$-space
 can now be found by extremizing this action
using our modified action principle. This yields
the equation
\begin{equation}
\frac{\delta {\cal A}}
  { \delta u^ \mu   } + {\cal F}_ \mu  = 0
\label{@}\end{equation}
where ${\cal F}_ \mu$ is the torsion force
\begin{equation}
{\cal F}_ \mu =  M \dot x^4 \dot S_{1 \mu 2}.
\label{@}\end{equation}
This is the equation for an autoparallel
in $\vec u$-space, which maps correctly back into
the equation of motion following from
the four-dimensional Lagrangian
(\ref{extendL@}).

When the solutions are restricted
by the constraint $\dot x^4=0$, the torsion force
disappears, so that it does not influence the classical
orbit at the end.
However, fluctuations make $\dot x^4$ nonzero
 so that the solution of the associated quantum system \cite{Kleinert4}
which describes a hydrogen atom
via a path integral is sensitive
to this force.

\section{Quantum Mechanics}

As mentioned in the Introduction, the nonholonomic mapping principle
was discovered when trying to solve
the  quantum-mechanical mechanical problem
of finding
the correct integration
measure for the path integral of the hydrogen atom.
Let us briefly sketch
the result.

In flat space, quantum mechanics may be defined
via path integrals as products
of ordinary integrals
over Cartesian coordinates on a grated time axis:
\begin{equation} \label{10.49}
({\bf x}\,t \vert {\bf x}'t') =
\frac{1}{\sqrt{2\pi i \epsilon\hbar/M}^D}\prod_{n=1}^{N}
\left[
\int_{-\infty}^{\infty} d \Delta x_n \right] \prod_{n=1}^{N+1} K_0^\epsilon
(\Delta
{\bf x}_n),
\end{equation}
where $K_0^\epsilon (\Delta {\bf x}_n)$ is an
abbreviation for the short-time amplitude
\begin{equation} \label{10.50}
K_0^\epsilon (\Delta {\bf x}_n) \equiv
\langle {\bf x}_n \vert\exp
\left( -\frac{i}{\hbar} \epsilon \hat{H}\right)
\vert {\bf x}_{n-1}\rangle
= \frac{1}{\sqrt{2\pi i \epsilon \hbar/M}^D}
\exp\left[{\frac{i}{\hbar}\frac{M}{2}\frac{(\Delta{\bf x}_n)^2}{\epsilon}}
\right]
\end{equation}
with $\Delta {\bf x}_n \equiv {\bf x}_n -{\bf x}_{n-1},\, {\bf x}
\equiv {\bf x}_{N+1},\,
{\bf x}' \equiv {\bf x}_0\,$.
A possible external
potential may be
 omitted since this would
contribute  in
an additive way,
uninfluenced by the space geometry.

The path integral may now be
transformed directly to spaces with
curvature and torsion
by applying the nonholonomic mapping formula
(\ref{LO})
to the
small but finite increments
$ \Delta {\bf x}$
in the action
as well as the measure of integration.
The correct result is found only by writing the initial measure in the
above form, and not in the form
\begin{equation}
\prod_{n=1}^{N}
\left[
\int_{-\infty}^{\infty} d x_n \right] ,
\label{@}\end{equation}
which in flat space is the same thing, but leads to a
wrong
measure in noneuclidean space.

There is a good reason
for having $ \Delta{\bf x}$ in the flat-space measure
at the start of
the nonholonomic transformation.
According to
Huygens' principle of wave optics, each point of a wave front
is a center of a new spherical wave propagating from that
point. Therefore, in a
 \ind{time-sliced path integral},
the differences
$\Delta x_n^i$ play a more fundamental role than the
coordinates themselves.

The details have been explained in the textbook
\cite{Kleinert4} and in my Carg\`ese lecture \cite{Kleinert7}, and need not be
repeated here.
As an important result,
we have {\em derived\/} for a nonrelativistic point particle of mass $m$ in a
purely
curved space the Schr\"odinger equation
\begin{equation}
-\frac{1}{2m}D^\mu D_\mu\psi(q,t)=i\hbar \partial _t   \psi(q,t),
\label{11.35bb}\end{equation}
which does not contain an extra $R$-term as in the earlier
literature on this subject  \cite{DEWITT}.
The operator
$ D^\mu D_\mu $ is equal to $\bar D^\mu \bar D_\mu-S^\mu\partial ^\mu$,
where
$\bar D^\mu \bar D_\mu$  coincides with the well-known Laplace-Beltrami
differential operator
$ \Delta=g^{-1/2}\partial _\mu g^{1/2} g^{\mu \nu}\partial _ \nu$

The appearance of the Laplace operator
$D_\mu D^\mu $ in the
Scr\"odinger equation (\ref{11.35bb})
is in conflict with
the traditional
physical \ind{scalar product} between two
wave functions $\psi _1(q)$ and
$\psi_2 (q)$:
\begin{equation} \label{11.58}
\langle \psi _2|\psi _1\rangle
\equiv \int d^Dq\sqrt{ g(q)}\psi_2 ^*(q)\psi _1(q).
\end{equation}
In such a scalar product, only the
 \ind{Laplace-Beltrami operator}
is a hermitian,
not the
 \ind{Laplacian} $D_\mu D^\mu $.
The bothersome term is the contracted torsion term
$-2S^\mu \partial _\mu \psi $ in $D^\mu D_\mu$.
This term ruins the hermiticity and thus
also the unitarity of the time evolution operator
of a particle in a space with curvature and torsion.

For presently known field theories of
elementary particles
the unitarity problem is fortunately absent.
There
the torsion field $S_{\mu \nu }{}^\lambda $
is generated by the \ind{spin current density} of the fundamental
matter fields.
The requirement of renormalizability restricts
 these fields
to
carry spin $1/2$.
However, the spin current density of spin-$1/2$ particles
happens to be a completely
antisymmetric tensor.\footnote{See, for example,
\aut{H.~Kleinert},
{\em Gauge Fields in Condensed Matter\/}, op.~cit.,
 Vol.~II, Part IV,  {\em Differential Geometry of Defects
and Gravity with Torsion\/}, p.~1432
.} This property
 carries over to the torsion tensor.
Hence, the torsion field in the universe satisfies $S^\mu =0$.
This implies that for a particle in a
universe with curvature and torsion,
the Laplacian always degenerates into the Laplace-Beltrami operator,
assuring unitarity after all.

The Coulomb system discussed in Section~\ref{@Coul} has
another
way of escaping the unitarity problem.
The path integral of this system
is solved by a transformation to
 a space with torsion  where
the physical
scalar product is
\cite{Kleinert4}
\begin{equation} \label{11.61}
\langle \psi _2|\psi _1\rangle _{\rm phys}
\equiv \int d^Dq \sqrt{g}\,w(q)\psi_2 ^*(q)\psi _1(q).
\end{equation}
with some scalar weight function $w(q)$.
This scalar product is different from
the naive
 scalar product
(\ref{11.58}). It is, however,
reparametrization-invariant, and $w(q)$ makes
the Laplacian $D_\mu D^\mu$
a Hermitian operator.

The characteristic property of torsion in this space
is that $S_ \mu(q) $
can be written as a gradient of a scalar function:
 $S_ \mu(q)= \partial _ \mu  s(q)  $
[see Eq.~(\ref{@tor})].
The same thing is true for any gradient torsion
(\ref{gradtor@})
with
\begin{equation}
s (q)=(1-D)\sigma(q)/2.
\label{@sfunc}\end{equation}
The weight-function is
\begin{equation}
w(q)=e^{-2 s (q)}.
\label{@weightf}\end{equation}
Thus, the physical scalar product can be expressed in terms of the
naive one
as follows:
\begin{equation} \label{11.61b}
\langle \psi _2|\psi _1\rangle _{\rm phys}
\equiv \int d^Dq \sqrt{ g(q)}e^ {-2\sigma(q)} \psi_2 ^*(q)\psi _1(q).
\end{equation}
Within this scalar product, the Laplacian $D_\mu D^\mu$ is, indeed, Hermitian.

To prove this, we observe that within the naive scalar product
(\ref{11.61}), a partial integration changes the covariant derivative $
-D_ \mu $ into
\begin{equation}
D_ \mu ^*\equiv (D_ \mu +2S_ \mu ).
 \label{11.partabl}\end{equation}
Consider, for example,
the scalar product
\begin{equation}
\int d^Dq \sqrt{g} U^{ \mu \nu_1\dots\nu_n}D_ \mu V_{ \nu _1\dots\nu_n}.
\label{@}\end{equation}
A partial integration of the derivative term $\partial _ \mu $ in $D_ \mu $
gives
\begin{eqnarray}
&&\hspace{-3cm}\mbox{surface term} ~-~ \int d^Ddq[(\partial _ \mu  \sqrt{g}
U^{ \mu \nu_1\dots\nu_n}) V_{ \nu _1\dots\nu_n}\nonumber \\
&& ~~~~~~~~~~~
- \sum_i \sqrt{g}U^{ \mu \nu_1\dots\nu_i\dots\nu_n}
 \Gamma_{ \mu  \nu _i}{}^ {\lambda _i}
V_{ \nu _1\dots \lambda _i\dots\nu_n}].
\label{gfcm3.47}\end{eqnarray}
Now we use
\begin{equation}
\partial _ \mu  \sqrt{g} =
 \sqrt{g}\;\bar \Gamma_{ \mu  \nu }{}^ \nu =
 \sqrt{g} (2S_ \mu +\Gamma_{ \mu  \nu }{}^ \nu ),
\label{@}\end{equation}
to rewrite
(\ref{gfcm3.47}) as
\begin{eqnarray}
&&\!\!\!\!\!\!\!\!\!\!\!\!\!\!\!\!\!\!\!\!\!\!\!\!\!\!\!\!\!\!\mbox{surface
term}
{}~-~
\int d^Dq \sqrt{g} \left[(\partial _ \mu
U^{ \mu \nu_1\dots\nu_n}) V_{ \nu _1\dots\nu_n}\right.\nonumber \\
&&~~~\hspace{-1pt}- \left.\sum_i  \Gamma_{ \mu  \nu _i}{}^ {\lambda _i}
U^{ \mu \nu_1\dots\nu_i\dots\nu_n} V_{ \nu _1\dots \lambda _i\dots\nu_n}\right.
\left.-
2S_ \mu U^{ \mu \nu_1\dots\nu_n} V_{ \nu _1\dots\nu_n}\right],
\label{@}\end{eqnarray}
which is equal to
\begin{eqnarray}
\mbox{surface term} &-& \int d^Dq \sqrt{g} (D_ \mu ^*
U^{ \mu \nu_1\dots\nu_n})V_{ \nu _1\dots\nu_n}.
\label{@}\end{eqnarray}
In the physical scalar product (\ref{11.61b}),
the corresponding operation is
\begin{eqnarray}
&&\!\!\!\!\!\!
\!\!\!\!\!\!\!\!\!\!\!\!
\int d^Dq \sqrt{g} e^{-2 \sigma (q)}U^{ \mu \nu_1\dots\nu_n}D_ \mu V_{ \nu
_1\dots\nu_n}
=
\nonumber \\
&&\!\!\!\!\!\!
{}~~~~=\mbox{surface term} - \int d^Dq \sqrt{g}(D_ \mu ^*
 e^{-2 \sigma (q)}
U^{ \mu \nu_1\dots\nu_n})V_{ \nu _1\dots\nu_n}\nonumber \\
&&\!\!\!\!\!\!
{}~~~~=
\mbox{surface term} - \int d^Dq \sqrt{g}
e^{-2 \sigma (q)}
(D_ \mu   \sqrt{g}
U^{ \mu \nu_1\dots\nu_n})V_{ \nu _1\dots\nu_n}.
\label{@}\end{eqnarray}
Hence, $iD_ \mu $
is a Hermitian operator, and so is
the Laplacian $D_\mu D^\mu$.


For spaces with  an arbitrary torsion, the
correct scalar product has yet to be found.
Thus the quantum equivalence principle is so far only
appicable to spaces with arbitrary curvature and gradient torsion.

\section{Relativistic Scalar Field in Space with Gradient Torsion}
\label{@RELFGT}
The scalar product in the above quantum mechanical system
is the key to the construction of an action
for a relativistic scalar field
whose particle trajectories are autoparallels.
{}From (\ref{@sfunc})
and (\ref{@weightf})
we must use a weigth factor
$w(q)=e^{-3 \sigma (q)}$ in the scalar product.
This scalar product introduced in \cite{Kleinert4}
has recently become the basis
of a series of studies in general relativity
\cite{saa,fizn}. In the latter work,
the action of a
relativistic free scalar field
$\phi$
was set up as follows:
\begin{equation}
{\cal A}[\phi]= \int d^4x\, \sqrt{-g} e^{-3 \sigma } \,
\left({\frac 1 2} g^{\mu \nu}| \nabla_\mu \phi \nabla_\nu \phi| -
\frac{m^2}{2}| \phi|^2 e^{-2 \sigma }\right).
\label{sfAF}
\label{@freef}\end{equation}
The associated Euler-Lagrange equation is
\begin{equation}
 D_\mu D^\mu
 \phi + m^2e^{-2 \sigma(x) } \phi =0,
\label{sfEG}
\end{equation}
whose eikonal approximation $\phi(x)\approx e^{i{\cal E}(x)}$
yields
the following equation for the phase ${\cal E}(x)$  \cite{fizn}:
\begin{equation}
e^{2 \sigma(x) }g^{\mu \nu } (x)[\partial_\mu {\cal E}(x)][ \partial_ \nu{\cal
E}(x)]   = m^2.
\label{@Eik}\end{equation}
Since $\partial_\mu {\cal E}$ is the momentum of the particle,
the replacement $\partial_\mu {\cal E}\rightarrow m\dot x_\mu$
shows that the eikonal equation (\ref{@Eik})
guarantees the constancy of the Lagrangian
\begin{equation}
 L=	 e^{ \sigma(x) }
\sqrt{g_{\mu \nu}(x)) \dot x^\mu \dot x^\nu}\equiv 1,~~~~\tau =s.
\label{@L}\end{equation}

{}From this constancy, in turn,
we easily derive the
autoparallel equation
(\ref{autopx}) with the gradient torsion
(\ref{gradtor@}),
corresponding the the
covariant
conservation law               (\ref{covLAWx})
for the energy-momentum tensor.

\section[Local Lorentz Frames versus Locally Flat Lorentz Frames]
{Local Lorentz Frames versus Locally Flat Lorentz Frames\label{MINF}}

For completeness, we clarify
in  some more detail the
difference
between the multivalued basis tetrads
$e^a{}_\mu(q)$ and the standard single-valued tetrads or vierbein fields
$h^ \alpha{}_\mu(q)$ introduced
in Eqs.~(\ref{LO}) and (\ref{4.4}), respectively.
For this
derive the minimal coupling of fields of arbitrary spin
to
gravity via the nonholonomic mapping principle,
generalizing the procedure in Subsections~\ref{Min} and
\ref{ssRiemCart}.
{}From the lesson learned in
Section \ref{Min}, we simply have to transform the flat-space
field theory
nonholonomically into the space with curvature and torsion, and this will
yield directly the correct field theory in that space.
This will produce, in particular, the covariant derivatives (\ref{covva})
needed to make the gradient terms in the Lagrange density
invariant under local translations  and local Lorentz transformations.
We introduce
a {\em fixed\/} set of Minkowski basis vectors ${\bf e}_a$ in the
flat space
and define intermediate local basis vectors
\begin{equation}
{\bf e}_ \alpha(q) =
{\bf e}_a (q) \frac{\partial x^ a }{\partial x^ \alpha }
= {\bf e}_a (q) e^ a{ }_{\alpha } (q)
     ,
\label{4.12}\end{equation}
as well as final ones
\begin{equation}
{\bf e}_\mu (q) = {\bf e}_a (q) \frac{\partial x^ \alpha }{\partial q^\mu }
     = {\bf e}_ \alpha (q) h^ \alpha {}_\mu (q).
\label{4.12}\end{equation}
An arbitrary vector field $ {\bf v}(q)$
can be expanded in either of these three bases as follows:
\begin{eqnarray}
 \!\!\!\!\!\! {\bf v} & \equiv  &
{\bf e}_a  v^a  =
	 {\bf e}_a e^a{}_\mu   v^ \mu  =
    {\bf e}_a e^a{}_ \alpha  (h^ \alpha {}_\mu   v^\mu )
	  = {\bf e}_a e ^{a \alpha } h_ \alpha {}^\mu   v_\mu
\equiv  {\bf e}_a e^a{}_ \alpha   v^ \alpha  \equiv
       {\bf e}_a e^{a \alpha }  v_ \alpha ,
\label{4.13}\end{eqnarray}
the last two expressions containing
the local Lorentz components $v_ \alpha(q)$ whose contravariant components
were
introduced
in (\ref{vcpmp})
and whose covariant derivatives are (\ref{covva}).
By changing the basis in the derivatives $\partial _a v_b$ we find
that that the
spin connection
$\Gamma\rms _{ \alpha  \beta }{}^ \gamma(q) $
is expressed
in terms of
$e^a{}_ \beta (q)$
in the same way as
the affine connection
was in (\ref{connection0}):
\begin{equation}
    \Gamma\rms_{ \alpha  \beta }{}^ \gamma
     \equiv e_a{} ^ \gamma \partial _ \alpha e ^a{}_ \beta  =
	      -e^a{}_ \beta \partial _ \alpha e_a{}^ \gamma .
\label{4.17}\end{equation}
Written out
 in terms of $e^a{}_\mu $ and $h_a{}^\mu $,
it reads
\begin{eqnarray}
   \Gamma _{ \alpha  \beta } {}^ \gamma  & = & e_a{}^ \lambda
      h^ \gamma {}_ \lambda  h_ \alpha {}^\mu  \partial _\mu
	     (e^a{}_ \nu  h_ \beta {}^ \nu )
\nonumber \\&=&   h^ \gamma {}_ \lambda h_ \alpha {}^\mu
	    h_ \beta {}^ \nu  \Gamma _{\mu  \nu }{}^ \lambda
	    + h^ \gamma {}_ \lambda  h_ \alpha {}^\mu
	      \delta ^ \lambda {}_ \nu  \partial _\mu
	     h_ \beta {}^ \nu
       =  h^ \gamma {}_ \lambda  h_ \alpha {}^\mu  h_ \beta  {}^ \nu
		( \Gamma {}_{\mu  \nu} {}^ \lambda + h^ \delta
		{}_ \nu \partial _\mu h_ \delta  {}^ \lambda )
		      \nonumber\\
      & = & h^ \gamma {}_ \lambda  h_ \alpha {}^\mu
		 h_ \beta {}^ \nu ( \Gamma _{\mu  \nu }{}^ \lambda
		       - \stackrel{h}{ \Gamma }_{\mu  \nu }{}^ \lambda
			  ),
\label{4.18}\end{eqnarray}
 where $\stackrel{h}{ \Gamma }_{\mu  \nu }{}^ \lambda  $
  is defined  in terms of $h$ in the same way
as $  \Gamma _{\mu  \nu }{}^ \lambda $
  is defined in terms of $e$ in (\ref{connection0}):
\begin{equation}
\label{connection1}
 \mathop\Gamma^h{}_{\mu \nu}{}^ \lambda(q)\equiv
 h_{ \alpha}^{\,\,\,\lambda} ( q )\partial_{\mu} \, h^{\, \alpha}_{\,\,\, \nu}
( q ) \, =-
  h^{\, \alpha}_{\,\,\, \nu} ( q ) \, \partial_{\mu} \,h_{
\alpha}^{\,\,\,\lambda} ( q )  .
\end{equation}
The second line in (\ref{4.18}) implies that $h_ \alpha {}^\mu  $
 satisfies identities like $e_a{}^\mu  $ in (\ref{2.47}):
\begin{equation}
  D_ \alpha h_ \beta {}^\mu = 0,~~~D_ \alpha  h^ \beta {}_\mu
	   = 0,
\label{4.20}\end{equation}
where
the covariant
derivative involves the connection for the Einstein index as
well as the spin connection for the local Lorentz
index
as in
Eqs.~(\ref{covderx}) and (\ref{covva}), respectively:
\begin{equation}
D_ \alpha h_ \beta{}^\mu= \partial _ \alpha
 h_ \beta{}^\mu  - \Gamma_{ \alpha \beta}{}^ \gamma h_  \gamma{}^\mu
+h_{ \alpha}{}^  \kappa\Gamma_ { \kappa \nu}{}^ \mu   h_ \beta{}^  \nu.
\label{x}\end{equation}

Since
 the metric is obtained from the tetrads $h^ \alpha{}_\mu$
by means of the relation [see also
(\ref{8.7a@})]
\begin{equation}
  g_{\mu  \nu } (q) = e^a{}_\mu  (q)
	   e^b{}_ \nu  (q) \eta_{ab}
	       \equiv h^ \alpha {}_\mu  (q)
		      h ^ \beta {}_ \nu (q) \eta_{ \alpha  \beta },
\label{4.21}\end{equation}
the right-hand side of
(\ref{connection1})
 can be rewritten
as in
(\ref{10.26}) replacing
$e_a{}^\mu$ by
 $h_\alpha{}^\mu$,
so that we  obtain
 a decomposition
completely analogous to (\ref{2.78})
for the affine  connection:
\begin{eqnarray}
\mathop \Gamma^h{} _{\mu \nu }{}^ \lambda
     & = &
 {\bar \Gamma}{} _{\mu \nu }{}^ \lambda+
\mathop {K}^h{} _{\mu \nu }{}^ \lambda
\label{4.19}\end{eqnarray}
where $ {\bar \Gamma}{} _{\mu \nu }{}^ \lambda$
is the Riemann connection
and
\begin{eqnarray}
   \mathop {K}^h{}_{\mu  \nu}{}^ \lambda     =
	    \mathop {S}^h{} _{\mu  \nu }{}^ \lambda  -
	       \mathop {S}^h{}_ \nu {}^ \lambda {}_\mu
	      +   \mathop {S}^h{}^ \lambda {}_{\mu  \nu }
\label{4.24a}\end{eqnarray}
an analog of the contortion tensor (\ref{contorttenb}).
The tensor  $   \displaystyle  \mathop {S}^h{}_{\mu  \nu }{}^ \lambda $
is the antisymmetric part of $    \displaystyle  \mathop { \Gamma}^h{}_{\mu
\nu }{}^ \lambda $,
and comparison with (\ref{AN}) and (\ref{KO10}) shows
that it yields the
object of anholonomy
(\ref{AN}) by a simple transformation of its indices:
\begin{equation}
\Omega_{ \alpha \beta}^{\,\,\,\,\,\, \gamma} =
\frac{1}{2}[h_ \alpha {}^\mu  h_ \beta {}^ \nu
  \partial _\mu {} h^ \gamma {}_ \nu
-(\mu\leftrightarrow   \nu)]=
 h^ \gamma {}_ \lambda  h_ \alpha {}^\mu
		h_ \beta {}^ \nu
		 \mathop S^{h}{}_{\mu  \nu }{} ^ \lambda.
\label{x}\end{equation}
If we now insert the decompositions
(\ref{4.19}) and (\ref{2.78}) into (\ref{4.18}),
the
Riemann connections
in
$ { \Gamma}_{\mu  \nu}{}^ \lambda $ and
$\displaystyle  \mathop { \Gamma}^h{}_{\mu  \nu}{}^ \lambda $
cancel each other,
and we obtain
\begin{eqnarray}
  \Gamma\rms _{ \alpha  \beta }{}^ \gamma
=  h^ \gamma {}_  \lambda  h_ \alpha {}^\mu  h_ \beta {}^ \nu
	   (K_{\mu  \nu }{}^ \lambda  - \stackrel{h}{K}_{\mu  \nu }{}^ \lambda
	    ),
\label{4.22}\end{eqnarray}
which is precisely the
spin  connection
(\ref{4.18aa}).

Note that due to (\ref{4.24a}) and the antisymmetry
of $ \stackrel{h}{S}_{\mu  \nu }{}^ \lambda$ in the first two indices,
 the
tensor
$\stackrel{h}{K}_{\mu  \nu \lambda}$
 is antisymmetric in the last two indices,
just as the contortion tensor $K_{\mu \nu \lambda}$
in Eq.~(\ref{antisK@}), so that also the
spin connection shares this antisymmetry.

It will be convenient to use $h^ \alpha {}_\mu , h_ \alpha {}^\mu  $
freely for changing indices $ \alpha $ into $\mu $,
for instance
\begin{eqnarray}  \label{4.25}
   K_{\alpha  \beta }{}^ \gamma  & \equiv  & h^ \gamma {}_ \lambda
	    h_ \alpha {}^\mu  h_ \beta {}^ \nu  K_{\mu  \nu }{}^ \lambda ,
   \\
  \stackrel{h}{ K}_{ \alpha  \beta }{}^ \gamma
      & = & h^ \gamma {}_ \lambda
	    h_ \alpha {}^\mu  h_ \beta {}^ \nu
             \stackrel{h}{K}_{\mu  \nu }{}^ \lambda .
\label{4.26}\end{eqnarray}
Observe that by introducing the tetrad fields
 $h^  \alpha{}_\mu
 ,
 h_ \alpha{}^\mu $,
the description of gravity effects  in terms of the 10
metric components $g_{\mu  \nu }$ and the 24 torsion
 components $K_{\mu  \nu }{}^ \lambda  $ has
 been replaced by 16 components $h_{\mu }{}^ \alpha $
 and the 24 $K_{\mu  \nu }{}^ \lambda $.
The additional 6 components are redundant, and this is the source of the
local Lorentz invariance of the theory which arises in addition to the
local translations of Einstein's general coordinate transformations.
Relation (\ref{4.21}) implies that
the tetrad fields $  h^  \alpha{}_\mu$
can be considered as another
 ``square root'' of the metric $g_{\mu  \nu }$ different from
 $e^{a}{}_{\mu }$. Obviously, such a ``square root''   is defined
 only up to an arbitrary local Lorentz transformation which
 accounts for the six additional degrees of freedom of the
 $h_ \alpha {}^\mu (q)$  with respect to the $g_{\mu  \nu }  (q)$
description. These six degrees of freedom
characterize the
local Lorentz transformations $\Lambda_ \alpha{}^a (q)\equiv
e_ \alpha{}^a(q) $
by which the intermediate
 basis vectors ${\bf e}_ \alpha(q)$ differ from
the fixed  orthonormal basis vectors ${\bf e}_a$:
\begin{equation}
 {\bf e}_ \alpha(q)\equiv
  {\bf e}_a.
\Lambda^a{}_ \alpha    .
\label{x}\end{equation}
The Lorentz properties
of  $\Lambda^a{}_ \alpha$ follow from the fact that
the basis vectors
${\bf e}_ \alpha(q)$
have the same Minkowskian scalar products as ${\bf e}_a$:
\begin{equation}
{\bf e}_ \alpha(q){\bf e}_  \beta(q)= \eta_{ \alpha \beta}.
\label{x}\end{equation}
As a consequence, the matrix  $\Lambda_ \alpha{}^a(q)$
satisfies [see (\ref{Loren@})]
\begin{equation}
 \eta_{ab}   \Lambda^a{}_ \alpha(q)
   \Lambda^b{}_  \beta(q)= \eta_{ \alpha \beta}.
\label{4.30}\end{equation}
 which is the defining equation of Lorentz transformations.
Since
these local Lorentz transformations
bring the good vierbein functions $h_ \alpha{}^\mu(q)$
satisfying (\ref{4.7})
to the multivalued functions $e_a{}^\mu(q)$
with noncommuting derivatives (\ref{RC}), the local Lorentz transformations
are also multivalued:
\begin{equation}
\left( \partial _\mu\partial _ \nu-
\partial _\nu\partial _ \mu \right)
 \Lambda_ a{}^{ \alpha} ( q )\neq 0.
\label{RC'}\end{equation}
 Note that, due to (\ref{4.17}), both  $e^a{}_ \alpha (q)$ and $
 e_a{}^ \alpha  (q)$ satisfy identities like
(\ref{2.47}), (\ref{4.20}),
\begin{equation}
  D_ \alpha  e^a{}_ \beta  = 0,~~~ D_ \alpha  e_a{}^ \beta = 0.
\label{4.32}\end{equation}
All derivatives in a reparametrization-invariant
theory can be recast in terms of nonholonomic coordinates
  $dx^ \alpha $, in which form it becomes invariant
under both local translations and local Lorentz transformations.
 Since the metric is $\eta ^{ \alpha  \beta }$,
  the invariant actions have the same form as those
 in a flat space except that the derivatives are replaced
 by covariant ones:
\begin{equation}
     \partial _ \alpha   v_ \beta  \rightarrow
       D_ \alpha  v_ \beta  = \partial _ \alpha
        v_ \beta   -  \Gamma\rms _{ \alpha  \beta }
    {}^ \gamma  v_ \gamma .
\label{x}\end{equation}

Nonholonomic volume elements are related
to true ones
by $d^4x=d^4 q \sqrt{g(q)}$.
An invariant action of a massless vector field is, for example,
\begin{equation}
     {\cal A} = \int d^4 x D_\alpha
	    v_ \beta  (q ) D^ \alpha v {}^ \beta
	 (q )
\label{4.33}\end{equation}
 It is the nonholonomic form of a generally covariant action. As we said
 in the beginning, the specification of space points
  must be made with the $q ^\mu $ coordinates.  For this
 reason the action is preferably written as
\begin{equation}
  {\cal A} = \int d^4 q    \sqrt{-g}  D_ \alpha
	        v_ \beta  (q ) D^ \alpha
	       v ^ \beta  (q ).
\label{4.34}\end{equation}
Under a general coordinate transformation
(\ref{coortrf@a}) {\em  \`a la\/} Einstein,
 $dq^\mu  \rightarrow dq'{} ^{\mu '} = dq^\mu   \alpha _\mu {}^{\mu '}$,
 the indices $ \alpha $ are unchanged. For instance, $h_ \alpha {}^\mu $
itself transforms as
\begin{equation}
  h_ \alpha {}^\mu  (q) \mathop{\rightarrow}_{E} h_ \alpha
     {}^{\mu '}(q') = h_ \alpha {}^\mu  (q)
        \alpha _\mu {}^{\mu '}(q).
\label{4.35}\end{equation}
 Vectors and tensors with indices $ \alpha , \beta,  \gamma,\dots \,$
 experience only changes of their arguments $q \rightarrow q
 + \xi$, so that their infinitesimal substantial changes
 are
\begin{eqnarray} \label{4.36}
     \delta _E  v_ \alpha  (q) & = & \xi^ \lambda
    \partial _ \lambda   v_ \alpha (q),\\
        \delta _E D_ \alpha    v_ \beta  (q)
      & = & \xi^ \lambda  \partial _ \lambda
	     D_ \alpha  v_ \beta (q).
\label{4.37}\end{eqnarray}
The freedom in choosing $h_ \alpha {}^\mu  (q)$ up to a local
Lorentz transformation when taking the ``square root''
 of $g_{\mu  \nu } (q)$ in (\ref{4.21}) implies
 that the theory should be invariant under
\begin{eqnarray}  \label{4.38}
  \delta _L dx^ \alpha  & = &  \omega ^ \alpha {}_ \beta
	(q) dx^ \beta ,      \\
    \delta _L h_ \alpha {}^\mu  (q) & = &  \omega _ \alpha {}^ \beta
		     (q) h_ \beta {}^\mu  (q).
\label{4.39}\end{eqnarray}
Here  $ \omega _ \alpha {}^{ \alpha '} (q)$ are the local
  versions of the infinitesimal Lorentz parameters.

Indeed the action (\ref{4.34}) is automatically invariant
 if every index $ \alpha $ is transformed accordingly:
\begin{eqnarray}\label{4.40}
      \delta _L  v _ \alpha (q)&  =&   \omega _ \alpha{} ^{ \alpha '}(q)
	 v_ {\alpha '} (q),         \\
     \delta _L D_ \alpha   v_ \beta  (q)
&=&  \omega _ \alpha {}^{ \alpha '}
       (q) D_ {\alpha '}  v_ \beta (q) +  \omega _ \beta {}^{ \beta '}
       (q) D_ \alpha   v_{ \beta '} (q).
\label{4.40'}\end{eqnarray}
 The variables $q^\mu $ are unchanged since
   (\ref{4.38}) refers only to the differentials
 $dx^ \alpha $ and leaves $dq^\mu $ unchanged.

It is useful to verify explicitly how the covariant
   derivatives guarantee local Lorentz invariance. From
(\ref{4.40})
we see that
the derivative $\partial _ \alpha   v_ \beta $
 transforms like
\begin{eqnarray}
    \delta _L \partial _ \alpha  v_\beta  & = &
    ( \delta _L \partial _ \alpha )  v _ \beta
     + \partial _ \alpha  ( \delta _L v_ \beta )\nonumber\\
     & = &
     \omega _ \alpha {}^{ \alpha '} \partial _ {\alpha '}
        v_ \beta  + \partial _ \alpha  ( \omega _ \beta {}^{ \beta '}
        v_{ \beta '})\nonumber\\
    & = &  \omega _ \alpha{}^{ \alpha '} \partial _{ \alpha '}
	 v_ \beta  +  \omega _ \beta {}^{ \beta '}
	\partial _ \alpha   v_{ \beta '} +
      (\partial _ \alpha   \omega _ \beta {}^{ \beta '})
	  v_{ \beta '}.
\label{4.42}\end{eqnarray}
The spin connection behaves as follows: Due to the factors
 $h_ \lambda {}^ \gamma  h_ \alpha {}^\mu  h_ \beta {}^ \nu $
 in (\ref{4.19}), the first piece of $ \Gamma\rms _{ \alpha  \beta }
   {}^ \gamma $, call it $  \Gamma\rms{}'_{\!\! \alpha  \beta }{}^ \gamma $,
transforms like a local Lorentz tensor:
\begin{equation}
  \delta _L   \Gamma\rms{} '_{\!\! \alpha  \beta }{}^ \gamma
      =   \omega _ \alpha {}^{ \alpha '}
       \Gamma\rms{} '_{ \alpha ' \beta } +
	 \omega _ \beta {}^{ \beta '}   \Gamma\rms{} '
	    _{ \alpha  \beta '}{}^ \gamma
	     +  \omega ^ \gamma {} _{ \gamma '}
	     \Gamma\rms{} '_{ \alpha  \beta }{}^ {\gamma '}.
\label{4.43}\end{equation}
 But from the second piece $ \stackrel{h}{ \Gamma }_{\mu  \nu }{}^ \lambda $
       there is a nontensorial derivative contribution
\begin{eqnarray}
        \delta _L \stackrel{h}{ \Gamma }_{\mu  \nu }{}^ \lambda
   & = &   ( \delta h_ \delta {} ^ \lambda ) \partial ^ \delta {}_ \nu
	   + h_ \delta {}^ \lambda \partial _\mu ( \delta h^ \delta {}_ \nu )
     \nonumber\\
     & = &  \omega _ \delta {}^{ \delta '} h_ {\delta '}
	       {}^  \lambda  \partial _\mu  h^ \delta {}_ \nu
	       + h_ \delta {}^ \lambda  \partial _\mu
		   ( \omega ^ \delta {}_{ \delta '} h^{ \delta '}
		   {}_ \nu )
	\nonumber\\
    & = &  \omega _ \delta {}^{ \delta '}h_ {\delta '}
	     {}^ \lambda  \partial _\mu
	      h^ \delta {}_ \nu  +   \omega ^ \delta {}_{ \delta '}
		h_ \delta {}^ \lambda \partial _\mu {} ^ \lambda
		\partial _\mu  h^{ \delta '}{}_ \nu
	     + \partial _\mu   \omega ^ \delta {}_{ \delta '}
		(h_ \delta {}^ \lambda  h^{ \delta '}
		  {}_ \nu)  \nonumber\\
     & = &  \partial _\mu   \omega ^ \delta {}_{ \delta '}
	      h^{ \delta '}{}_ \nu   = - \partial _\mu
	        \omega _ {\delta '} {}^ \delta h_ \delta
		{}^ \lambda  h^{ \delta '}{}_ \nu      ,
\label{4.44}\end{eqnarray}
the cancellation in the third line being due to the
antisymmetry of $  \omega _ \delta {}^{ \delta '} =
 - \omega ^{ \delta '}{}_ \delta $.
  Thus we arrive at
\begin{eqnarray}
   \delta _L \stackrel{h}{ \Gamma } _{\mu  \nu }{}^ \lambda
     &=&  \partial _\mu   \omega ^ \delta {}_{ \delta '}
	 \,h_ \delta {}^ \lambda  h^{ \delta '}{}_{ \nu '}
       , ~~~~~                \nonumber \\
	  \delta _L  \Gamma\rms_{ \alpha  \beta }{}^ \gamma
	   &=& \delta _{L_0}  \Gamma\rms _{ \alpha  \beta }{}
       ^ \gamma  + \partial _ \alpha  \omega _ \beta {}^ \gamma ,
\label{lastterm}\end{eqnarray}
   where $ \delta _{L_0}$ abbreviates
the three tensor-like
 transformed terms
corresponding to
 (\ref{4.43}):
\begin{equation}
  \delta _{L_0}   \Gamma\rms{} _{\!\! \alpha  \beta }{}^ \gamma
      =   \omega _ \alpha {}^{ \alpha '}
       \Gamma\rms{} _{ \alpha ' \beta } +
	 \omega _ \beta {}^{ \beta '}   \Gamma\rms{}
	    _{ \alpha  \beta '}{}^ \gamma
	     +  \omega ^ \gamma {} _{ \gamma '}
	     \Gamma\rms{} _{ \alpha  \beta }{}^ {\gamma '}.
\label{4.43a}\end{equation}
   The last term in (\ref{lastterm}) is precisely what is required to cancel
     the last nontensorial piece of (\ref{4.42}), when
    transforming $D_ \alpha   v _ \beta $,
       so that we indeed obtain the covariant transformation
 law (\ref{4.40'}).

Armed with these transformation laws it is now
     straightforward to introduce spinor fields into a gravity
   theory. In a local    inertial frame
 (such as a freely falling elevator), a spinor field $\psi(q)$ transforms
    like
\begin{equation}
      \delta _L \psi (q) = -\frac{i}{2}  \omega ^{ \alpha  \beta }
	   (q) \Sigma_{ \alpha  \beta } \psi (q),
\label{4.46}\end{equation}
 when locally changing from one such frame of reference to another
   Lorentz transformed one. The matrices  $\Sigma_{ \alpha  \beta }$
      represent the local Lorentz group on the $\psi$-fields.
  They are antisymmetric in their indices, and have the nonzero
commutation relations
\begin{equation}
    \left[  \Sigma _{ \alpha  \beta } ,  \Sigma _{ \alpha  \gamma }
   \right] = - i\eta_{ \alpha  \alpha }  \Sigma _{ \beta  \gamma }~~~~~{\rm
no~sum~over }~\alpha{}.
\label{4.47}\end{equation}
  For vector representations, they are given explicitly by
\begin{eqnarray}
   \left( \Sigma _{ \alpha  \beta }\right)_{ \alpha ' \beta '} & = &
     i\left( \eta_{ \alpha  \alpha '} \eta_{ \beta  \beta '} -
      \left( \alpha \leftrightarrow  \beta \right)\right).
\label{4.48}\end{eqnarray}
 Replacing
$\psi$ by $v$ and writing out the Lorentz indices,
Eq.~(\ref{4.46}) reduces to (\ref{4.39}):
\begin{equation}
    \delta _L  v _ \alpha  = - \frac{i}{2}  \omega ^{ \gamma  \delta }
     \,  i \left(\eta_{ \gamma  \alpha }  \eta_{\delta  \beta}
	   - (  \gamma  \leftrightarrow  \delta )\right) v^ \beta
	 =  \omega _ \alpha {}^ \beta   v _ \beta .
\label{4.49}\end{equation}
    For Dirac fields,  the representation matrices
$\Sigma_{ \alpha  \beta  } $ are expressed
   products  of Dirac matrices:
\begin{equation}
  \Sigma _{ \alpha  \beta } = \frac{i}{4} \left[  \gamma _ \alpha ,
	 \gamma _ \beta \right] .
\label{4.50}\end{equation}
 The derivative of $ \psi $ changes as
\begin{eqnarray}
   \delta _L \partial _ \alpha  \psi & = &   \omega _ \alpha {}^{ \alpha '}
      \partial _{ \alpha '} \psi + \partial _ \alpha
	    \delta _L \psi
 =   \omega _ \alpha {}^{ \alpha '} \partial _{ \alpha '}
	 \psi - \frac{i}{2} \partial _ \alpha
	\left( \omega ^{ \beta  \gamma }\Sigma_{ \beta  \gamma }
	    \right)\psi \nonumber \\
    & = &  \omega _ \alpha {}^{ \alpha '}  \partial_{ \alpha '}
	 \psi - \frac{i}{2}   \omega ^{ \beta  \gamma }
	\Sigma_{ \beta  \gamma } \partial _ \alpha \psi
	     - \frac{i}{2} \left( \partial _ \alpha
	  \omega ^{ \beta  \gamma }\right) \Sigma_{ \beta  \gamma }
	  \psi .
\label{4.51}\end{eqnarray}
The first two terms describe the normal Lorentz
behavior of $ \partial _ \alpha  \psi$. The last
   term accounts for the $q$-dependence of the angles
 $ \omega ^ { \beta  \gamma } (q)$.
   It does not appear if we go over to the
 covariant derivative
\begin{equation}
  D_  \alpha  \psi (q) \equiv  \partial _ \alpha \psi (q)
	     + \frac{i}{2} \Gamma\rms _{ \alpha  \beta }
     {}^ \gamma  \,\Sigma^ \beta {}_ \gamma  \psi (q).
\label{4.52}\end{equation}
 Indeed, when forming
\begin{equation}
   \delta _L\frac{i}{2}   \Gamma\rms _{ \alpha  \beta }
       {}^ \gamma  \Sigma^ \beta {}_ \gamma \psi
	 (q)
 =  \frac{i}{2}  \delta _L \Gamma\rms_ {\alpha  \beta }
	{}^ \gamma   \Sigma^ \beta {}_ \gamma  \psi
	+ \frac{i}{2} \Gamma\rms_{ \alpha  \beta }{}^ \gamma
       \Sigma^ \beta {}_ \gamma   \delta _L \psi,
\label{4.55}\end{equation}
we obtain two terms. The first of these corresponds
to a tensor transformation law,
 being  equal to
\begin{equation}
   \delta _{L_0} \frac{i}{2}   \Gamma\rms _{ \alpha  \beta }{}^ \gamma
	\Sigma^ \beta {}_ \gamma  \psi = - \frac{i}{2}
        \omega ^{ \s \tau }  \Sigma _{ \s \tau }
	 \left(\frac{i}{2} \Gamma\rms_{ \alpha  \beta }{ }^ \gamma
	  \Sigma ^{ \beta }{}_{ \gamma } \psi\right).
\label{4.54}\end{equation}
It is obtained by inserting into (\ref{4.55})
the equations (\ref{lastterm}) and (\ref{4.46}), and
 applying the commutation rule (\ref{4.47}). The second, nontensorial
 term arises from $\partial _ \alpha   \omega _ \beta {}^ \gamma $
in (\ref{lastterm}):
\begin{equation}
   \frac{i}{2}  \partial _ \alpha   \omega _ \beta {}^ \gamma
	 \Sigma^ \beta {}_ \gamma \psi,
\label{4.56}\end{equation}
 and cancels the last term in (\ref{4.51}).
  Thus $D_ \alpha \psi $ behaves like
\begin{equation}
   \delta _L D_ \alpha \psi =  \omega _ \alpha {}^{ \alpha '}
       (q) D_{ \alpha '} \psi - \frac{i}{2}  \omega ^{ \beta  \gamma }
       (q) \Sigma_{ \beta  \gamma } D_ \alpha \psi
\label{4.57}\end{equation}
 and represents, therefore, a proper covariant derivative which
 generalizes the standard Lorentz transformation behavior
    to the case of local transformations $ \omega _ \alpha {}^ \beta (q)$.

We can now immediately construct the spin$-\frac{1}{2}$ action
 for a Dirac particle in a gravity field
\begin{eqnarray}
   {\cal A}_m [h, K, \psi] & = & \frac{1}{2}
   \int d^4q  \sqrt{-g}  \bar \psi ( \gamma ^ \alpha D_ \alpha
	      -m) \psi (q) + {\rm h.c.} \\
	& \equiv  & \frac{1}{2}  \int d^4q  \sqrt{-g} \bar \psi
	    \gamma ^a \left(\partial _  \alpha  +
	  \frac{i}{2}    \Gamma _{ \alpha  \beta }{}^ \gamma
        \Sigma ^ \beta {}_ \gamma \right)\psi (q)
	    + {\rm h.c.}
\label{4.58}\end{eqnarray}
If we wish, we may change the derivatives from
 $ \partial _ \alpha $ to $\partial _\mu $  by using
 $\partial _ \alpha  = h_ \alpha {}^\mu  \partial _\mu $
 and $ \gamma ^ \alpha  = h^ \alpha {}_\mu  (q)  \gamma ^\mu (q)$
 so that $ \Sigma _{ \alpha  \beta } (q) = (i/4) [ \gamma _ \alpha (q)$,
 $  \gamma _ \beta (q)]$ and, expressing
  $\Gamma\rms_{ \alpha  \beta }{}^ \gamma $ by
   (\ref{4.19}), the action reads
\begin{eqnarray}
    {\cal A}_m [h, K, \psi] & = & \frac{1}{2} \int d^4q  \sqrt{-g}
	\bar\psi (q)
 \left\{  \gamma ^\mu (q) \left[ \partial _\mu
	   + \frac{i}{2} (K_{\mu  \nu }{}^ \lambda  -
	       \stackrel{h}{K}_{\mu  \nu }{}^ \lambda  )
            \Sigma ^ \nu {}_ \lambda
		 \right] -m\right\} \psi (q) + {\rm h.c.}
\label{4.59}\end{eqnarray}
  Due to the $x$ dependence
  of $  \gamma ^\mu $ and $  \Sigma ^{\mu  \nu }{} _ \lambda $,
   this form is, however, not very convenient
for calculatons.
This minimal type of coupling
bewteen spin and gravity
can easily be generalized to higher-spin fields if desired.
%

\section{Conclusion and Outlook}

We have pointed out that
the nonholonomic mapping principle,
supplies us with a perfect tool
for deriving
physical laws in spaces with curvature and torsion
by means of
multivalued coordinate transformations.
As mentioned earlier, there are other evidences
for the correctness of this
principle, one of them being deduced
 from the solution of the path
integral of the Coulomb system.
As a particular result we have derived from
this principle a new variational procedure
for Hamilton's action principle
which has led to the surprising result that
trajectories of spinless point particles
are autoparallels,
not
geodesics as commonly believed \cite{Hehl1,Hehl2,Hehl3}.

Since spinless particles experience
a torsion force, we expect them to be also the
source of torsion.
Under the assumption that torsion propagates we may add
to the gravitational action
a gradient
term involving the torsion,
and
will be able to derive
deviations from Einstein's gravity
effects (deflection of light, gravitational red shift, perihel precession of
mercur)
\cite{Sabbata}.
The experimental smallness of such deviations
will provide us with limits on the
coupling constant in front of the gradient term.

Up to now, it is doubtful
 whether the minimally coupled field theories
described in Section \ref{MINF} are physically correct.
A proper construction
will require a
field version
of the new variation formula
(\ref{198@}) caused by the closure failure.
This is in fact necessary
and nontrivial even for the Schr\"odinger action (\ref{LD@}).
In the semiclassical limit (eikonal approximation),
Schr\"odinger wave functions have to describe
autoparallel particle trajectories.
Except for the case of gradient torsion discussed in Section~\ref{@RELFGT}
 it is unknown how to achieve this.

It should be pointed out that conventional gravity
in which the torsion
field is coupled only to spin has the
severe consistency problem
that spin  and orbital angular momentum
are distinguishable which contradicts the
universality principle of these in elementary particle theory \cite{UNI}.
Only a modified trosion coupling which leads to autoparallel trajectories
has a chance of being consistent.

Another problem with
conventional field theory of gravity theory with torsion
arises in the context of
electroweak gauge theories,
In the theory,
massless and massive
vector bosons are coupled differently to torsion.
This, however, is incompatible with the Higgs mechanism
according to which the vector meson masses
arise from a spontanoeus symmetry breakdown
of a scalar field theory.
As shown recently in \cite{HIGGS},
only a Higgs field
whose particles run along
autoparallel trajectories can
remove this problem.
This problem should be
studied in more detail
using the field thery of gravity with gradient torsion
in Ref.~\cite{hojman}
where massless vector mesons
do couple to torsion without violating gauge invariance.

The motion of particles with spin (see e.g.\
Ref.\ \cite{Fuchs}) will also be altered by a
generlization of the arguments
given above.

\section{Acknowledgement}

I am grateful to Dr. A. Pelster
for many
stimulating discussions,
and to Drs. G. Barnich, H. von Borzeskowski, C.
Maulbetsch, and S.V. Shabanov useful comments.

\def\thesection{Appendix A}
\setcounter{equation}{0}
\def\theequation{A\arabic{equation}}
\section{Partial Integration in Spaces with Curvature and Torsion}
Take any tensors $U^{\mu \dots \nu }, V_{\dots
 \nu \dots}$ and consider an invariant volume  integral
\begin{eqnarray}
  \int dx \sqrt{ -g} U^{\mu \dots \nu \dots} D_\mu V_{\dots \nu \dots}.
\label{5.46}\end{eqnarray}
A partial integration gives
\begin{eqnarray}
  - \int dx \left[ \partial _\mu
           \sqrt{ -g} U^{\mu \dots \nu \dots} V_{\dots  \nu  \dots}
           +\sum _{i} U^{\mu \dots  \nu _i \dots} \Gamma _{\mu \nu
_i}{}^{\lambda _i}
           V_{\dots \lambda _i \dots}\right]+{\rm surface~terms},
\label{5.47}\end{eqnarray}
where the sum over $i$
 runs over all
indices
 of $V_{\dots \lambda _i\dots }$, linking them via the affine connection
with the corresponding
indices of $U^{\mu \dots \nu _i \dots}$.
Now we use the relation
\begin{eqnarray}
   \partial _\mu \sqrt{ -g} =
 \sqrt{ -g} \bar\Gamma _{\mu \kappa }{}^\kappa
= \sqrt{ -g} \Gamma _{\mu \kappa }{}^\kappa
       = \sqrt{ -g} \left( 2S_\mu + \Gamma _{\kappa \mu }{}^\kappa \right)
\label{@}\end{eqnarray}
and (\ref{5.47}) becomes
\begin{eqnarray}
  &   &
 - \int dx \sqrt{ -g}
          \left[( \partial _\mu  U^{\mu \dots \lambda _i \dots}
 -\Gamma _{\kappa \mu }{}^\kappa U^{\mu \dots \lambda _i \dots}
      + \sum _{i} \Gamma _{\mu \nu _i}{}^{\lambda _i  }
            U^{\mu \dots \nu _i \dots }
 ) V_{\dots \lambda _i \dots }
\right .
           \nonumber \\
         &  & \left . ~~~~~~~~  ~~~~~~
+ 2 S_\mu \sum_i U^{\mu \dots \lambda _i \dots}
                V_{\dots \lambda _i \dots} \right] +{\rm surface~terms}.
 \label{q}\end{eqnarray}
Now, the terms in parentheses are just the covariant
 derivative of $U^{\mu \dots \nu_i \dots}$ such that we
 arrive at the rule
\begin{eqnarray}
 \int dx \sqrt{ -g} U^{\mu \dots \nu \dots} D_\mu V_{\dots \nu \dots}
= -\int dx \sqrt{ -g} D_\mu {}^* U^{\mu \dots \nu  \dots}
V_{\dots \nu  \dots}+{\rm surface~terms},
\label{5.48}\end{eqnarray}
where $D^{*}_\mu $ is defined as
\begin{eqnarray}
   D_\mu ^* \equiv D_\mu + 2S_\mu,
\label{5.45}\end{eqnarray}
abbreviating
\begin{eqnarray}
      S_\kappa  \equiv  S_{\kappa \lambda }{}^\lambda , ~~
                  S^\kappa \equiv S^\kappa {}_\lambda {}^\lambda.
\label{5.37}\end{eqnarray}

It is easy to show that the operators $D_ \mu $ and $D^*_ \mu $
can be interchanged in the rule (\ref{5.48}), i.e.,
there is also the rule
\begin{eqnarray}
 \int dx \sqrt{ -g}   V_{\dots \nu  \dots}
 D_\mu {} U^{\mu \dots \nu  \dots} =
         -\int dx \sqrt{ -g} U^{\mu \dots \nu \dots}
D^*_\mu V_{\dots \nu \dots}
      +{\rm surface~terms}.
\label{5.48c}\end{eqnarray}
For the particular case that
$V_{\dots \nu  \dots}
$  is equal to $1$,
the second rule yields
\begin{eqnarray}
 \int dx \sqrt{ -g}
 D_\mu {} U^{\mu } =
         -\int dx \sqrt{ -g}\,2 S_ \mu U^{\mu}
      +{\rm surface~terms} .
\label{5.48cc}\end{eqnarray}

\end{document}